\documentclass[final,5p,times]{elsarticle}

\usepackage{graphicx}
\usepackage{GRAF}

\newcommand{\degree}{\ensuremath{^\circ}}
\newcommand{\CesrTA}{\textsc{CesrTA}}

\def\gappeq{\mathrel{ \rlap{\raise.5ex\hbox{$>$}}
                      {\lower.5ex\hbox{$\sim$}}  } }
\def\lappeq{\mathrel{ \rlap{\raise.5ex\hbox{$<$}}
                      {\lower.5ex\hbox{$\sim$}}  } }

\makeatletter
\def\appendixname{Appendix}
\renewcommand{\appendix}{\par
  \setcounter{section}{0}%
  \setcounter{subsection}{0}%
  \gdef\thesection{\appendixname~\@Alph\c@section}%
  \gdef\thesubsection{\@Alph\c@section.\@arabic\c@subsection}
\renewcommand\section{%
           \addtocontents{toc}{\string\let\string\numberline\string\tmptocnumberline}{}{}%
           \@startsection {section}{1}{\z@}%
           {18\p@ \@plus 6\p@ \@minus 3\p@}%
           {9\p@ \@plus 6\p@ \@minus 3\p@}%
           {\normalsize\bfseries\boldmath}}%
\renewcommand\subsection{%
           \addtocontents{toc}{\string\let\string\numberline\string\regnumberline}{}{}%
           \@startsection{subsection}{2}{\z@}%
           {12\p@ \@plus 6\p@ \@minus 3\p@}%
           {3\p@ \@plus 6\p@ \@minus 3\p@}%
           {\normalfont\normalsize\itshape}}%
}
\def\regnumberline#1{\hb@xt@\@tempdima{#1\hfil}}
  \renewcommand\bibsection{%
   \section*{\refname\@mkboth{\MakeUppercase{\refname}}{\MakeUppercase{\refname}}}%
   \addcontentsline{toc}{section}{\refname}%
  }%
\long\def\@makecaption#1#2{%
  \vskip\abovecaptionskip\footnotesize
  \sbox\@tempboxa{{\bf #1.} #2}%
  \ifdim \wd\@tempboxa >\hsize
    {\bf #1.} #2\par
  \else
    \global \@minipagefalse
    \hb@xt@\hsize{\hfil\box\@tempboxa\hfil}%
  \fi
  \vskip\belowcaptionskip}
\long\def\@address#1{\g@addto@macro\elsauthors{%
    \def\baselinestretch{1}%
    \addsep\normalsize\itshape#1\def\addsep{\par\vskip6pt}%
    \def\authorsep{\par\vskip8pt}}}
\def\ps@pprintTitle{%
     \let\@oddhead\@empty
     \let\@evenhead\@empty
     \def\@oddfoot{\ifnopreprintline\relax\else\footnotesize\itshape
     \hfill\@date\fi}%
     \let\@evenfoot\@oddfoot}
\makeatother

\biboptions{sort&compress}

\journal{arXiv}

\usepackage{ifthen}

\usepackage{array}

\RequirePackage[colorlinks=true,hyperfootnotes=false,bookmarksnumbered=true]{hyperref}
\RequirePackage[all]{hypcap}
\RequirePackage[capitalise,nameinlink,noabbrev]{cleveref}

\hypersetup{pdftitle={Report on Instrumentation \& Methods for In-Situ Measurements of SEY in an Accelerator Environment},
            pdfauthor={W. Hartung et al.},
            pdfkeywords={secondary emission, electron cloud, beam scrubbing, storage ring, damping ring}}

\renewcommand{\today}{\number\day\space \ifcase\month\or
  January\or February\or March\or April\or May\or June\or
  July\or August\or September\or October\or November\or December\fi
  \space \number\year}

\begin{document}

\begin{frontmatter}



\title{Report on Instrumentation and Methods for In-Situ Measurements
  of the Secondary Electron Yield in an Accelerator Environment}


\author{W.~H.~Hartung}
\author{D.~M.~Asner\fnref{pnnl}}
\fntext[pnnl]{Present address: Pacific Northwest National Laboratory, Richland, WA}
\author{J.~V.~Conway}
\author{C.~A.~Dennett\fnref{mit}}
\fntext[mit]{Present address: Department of Nuclear Science and Engineering, Massachusetts Institute of Technology, Cambridge, MA}
\author{S.~Greenwald}
\author{J.-S.~Kim\fnref{prince}}
\fntext[prince]{Present address: Department of Electrical Engineering, Princeton University, Princeton, NJ}
\author{Y.~Li}
\author{T.~P.~Moore}
\author{V.~Omanovic}
\author{M.~A.~Palmer\fnref{fnal}}
\fntext[fnal]{Present address: Fermi National Accelerator Laboratory, Batavia, IL}
\author{C.~R.~Strohman}

\address{Cornell Laboratory for Accelerator-based ScienceS and
  Education, Cornell University, Ithaca, New York, USA}

\begin{abstract}
The achievable beam current and beam quality of a particle accelerator
can be limited by the build-up of an electron cloud (EC) in the vacuum
chamber.  Secondary electron emission from the walls of the vacuum
chamber can contribute to the growth of the electron cloud.  An
apparatus for in-situ measurements of the secondary electron yield
(SEY) of samples in the vacuum chamber of the Cornell Electron Storage
Ring (CESR) has been developed in connection with EC studies for the
CESR Test Accelerator program (\CesrTA).  The \CesrTA{} in-situ system,
in operation since 2010, allows for SEY measurements as a function of
incident electron energy and angle on samples that are exposed to the
accelerator environment, typically 5.3~GeV counter-rotating beams of
electrons and positrons.  The system was designed for periodic
measurements to observe beam conditioning of the SEY with
discrimination between exposure to direct photons from synchrotron
radiation versus scattered photons and cloud electrons.  The SEY
chambers can be isolated from the CESR beam pipe, allowing us to
exchange samples without venting the CESR vacuum chamber.
Measurements so far have been on metal surfaces and EC-mitigation
coatings.  The goal of the SEY measurement program is to improve
predictive models for EC build-up and EC-induced beam effects.  This
report describes the \CesrTA{} in-situ SEY apparatus, the measurement
tool and techniques, and iterative improvements therein.
\end{abstract}

\end{frontmatter}

\tableofcontents
%
\newlength{\narrowwidth}%
\newlength{\middlewidth}%
\newlength{\widewidth}%
\newlength{\widestwidth}%
\newlength{\heightone}%
\setlength{\heightone}{0.3\textheight}%
\newlength{\heighttwo}%
\setlength{\heighttwo}{0.35\textheight}%
\ifthenelse{\lengthtest{\columnwidth=\textwidth}}
{\setlength{\narrowwidth}{0.6\textwidth}
\setlength{\middlewidth}{0.7\textwidth}
\setlength{\widewidth}{0.85\textwidth}
\setlength{\widestwidth}{\textwidth}
\setlength{\heightone}{0.8\heightone}
\setlength{\heighttwo}{0.8\heighttwo}}
{\setlength{\narrowwidth}{\columnwidth}
\setlength{\middlewidth}{\columnwidth}
\setlength{\widestwidth}{\columnwidth}
\setlength{\widewidth}{\columnwidth}}


\section{Introduction\label{S:intro}}

Ideally, the beams in a particle accelerator propagate through a
perfectly evacuated chamber.  In reality, the vacuum chamber contains
small amounts of residual gas, ions, and low-energy electrons.  A
number of processes can contribute to the build-up of the low-energy
electrons: synchrotron-radiated photons striking the wall of the
chamber can produce electrons by photo-emission; in the absence of
synchrotron radiation, electrons can be produced by bombardment of the
wall by the beam halo or ionisation of the residual gas by the beam.
The electron population grows if the electrons hit the wall and
produce secondary electrons with a probability greater than unity.  In
extreme cases, a large density of electrons can build up inside the
beam chamber, causing disruption of the beam, heating of the chamber
walls, and degradation of the vacuum.  This is referred to as an
``electron cloud'' (EC).

Electron cloud effects were first observed in accelerators in the
1960s \cite{ECLOUD04:9to13}.  Positively charged beams are typically
more prone to EC effects.  Adverse effects from EC that have been
observed in recent years include beam instabilities
\cite{PRSTAB6:034402, PRSTAB7:094401, PRL79:3186, PRSTAB9:012801,
PRL74:5044, PRSTAB11:041002, PRSTAB5:094401, PRSTAB11:010101,
EPAC08:TUPP043}, degradation in the beam quality
\cite{PRSTAB16:051003, PAC01:TPPH100, PRSTAB11:041002}, and excess
load to the cryogenic system of cold-bore vacuum chambers
\cite{PRSTAB13:073201}.  Several accelerators have been modified to
reduce the cloud density \cite{PRSTAB9:012801, PAC01:TPPH100,
PRSTAB11:041002}.  EC concerns led to EC mitigation features in
the design of recent accelerators \cite{EPAC08:TUPP043,
ECLOUD12:Wed0830} and proposed future accelerators
\cite{JVSTA30:031602, IPAC10:WEPE097, EPAC08:MOPP050}.  Additional
information on EC issues can be found in review papers such as
\cite{ECLOUD04:9to13, ECLOUD12:Wed0830, ECLOUD12:Tue1815}.

The Cornell Electron Storage Ring (CESR) provides x-ray beams for
users of the Cornell High Energy Synchrotron Source (CHESS) and serves
as a test bed for future accelerators through the CESR Test
Accelerator program (\CesrTA) \cite{PAC09:FR1RAI02, ECLOUD10:OPR06,
CLNS:12:2084}.  Major goals of the \CesrTA{} program are to develop
tools and techniques for operation at low beam emittance and to better
understand electron cloud effects and their mitigation.  The EC
density is measured with multiple methods, including retarding field
analyzers \cite{PRSTAB17:061001}, shielded button electrodes
\cite{NIMA749:42to46}, and microwave excitation
\cite{NIMA754:28to35}.  The effectiveness of several types of
coatings for EC mitigation has been measured by installing coated and
instrumented chambers \cite{PRSTAB17:061001, IPAC13:THPFI088}.

In the presence of a stored beam and synchrotron radiation (SR), three
surface phenomena are important in determining the build-up of the
electron cloud: photo-emission of electrons; secondary emission of
electrons; and scattering of photons.  As indicated above, secondary
emission is particularly important---since it is possible for a
surface to release more electrons than are incident, secondary
emission can make the electron cloud density grow, even without
additional photons.

Surface properties are known to change with time in an accelerator
vacuum chamber: this is referred to as ``conditioning'' or ``beam
scrubbing,'' and is thought to be due to removal of surface
contaminants by surface bombardment.  The likely ammunition for
surface bombardment includes SR photons radiated by the stored beam,
scattered photons, electrons from the electron cloud, ions, and beam
halo.

Because of the importance of secondary emission for electron cloud
effects, a system was developed for in-situ measurements of the
secondary electron yield (SEY) as a function of the energy and angle
of the incident primary electrons.  The goals of the \CesrTA{} in-situ
SEY studies included (i) measuring the SEY of surfaces that are
commonly used for beam chambers; (ii) measuring the effect of beam
conditioning; and (iii) comparing different materials and mitigation
coatings.  Samples were made from the same materials as one would find
in an accelerator vacuum chamber, with similar surface preparation
(sometimes called ``technical surfaces'' in the literature), as
opposed to the pure materials and
single-crystal samples which would be used for studies of intrinsic
properties of solids.

The effect of exposure to an accelerator environment on the SEY of
surfaces has been studied at Argonne \cite{JVSTA21:1625to1630}, CERN
\cite{EPAC00:THXF102, ECLOUD02:17to28, EPAC02:WEPDO014,
PRSTAB14:071001, IPAC11:TUPS028}, KEK \cite{ECLOUD07:72to75,
ANTIECLOUDCOAT09:24}, and SLAC \cite{NIMA621:47to56}.  In-situ
studies have been supplemented by bench measurements of conditioning
by an electron beam \cite{JVSTA21:1625to1630, EPAC00:THXF102,
ECLOUD02:17to28, EPAC02:WEPDO014, PRL93:014801, PRSTAB16:011002,
PRSTAB16:051003, JVSTA25:675to679, ECLOUD07:82to85,
ANTIECLOUDCOAT09:24, SHINKU48:118to120, NIMA469:1to12,
NIMA551:187to199, JVSTA23:1610to1618}.  Additionally, sources of
systematic error in SEY measurements and countermeasures have been
studied at SLAC \cite{ECLOUD04:107to111, SLAC:PUB10541}.

In some of the other accelerator SEY conditioning studies, the samples
were installed into the beam pipe for an extended period and then
moved to a laboratory apparatus for SEY measurements.
At Argonne, the removal of the samples required a brief exposure to
ambient air \cite{JVSTA21:1625to1630}.  At PEP-II, the samples were
moved under vacuum using a load-lock system \cite{NIMA621:47to56}.

Studies at CERN and KEK, on the other hand, used in-situ systems for
the SEY measurements, so that the samples did not have to be removed
from the tunnel \cite{ECLOUD02:17to28, EPAC02:WEPDO014,
  ECLOUD07:72to75, ANTIECLOUDCOAT09:24}.  The in-situ SEY systems
allow for more frequent measurements with fewer concerns about
recontamination of the surface between beam exposure and the SEY
measurement, but require a more elaborate system in the accelerator
tunnel.

The SEY apparatus developed for \CesrTA{} was based on the system used
in PEP-II at SLAC \cite{NIMA621:47to56}.  In lieu of the load-lock
system used at SLAC, a more advanced vacuum system was designed,
incorporating electron guns for in-situ SEY measurements.  The
measurements at \CesrTA{} are similar to the in-situ measurements at
CERN and KEK, but with several differences: (i) we have studied a
wider variety of materials than measured at CERN; (ii) we have done
more frequent measurements than done at KEK to get a more complete
picture of SEY conditioning as a function of time and beam dose;
(iii) we have measured the dependence of SEY on position and 
  angle of incidence.
Systems similar to the \CesrTA{} stations were recently sent to
Fermilab for EC studies in the Main Injector \cite{IPAC12:MOPPC019}.

The \CesrTA{} in-situ samples are typically measured weekly during a
regularly-scheduled 6-hour tunnel access.  The SEY chamber design
allows for samples to be exchanged rapidly; this can be done during
the weekly access if needed.  As was the case for the PEP-II studies,
there are 2 samples at different angles, one in the horizontal plane,
the other 45\degree{} below the horizontal plane.  This allows us to
compare conditioning by bombardment from direct SR photons in the
middle of the horizontal sample versus bombardment by scattered
photons and EC electrons elsewhere.  Because the accelerator has down
periods twice a year, we have been able to keep some samples in
ultra-high vacuum after beam conditioning and observe the changes in
SEY over several weeks, without exposure to air.

Models have been developed to describe the SEY as a function of
incident energy and angle (for example, the probabilistic model of
M. Furman and M. Pivi \cite{PRSTAB5:124404}).  In the models, the
secondary electrons are generally classified into 3 categories: ``true
secondaries,'' which emerge with small kinetic energies; ``rediffused
secondaries,'' whose energies are distributed from low to high, up to
the energy of the incident primary electron; and ``elastic
secondaries,'' which emerge with the same energy as the incident
primary.  The SEY models are used to predict the EC density and its
effect on the accelerator beam.  Our in-situ SEY measurement program
is ultimately oriented toward developing more realistic SEY model
parameters to allow for more accurate predictions of EC effects.

This report describes the apparatus and techniques developed for the
in-situ SEY measurements, including the issues that were encountered
and improvements that were made.  For clarity, we will divide the
stages of the measurement program chronologically into two parts,
Phase I and Phase II, and further subdivide the latter into Phase IIa
and Phase IIb.  We describe the in-situ apparatus in \autoref{S:app}.
We discuss the basic features of the SEY measurements in
\autoref{S:basics}.  In Phase I, samples of 3 different materials were
measured, starting in January 2010; the Phase I measurement techniques
are summarised in \autoref{S:phasei}.  Improvements were made to the
hardware and measurement techniques between January 2011 and August
2011.  In Phase II, additional materials were measured in parallel
with additional improvements to the techniques, with measurements
starting in September 2011; the Phase II improvements are discussed in
\autoref{S:phaseii}.  Data analysis methods are discussed in
\autoref{S:analysis}.  Examples of results are given in
\autoref{S:exam}.  The information herein is presented in a more
compact form in a separate paper \cite{SUBMITTED:SEY1}.  More details
on the results for metals (aluminum, copper, and stainless steel) and
films for EC mitigation (titanium nitride, amorphous carbon, and
diamond-like carbon) can be found in other papers \cite{CLNS:12:2084,
  ECLOUD10:PST12, PAC11:TUP230, IPAC13:THPFI087}.

\section{Apparatus\label{S:app}}

There are two SEY stations to allow exposure of two samples to the
accelerator environment.  The SEY measurements are done in the
accelerator tunnel with an electron gun while the samples remain under
vacuum.  Magnetic shielding is included to ensure that low-energy
electrons from the gun are not deflected by stray magnetic fields.
The samples are typically exchanged without removal of the SEY
stations from the tunnel.  An additional station outside the tunnel is
used for supplementary measurements.

\subsection{Storage Ring Environment}

The Cornell Electron Storage Ring has a circumference of 768 meters.
Electrons and positrons travel in opposite directions through a common
beam pipe; both species can therefore affect the build-up of the
electron cloud in the vacuum chamber.  The in-situ SEY system is
installed in a straight section called ``L3,'' which originally was
the site of a detector for high-energy physics.  The beam pipe in L3
is stainless steel and has a circular cross-section, with an inner
diameter of 89 mm.  The SEY beam pipe includes a retarding field
analyser for measurements of the electron cloud density and energy
distribution.

The SEY system is installed at the East end of L3; nearby bending
magnets are located such that the SEY samples are exposed
predominantly to SR from the electron beam, the closest bending magnet
being about 6~m away.  The photon flux at the SEY stations is lower
than the ring-wide average.  Most of the beam exposure of the SEY
samples happens with CHESS conditions: a beam energy of 5.3~GeV, with
beam currents of $\sim 200$~mA for both electrons and positrons.

In an imperfect vacuum environment, beam scrubbing can be counteracted
by recontamination of surfaces from the residual gas.  Cold cathode
ionisation gauges are used to monitor the pressure in the CESR beam
pipe; the closest gauge is about 1~m from the SEY stations.  The base
pressure is generally $\lappeq 1.3 \cdot 10^{-7}$~Pa.  With CHESS
beams, the pressure is typically $\lappeq 6 \cdot 10^{-7}$~Pa after
beam conditioning; after venting, the pressure can reach as high as
$\sim 10^{-4}$~Pa during initial beam conditioning.  After beam
conditioning, H$_2$ is the dominant residual gas.

\subsection{In-Situ SEY Stations}

\begin{figure*}[tb]
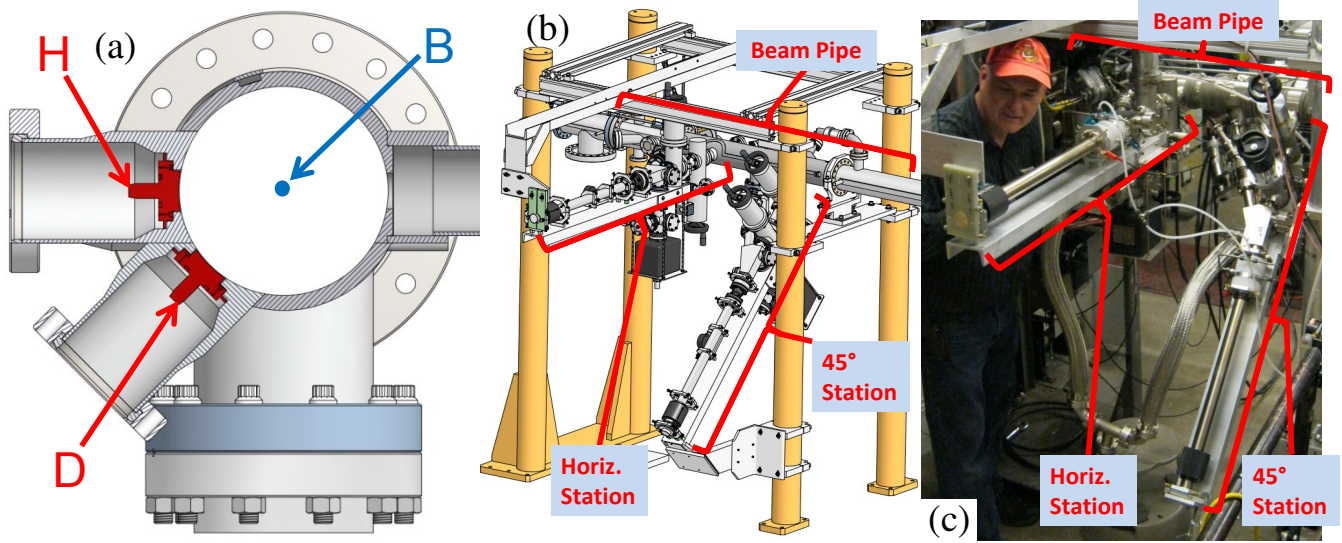

\makebox[\textwidth]{%
\GRAFheight[\heightone]{250}{280}
\GRAFlabelcoord{50}{250}
\incGRAFlabel{fig_samples_bp}{(a)}%
\GRAFheight[\heightone]{440}{540}
\GRAFlabelcoord{45}{500}
\incGRAFlabel{fig_iso_station_l3}{(b)}%
\GRAFheight[\heightone]{420}{540}
\GRAFlabelcoord{0}{4}
\GRAFlabelbox{40}{30}
\incGRAFboxlabel{fig_pic_station_l3}{(c)}%
}

\caption{(a) ``Beam's eye'' view of the SEY stations showing the beam
  (B), horizontal sample (H), and 45\degree{} sample (D\@).  (b)
  Isometric drawing of the SEY stations, CESR beam pipe, and supports.
  (c) Photograph of the SEY stations in the tunnel.  Note that (a)
  does not show the longitudinal separation of about 0.4~m between the
  samples, though this can be seen in (b) and (c).\label{F:sys}}

\end{figure*}

\begin{figure*}[tb]

\makebox[\textwidth]{%
\GRAFheight[\heighttwo]{622}{542}
\GRAFlabelcoord{450}{200}
\incGRAFlabel{fig_iso_stat_sngl}{(a)}%
\hspace*{\fill}%
\setlength{\unitlength}{\heighttwo/960}%
\begin{picture}(920,960)(0,0)
\put(0,0){\includegraphics[height=\heighttwo]{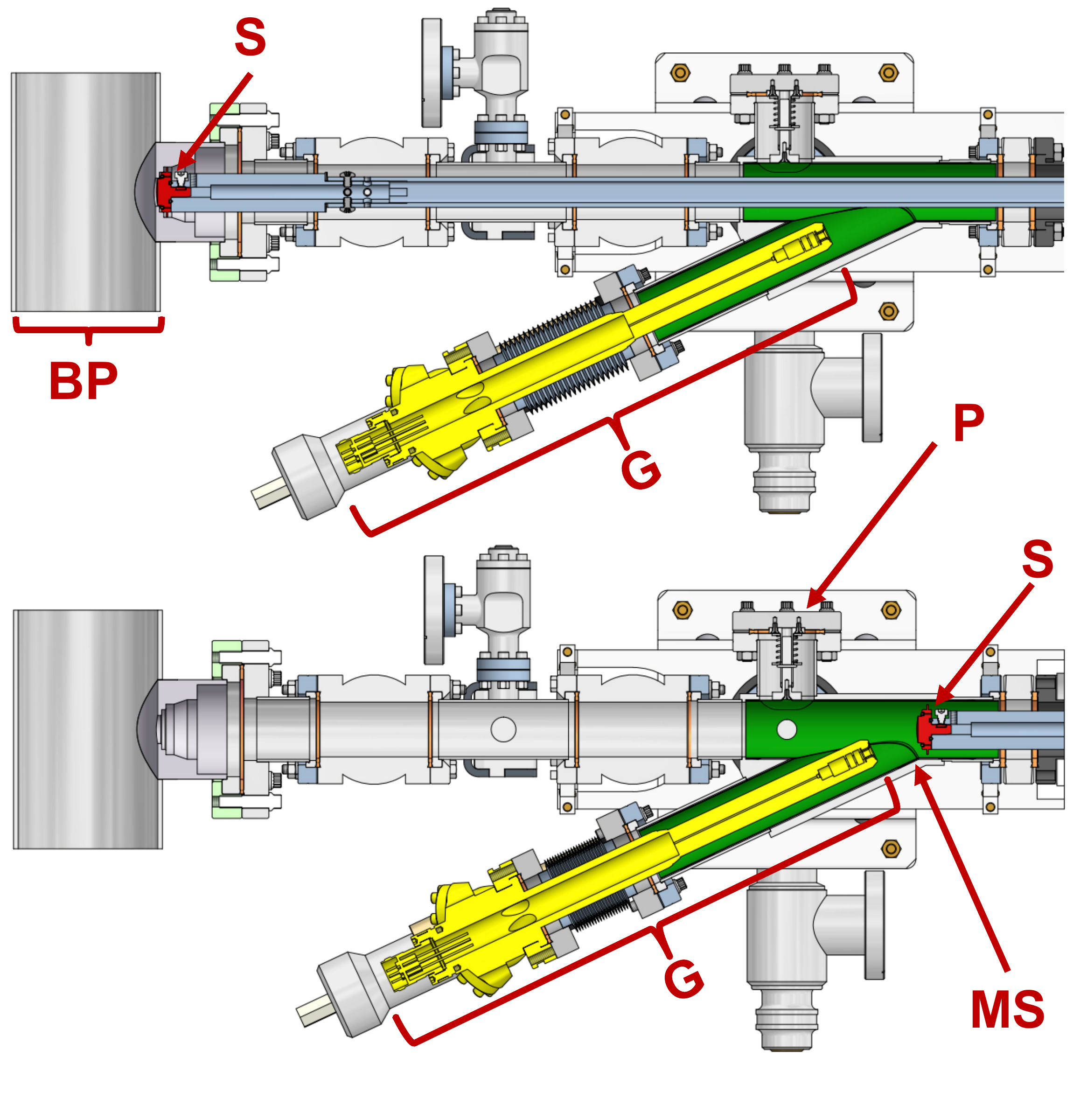}}
\put(800,900){\Large (b)}
\put(725,50){\Large (c)}
\end{picture}%
}

\caption{(a) Isometric view of one SEY station; the beam pipe and
  second gate valve are not shown.  Cross-sectional views of in-situ
  station with (b) sample inserted in beam pipe and (c) sample
  retracted for SEY measurements.  S: sample (red); G: electron gun
  (yellow); MS: magnetic shield (green); BP: beam pipe; P: port for
  sample exchange; C: vacuum crotch; B: ceramic break; SA: sample
  actuator; GV: gate valve; IGP: ion pump.\label{F:arm}}

\end{figure*}

As shown in \autoref{F:sys}a, the samples have a curved surface to
match the circular beam pipe cross-section.  Both samples are
approximately flush with the inside beam pipe, with one sample
positioned horizontally in the direct radiation stripe, and the other
sample positioned at $45^\circ$, beneath the radiation stripe.

\autoref{F:sys}b shows the SEY stations, with the equipment for moving the
samples under vacuum and measuring the SEY;
\autoref{F:sys}c shows a photograph
of the SEY stations in CESR.

More detailed drawings of one SEY station are shown in
\autoref{F:arm}.  A custom-designed vacuum ``crotch'' (made from
316LN stainless steel) provides an off-axis port for an electron gun,
a pumping port, and a side port for sample exchange.  The sample is
mounted on a linear positioner with a magnetically-coupled manual
actuator.\footnote{Model DBLOM-26, Transfer Engineering, Fermont, CA.}
The sample and sample positioner are electrically isolated from the
grounded beam pipe by a ceramic break.\footnote{Model BRK-VAC5KV-275,
Accu-Glass Products, Inc., Valencia, CA.}  The electron gun is at an
angle of 25\degree{} from the axis of the sample positioner.  The gun
is mounted on a compact linear positioner\footnote{Model LMT-152, MDC
Vacuum Products, LLC, Hayward, CA.}  so it can move out of the
sample positioner's path when the sample is inserted into the beam
pipe (\autoref{F:arm}b).

When the sample is in the beam pipe for exposure to SR and the
electron cloud (\autoref{F:arm}b), force is applied to the actuator
to ensure that the sample remains well seated.  When the sample is in
the SEY measuring position (\autoref{F:arm}c), the gun is moved
forward to make the gun-to-sample distance nominally 32.9~mm for the
SEY measurements.  
Moving the gun forward allows for a smaller beam spot size on the
sample and a larger range of incident angles.

The instrumentation for the SEY measurement is the same as that used
in previous studies at SLAC \cite{NIMA551:187to199}.  A
picoammeter\footnote{Model 6487, Keithley Instruments, Inc.,
Cleveland, OH.} is used to measure the current from the sample; the
sample dc bias is provided by a power supply internal to the
picoammeter.

The vacuum in the SEY stations is maintained by ion pumps.  The
chambers also include titanium sublimation pumps in case additional
pumping is needed.  During the SEY measurements, one or both of two
gate valves are closed to isolate the CESR vacuum system from the SEY
chambers.  Initially, hot-filament ionisation gauges were used to
monitor the pressure in the SEY chambers.  These were removed after
the first few measurements on Al, as it appeared that the out-gassing
from the filament might be affecting the SEY\@.  Subsequently, we have
used the ion pump current read-backs to infer the pressure in the SEY
chambers.  (We elected not to use cold cathode gauges, as these would
have introduced stray magnetic fields if installed close enough to
provide an accurate measurement of the pressure in the SEY chambers.)
The ion pump read-backs indicate that the base pressure is less than
$10^{-7}$~Pa.  With the electron gun on, the pressure increases
slowly, but typically remains below $10^{-6}$~Pa.

\subsection{Samples\label{S:sam}}

\begin{figure}[tb]
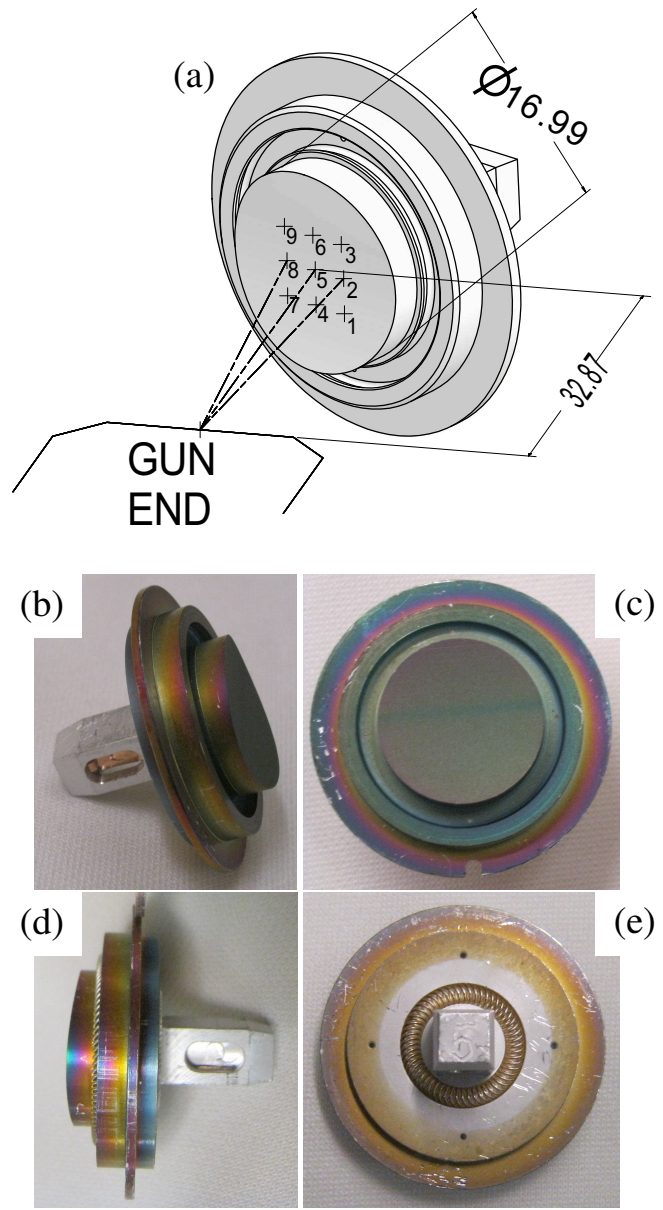

\centering
\GRAFwidth[\narrowwidth]{1038}{842}
\GRAFlabelcoord{275}{700}
\incGRAFlabel{fig_gun_align}{(a)}\\[3ex]%
\makebox[\narrowwidth]{%
\GRAFlabelbox{200}{150}
\GRAFlabelcoord{-100}{825}
\GRAFheight[0.175\textheight]{833}{1000}
\incGRAFboxlabel{fig_pic_sam_iso}{(b)}
\GRAFlabelcoord{925}{825}
\GRAFheight[0.175\textheight]{1000}{1000}
\incGRAFboxlabel{fig_pic_sam_front}{(c)}
}\\
\makebox[\narrowwidth]{%
\GRAFlabelbox{200}{150}
\GRAFlabelcoord{-100}{825}
\GRAFheight[0.175\textheight]{833}{1000}
\incGRAFboxlabel{fig_pic_sam_side}{(d)}
\GRAFlabelcoord{925}{825}
\GRAFheight[0.175\textheight]{1000}{1000}
\incGRAFboxlabel{fig_pic_sam_back}{(e)}
}\\

\caption{(a) Isometric drawing of the SEY sample, showing the 3 by 3
  grid of points where the SEY was measured in Phase I\@.  The axis of
  the electron gun is inclined by 25\degree{} relative to the
  sample's surface normal.  The diameter of the sample face and
  distance from the gun to the middle of the sample are indicated
  (dimensions in mm).  (b-e) Photographs of a sample with a TiN
  coating: (c) front, (d) side, (e) back.\label{F:grid}}

\end{figure}

\autoref{F:grid} shows a drawing and photographs of the SEY sample.
The samples are machined from bulk material;
The design includes a groove in the
back of the sample with a coiled ring spring (see
\autoref{F:grid}e) to ensure good electrical contact with the
positioner rod which holds the sample.

The samples were solvent cleaned without mechanical polishing or
etching, as typically surfaces in the CESR vacuum chamber are not
polished prior to installation.  Coatings (if any) were applied to the
finished samples after solvent cleaning.

\subsection{Sample Exchange}

As shown in \autoref{F:sys}, there are two gate valves between the
beam pipe and the SEY chamber so that the sample can be changed
without venting the beam chamber.  (The second gate valve allows for
the SEY chamber to be removed from the tunnel with the beam pipe and
the sample still under vacuum).  The SEY chamber has a special port
for changing the samples in the tunnel (see \autoref{F:arm}), with
a custom-designed hole and patch in the magnetic shield.  To minimize
exposure to tunnel air, water vapor, and dust, nitrogen gas is flowed
through the system when samples are exchanged; the exchange can be
done with the flanges open for only a few minutes.  When samples are
exchanged with the N$_2$ gas purge, the ultra-high vacuum recovers
sufficiently to resume measurement within 24~hours.  This makes it
possible to change samples during a scheduled tunnel access over a
regular CHESS running period.

\subsection{Magnetic Shielding\label{S:shield}}

At low energy (up to about 100~eV), the electrons can be deflected by
up to a few millimeters by stray magnetic fields.  To mitigate this
problem, the electron gun and SEY sample are surrounded by a
custom-made magnetic shield, shown in green in \autoref{F:arm}.  The
shield is inside the vacuum chamber and includes intersecting tubes
for shielding of the sample positioner tube and the electron gun side
port.  The shield has a hole for the ion pumping port and, as
described above, a patch for the sample exchange port.  An internal
shield has the advantage of being smaller, simpler and less
susceptible to accidental damage than an external shield, at the cost
of making the vacuum system more complicated.  The shield was
fabricated from nickel alloy mu-metal sheet of thickness 0.5~mm.  The
machining, forming, welding, and final heat treatment were done by a
vendor\footnote{MuShield, Inc., Londonderry, NH.} to our
specifications.  Metal finger stock was spot-welded to the outside of
the shield for electrical grounding.

Measurements with a field probe indicated that the shield reduces the
stray magnetic field to about 10~$\mu$T or lower.  To check the
deflection with the shield present, we measured the transmission
through a collimation electrode with a 1~mm vertical slit in front of
the sample.  At each electron beam energy, the beam was scanned across
the slit using the gun's horizontal deflection electrode to determine
whether compensation was needed to maximise the current through the
slit.  These measurements confirmed that the stray magnetic field is
well shielded.  Additional information about the collimation
measurements can be found in \ref{S:collim}.

The SEY station includes a metal rod to hold the sample (shown in
light blue in Figures~\ref{F:arm}b and \ref{F:arm}c).  The rod travels
through the magnetic shield (as can be seen in \autoref{F:arm}).
We use an aluminum rod because we found that a stainless steel rod
produced a small residual magnetic field.

\subsection{Electron Gun\label{S:gun}}

The electron gun\footnote{Gun: Model ELG-2; power supply: Model
EGPS-1022C and EGPS-1022D; Kimball Physics, Inc., Wilton, NH\@.  The
insertion length of the electron gun is custom; the power supply
design was modified for more stable output at low current.}
provides a dc beam with electrostatic acceleration to a maximum energy
of 2~keV\@.  The gun energy, current, deflection, and focusing are
adjustable.  The deflection is produced via horizontal and vertical
electrostatic fields from paired electrodes.  The focusing is produced
by biasing a ring electrode.  The focusing and deflection elements are
internal to the electron gun.  The gun's electron beam is produced by
thermionic emission.  Additional information on the control of the
electron gun current can be found in \autoref{S:gunctl}.

\subsection{Off-Line SEY Station}

A duplicate SEY station was deployed outside the tunnel.  This station
is basically the same as the stations in L3, but it is not attached to
a beam pipe and has the advantage of being accessible when the
accelerator is running.  This makes the off-line station useful for
supplemental SEY measurements and debugging of measurement
techniques.

\section{SEY Measurement: Basics\label{S:basics}}

In order to keep the stations compact enough for deployment in the
tunnel, we use an indirect method to measure the secondary electron
yield.  Our basic measurement method is the same as has been used by
SLAC \cite{NIMA551:187to199, NIMA621:47to56, ECLOUD04:107to111} and
other groups.

We measure the dependence of the SEY on (i) incident energy $K$, (ii)
incident angle $\theta$, and (iii) impact position of the primary
electrons ($\theta$ = angle from the surface normal).  For (i), we
scan the electron gun energy.  With the compact in-situ system, we
cannot independently vary the angle and position.  However, because of
the curvature of the sample, scanning the beam spot vertically changes
the position with little change in the incident angle, while scanning
horizontally changes both the position and the angle (see
\autoref{F:grid}a).  Hence, for (ii) and (iii), we scan the vertical
and horizontal deflection of the electron gun.  We make the beam spot
size on the sample as small as possible for good position and angle
resolution.

\subsection{Secondary Electron Yield}

The secondary electron yield is defined as the number of secondary
electrons released from a surface divided by the number of incident
primary electrons.
In terms of current,
\begin{equation}
{\rm SEY} = -\frac{I_s}{I_p} \; ,\label{E:seys}
\end{equation}
where $I_p$ is the current of the primary electrons incident on the
sample and $I_s$ is the current of the secondary electrons released by
the bombardment of primary electrons.  The minus sign in
\cref{E:seys} is included because the primary and secondary
electrons travel in opposite directions relative to the sample and
hence we use a convention in which $I_p$ and $I_s$ have opposite
signs.  The SEY depends on the energy and angle of incidence of the
primary electron beam.

\subsection{Indirect SEY Measurement}

The primary current $I_p$ is measured by firing electrons at the
sample with the electron gun and measuring the current from the sample
with a positive bias voltage.  A high positive bias voltage,
$V_b = +150$~V, is used to recapture secondaries produced by the primary
beam, so that the net current due to secondaries is zero in the ideal
case.

The current $I_s$ due to secondary electrons is measured indirectly.
The total current $I_t$ is measured by again firing electrons at the
sample, but with a low negative bias ($V_b = -20$~V) on the sample to
repel the secondaries.  Since $I_t = I_p + I_s$, we calculate SEY as
\begin{equation}
{\rm SEY} = - \frac{I_t-I_p}{I_p} = 1 - \frac{I_t}{I_p} \; .\label{E:seyt}
\end{equation}
If $|I_p| > |I_s|$, then $I_p$ and $I_t$ have the same sign, SEY $<$
1, and we observe a net flow of electrons from the gun to the sample
when measuring $I_t$.  If $|I_p| < |I_s|$, then $I_p$ and $I_t$ have
opposite signs (because $I_s$ and $I_p$ have opposite signs), SEY $>$
1, and we observe a net flow of electrons away from the sample.  In
either case, current from the power supply flows through the
picoammeter as needed to maintain the $-20$~V bias on the sample.

A complication is that some secondary electrons may hit the walls of
the vacuum chamber and produce additional electrons by secondary
emission; hence, the negative sample bias should be large enough to
prevent these electrons from returning to the sample.  We chose $V_b =
-20$~V based on past measurements at SLAC \cite{ECLOUD04:107to111}.

Some SEY systems include an additional electrode to allow for a more
direct measurement of $I_s$, for example the conical electrode system
at KEK \cite{IPAC10:TUPD043} and the energy analyser system used at
SLAC \cite{ECLOUD04:107to111}.  Our in-situ setup cannot accommodate
an extra electrode, so we cannot use such a method.  We should also
note that the positive bias for the $I_p$ measurement in our indirect
method is not able to retain elastic secondaries, so that the elastic
contribution to the SEY is not fully accounted for, as has been
pointed out previously \cite{ECLOUD04:107to111}.

\subsection{Parasitic Conditioning During the SEY Measurement\label{S:park}}

To measure the SEY, we bombard the sample with electrons,
which conditions the surface and changes the SEY\@.  Since the goal is
to observe conditioning by SR photons and the electron cloud, it is
best to minimise conditioning by the electron gun during the SEY
measurement \cite{ECLOUD04:107to111}.  A low electron gun current and
a rapid measurement help to mitigate this problem, but the current
must be large enough to measure and, as will be discussed below (see
\autoref{S:trans} in particular), waiting times are needed for
stable conditions.  As a result, we ``park'' the beam at a known
position on the sample with a small beam spot size when we are not
measuring $I_t$.  We will discuss the issue of parasitic conditioning
further in \autoref{S:charge}.

\subsection{Electron Gun Warm-Up; Gun Current\label{S:warm}}

Before starting the SEY measurement, we warm up the electron gun
cathode, typically for 30 to 60 minutes.  During the warm-up period,
we set the gun energy to zero and the deflection to maximum to prevent
the electron beam from reaching the sample.  After warming up, we set
the gun energy to 300~eV, deflect the beam to the parking point, and
adjust the gun parameters (see \autoref{S:gunctl}) to get the
desired value of $I_p$.  We typically observe a change in $I_p$ with
energy and a drift in $I_p$ with time (see \autoref{F:IpE} below).
The latter is presumably due to the cathode temperature still changing
slowly with time after the warm-up period (our choice of warm-up time
is a compromise between the need for stable current and the need to
finish the measurements in the available access time).

\subsection{Electron Gun Deflection and Spot Size\label{S:spot}}

Both the energy and the deflection of the electron gun are varied in
the SEY scan.  The deflection voltages must be scaled with energy to
produce the same deflection angles for different energies (see
\autoref{S:defl} for more information).

The focusing is set to minimize the beam spot size at the sample.
The focusing voltage must be adjusted with energy, but the
relationship is not linear.  Separate measurements were done to
find the parameters to produce the minimum spot size as a function of
beam energy for the nominal gun-to-sample spacing (see \autoref{S:focus}
for more information).

With the focus adjusted to minimise it, the estimated beam spot sizes
for different energy ranges are as follows: slightly larger than 1~mm
between 20~eV and 200~eV; $\leq 0.75$~mm from 250~eV to 700~eV; about
1.2~mm at 1500~eV (increasing with energy between 800~eV and
1500~eV)\@.  For the 3 by 3 grid of Phase I measurements
(\autoref{S:phasei}), the approximate distance between adjacent
grid points is 3.7~mm, so there is no overlap between grid points.
For the higher resolution double grid developed in Phase II there is
overlap between grid points over part of the energy range (as will be
discussed in \autoref{S:pts}).

Collimation measurements provided the basis for the focus set point as
a function of energy and the beam spot size estimates, as described in
\ref{S:collim}.

\subsection{Electrical System\label{S:elec}}

\begin{figure}[b]
\centering
\includegraphics[width=\widewidth]{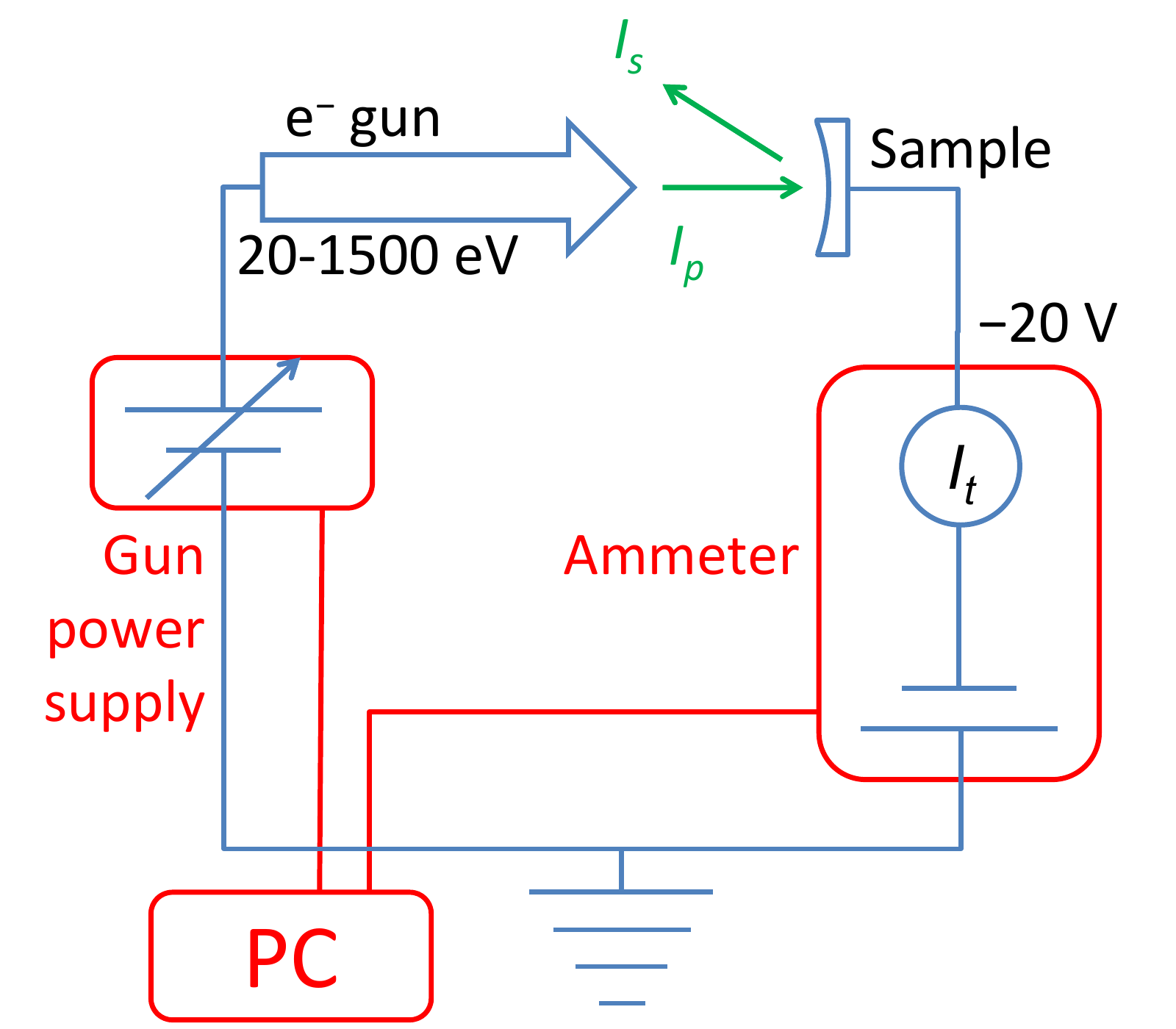}

\caption{Electrical and data acquisition schematic, showing the
  sample with a negative bias to measure the total current ($I_t = I_p
  + I_s$).\label{F:ElecSys}}

\end{figure}

\begin{figure*}
\centering
\includegraphics[width=\textwidth]{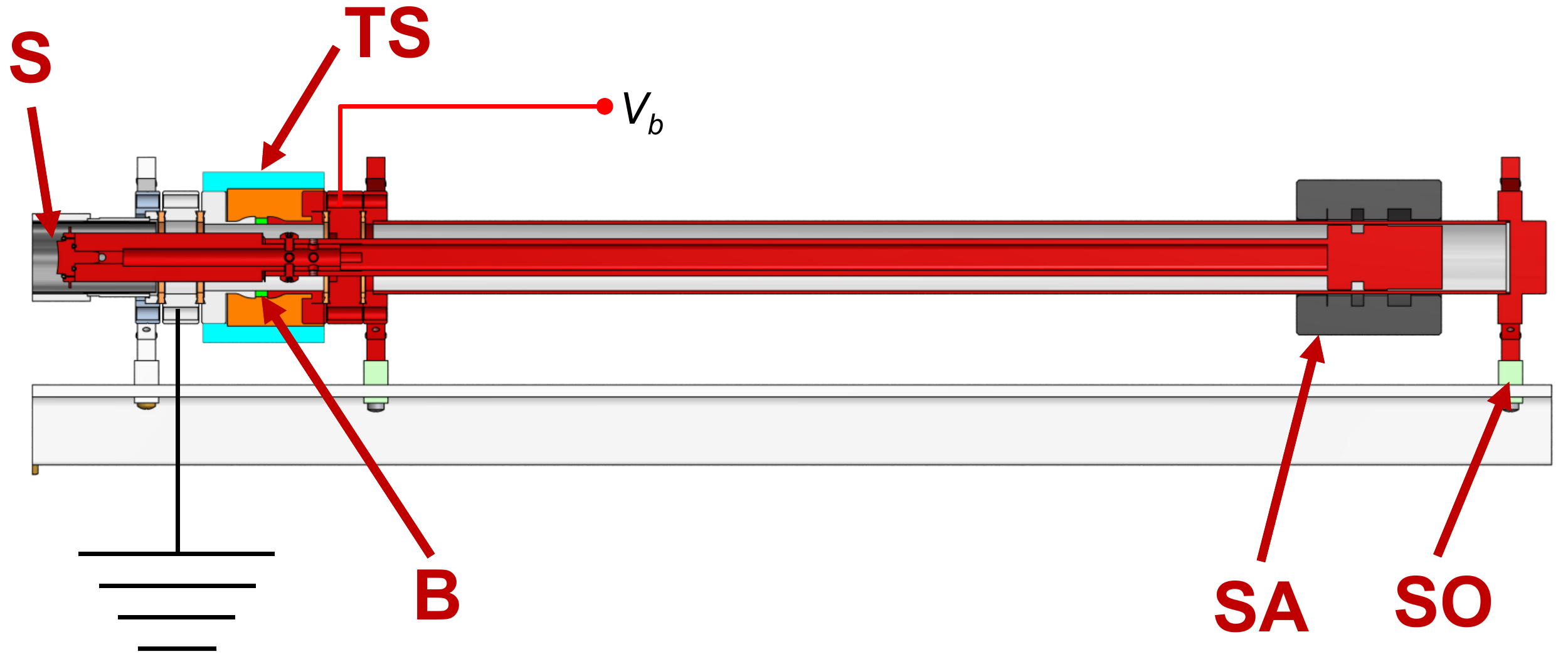}

\caption{Side view of the SEY station, indicating portions which are
  biased in red.  The chamber on the left is connected to the grounded
  beam pipe and support beam.  The orange region represents the
  nitrogen gas blanket around the ceramic break.  S: sample; B:
  ceramic break (green); SA: sample actuator; SO: stand-off (light
  green); TS: Teflon shell (light blue).\label{F:ElecBias}}

\end{figure*}

An electrical schematic of the system is shown in
\autoref{F:ElecSys}.  As illustrated in \autoref{F:ElecBias}, the
bias voltage is applied to the sample and positioner arm, which are
separated by the ceramic break from the grounded SEY chamber, magnetic
shield, and beam pipe.  A nitrogen gas blanket is also shown in
\autoref{F:ElecBias}; this will be discussed in \autoref{S:LCmit}.

Low-noise triaxial cables are used to bring the signals from the
sample positioner arms to the picoammeters.  The middle and outer
conductors of the triaxial cable are connected to the SEY station
ground and the inner conductor is connected to the sample.  The
picoammeter provides the biasing voltage, in addition to measuring the
current: a small shielded circuit (connected to the picoammeter
through another short triaxial cable) is used to connect the bias
voltage from the picoammeter power supply.  The outer conductors of
the triaxial cables provide a shield for the signals carried by the
middle and inner conductors.  To avoid a ground loop, the outer
conductors of the long and short triaxial cables are not connected to
each other.

As can be seen in \autoref{F:ElecBias}, the sample positioner arms are
not electrically shielded.  As a result, activity that disturbs the
air near the SEY stations produces noise in the current signals.  In
the tunnel, the area adjacent to the stations is roped off when SEY
measurements are done to discourage visitors.  With the off-line
station, we minimise the presence of personnel in the room when doing
SEY measurements.

As shown schematically in \autoref{F:ElecSys}, each station operates
independently with a dedicated picoammeter, electron gun, electron gun
power supply, and CPU, so that measurements on the horizontal and
45\degree{} sample can be done in parallel.  The electronics are
installed on a mobile equipment rack so that they can be removed when
the accelerator is operating.

\subsection{Data Acquisition}

The SEY scans are automatic and are controlled by a data acquisition
program (DAQP) implemented in LabVIEW\@.\footnote{Version 8.2,
  National Instruments, Austin, TX.}  The LabVIEW program incorporates
existing software from Kimball Physics and Keithley for control and
readout of the electron gun and picoammeter, respectively.
Communication with the picoammeter is via an RS-232 serial connection;
communication with the electron gun power supply is via PCI
cards.\footnote{PCI-6034E and PCI-6703, National Instruments, Austin,
  TX.}  We developed and implemented the algorithms to load the
desired gun and picoammeter setting, pause for the necessary settling
times, and record the signals for the SEY scans \cite{ECLOUD10:PST12}.
Development of the DAQP has been an important part of our SEY
measurement program, resulting in a relatively sophisticated tool for
control of SEY scans.  The DAQP is identical for the $45^\circ$
system, horizontal system, and off-line system.

\section{Measurement Method: Phase I\label{S:phasei}}

Phase I measurements with the SEY stations in L3 began in January 2010
and ended when the stations were removed in January 2011.  The
techniques used for the Phase I measurements have been reported
previously \cite{CLNS:12:2084, ECLOUD10:PST12, PAC11:TUP230}.  The SEY
was measured on a 3 by 3 grid, as shown in \autoref{F:grid}a.  The
energy was scanned from 20~eV to 1500~eV with a step of 10~eV.

In the Phase I algorithm, the program scanned through the energies and
deflections with a constant sample bias, and the process was repeated
after changing the bias.

The first scan was done with positive sample bias ($V_b = 150$~V) to
measure $I_p$, with gun settings for $I_p \approx 2$~nA\@.  This
measurement was done with the deflection set to put the beam in a
parking point between Point 5 and Point 9 of the grid
(\autoref{F:grid}a) to reduce conditioning at the measurement grid
points.

The second scan stepped through the same gun energies with a negative
bias ($V_b = -20$~V) on the sample to measure $I_t$.  At each gun
energy, the beam was rastered across all 9 grid points while the
program recorded the current for each point.

\begin{figure}[htb]
\centering
\includegraphics[width=\widestwidth]{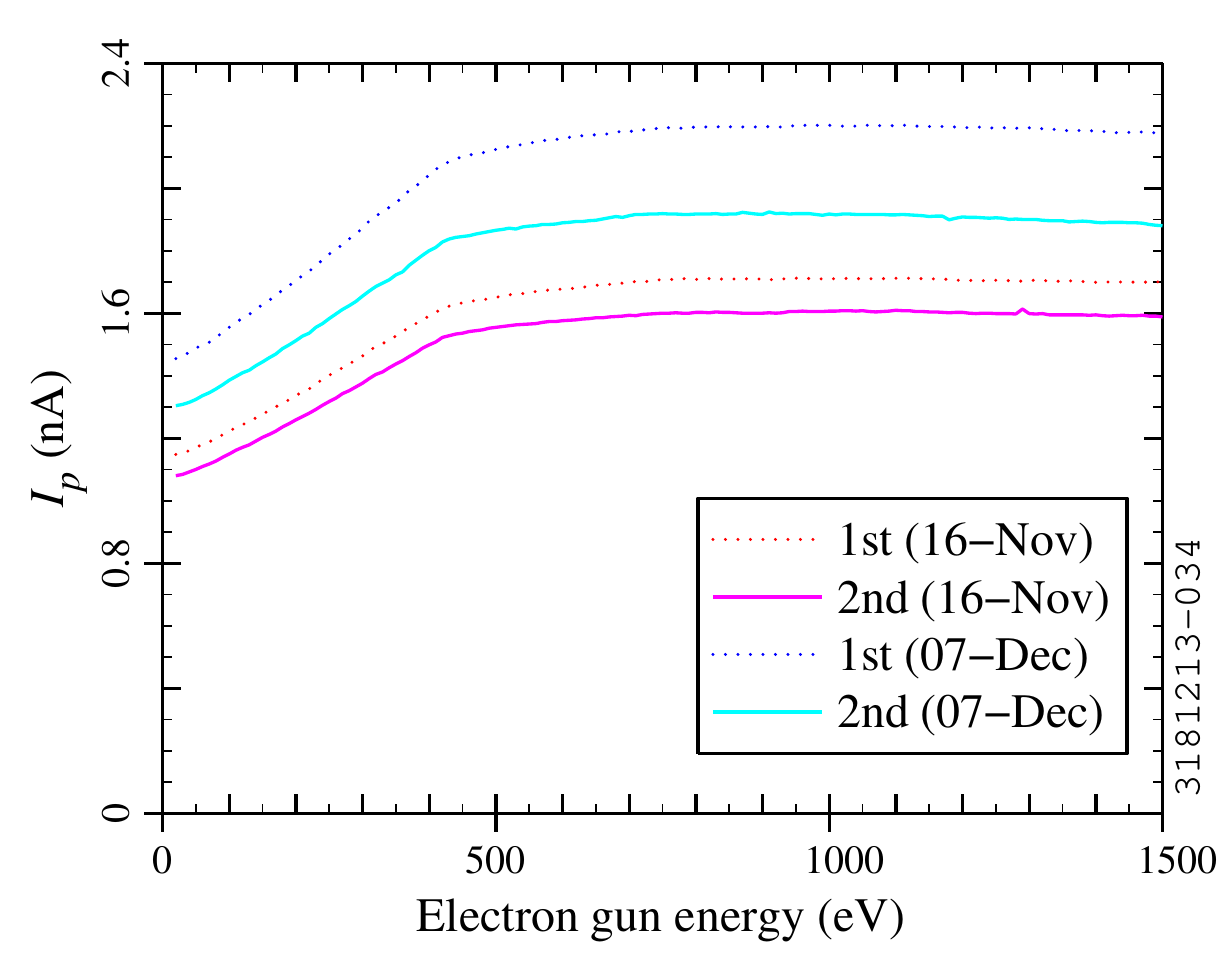}\\[-2ex]

\caption{Repeated scans of primary current as a function of gun energy
  for the horizontal amorphous carbon sample (measured in 2010).  The
  measurements labelled ``1st'' and ``2nd'' were done before and after
  an $I_t$ scan (the $I_t$ scan takes about 15 minutes).  Ideally,
  $I_p$ should be constant at 2~nA, but, in reality, $I_p$ depends on
  the gun energy and varies from one scan to another.\label{F:IpE}}

\end{figure}

As indicated above, the gun output current varies with gun energy and
drifts in time.  To minimize the error due to the drift in gun
current, we did a second $I_p$ scan after the $I_t$ scan.  The first
and second $I_p$ values for a given energy were averaged for
calculation of the SEY as a function of energy and grid point.
\autoref{F:IpE} shows examples of ``before'' (dotted lines) and
``after'' (solid lines) scans of $I_p$.

\section{Measurement Method: Issues and Phase II Improvements\label{S:phaseii}}

\begin{figure}[htb]
\centering
\includegraphics[width=\widewidth]{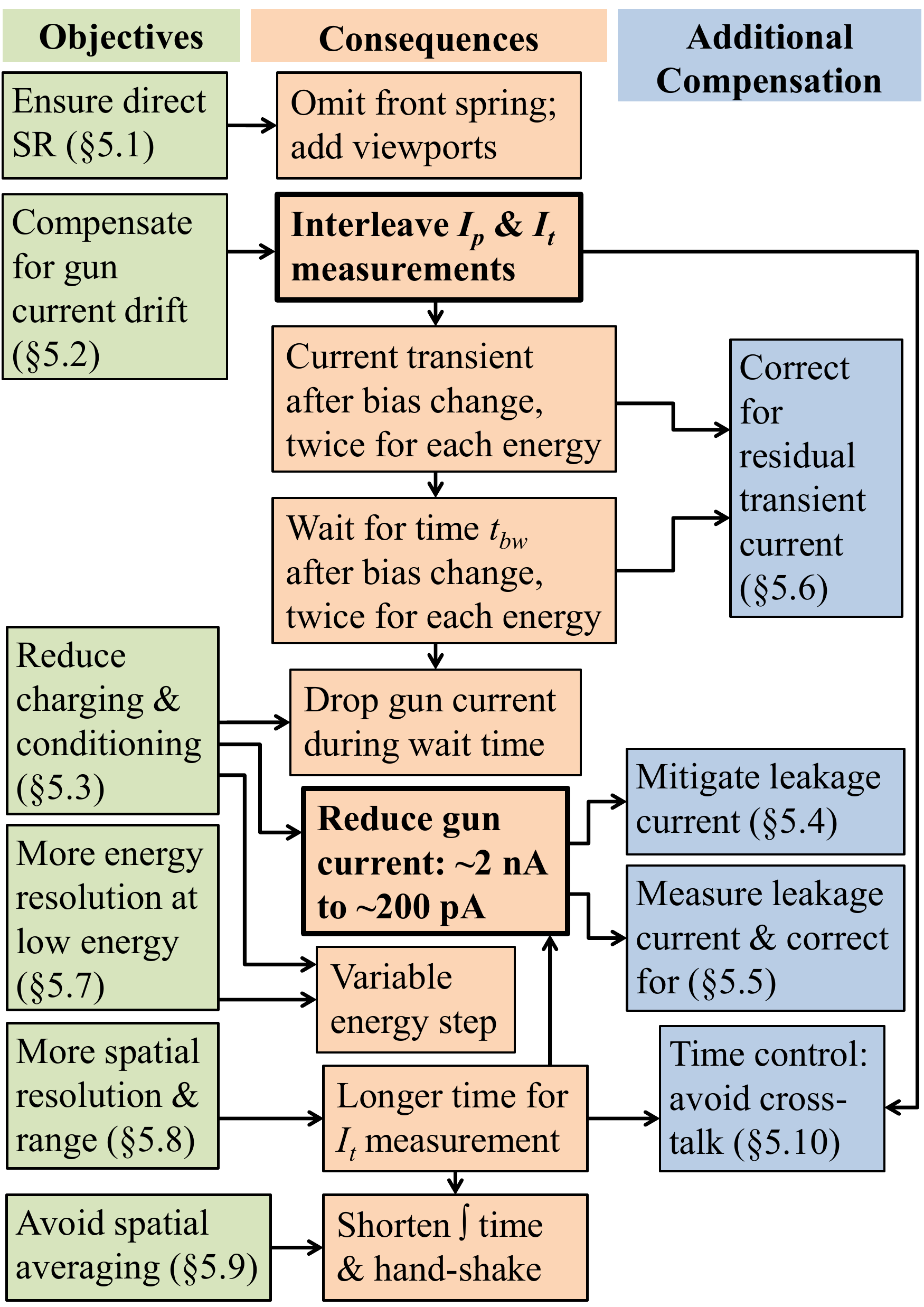}

\caption{Flowchart showing the modifications to the SEY measurement
  method for Phase II and the interrelationships amongst various
  modifications.  Changes with major ramifications are
  highlighted in bold type.\label{F:flowchart}}

\end{figure}

Our experience in Phase I led to iterations in the measurement method.
Modifications for Phase II are described in this section.  The
modifications are outlined in \autoref{F:flowchart}, along with the
causal connections amongst them.

\begin{table}[htb]

\caption{SEY measurements in L3 and improvements in
  techniques.\label{T:hist}}
\vspace*{-2ex}
\begin{center}
\ifthenelse{\lengthtest{\columnwidth=\textwidth}}
{\addtolength{\extrarowheight}{-5pt}}
{\relax}
\begin{tabular}{lll} \hline
Samples & Dates & Comments \\ \hline\hline
\multicolumn{3}{r}{\bf Phase I Measurements} \\ \hline\hline
TiN & Jan 2010- & Commission systems \\
1st pair & Aug 2010 & \\ \hline
Al & Aug 2010- & Remove vacuum gauges \\
6061-T6 & Nov 2010 & \\ \hline
Amor- & Nov 2010- & \\
phous C & Jan 2011 & \\ \hline \hline
\multicolumn{3}{r}{\bf Phase IIa Development} \\ \hline\hline
(Stations & Jan 2011- & Ensure exposure to direct\\
 out of L3) & Aug 2011 & SR (\S \ref{S:protrude}); mitigate gun current\\
 & &  drift (\S \ref{S:Istab}), charging (\S \ref{S:charge}), \& \\
 & & leakage current (\S \ref{S:LCmit}--\ref{S:trans}) \\ \hline \hline
\multicolumn{3}{r}{\bf Phase IIa Measurements} \\ \hline\hline
Diamond- & Sep 2011-& Investigate spatial resolution \\
like C & Nov 2011 & \\ \hline
TiN & Nov 2011 & Improve spatial resolution (\S \ref{S:hshake}); \\
2nd pair & -Mar 2012 & variable energy step (\S \ref{S:EngRes});  \\
 & & investigate spatial resolution \& \\
 & & range \\ \hline
Cu & Mar 2012 & Improve spatial resolution \& \\
OFHC & -Jul 2012 & range \\ \hline\hline
\multicolumn{3}{r}{\bf Phase IIb Measurements} \\ \hline\hline
Stainless & Aug 2012- & Full spatial resolution \& range \\
steel & Sep 2012 &   (\S \ref{S:pts}); mitigate parasitic \\
316 &  &  conditioning (\S \ref{S:charge}) \& cross- \\
 &  & talk (\S \ref{S:cross})\\ \hline
TiN & Oct 2012- & Recondition after exposure to air \\
2nd pair & Jan 2013 & \\ \hline
Al 6063 & Jan 2013- & \\ \hline
\end{tabular}
\end{center}
\end{table}

\begin{table}[htb]

\caption{Timing parameters for Phase IIb SEY
  scans.\label{T:SEYtimepar}}

\begin{center}
\begin{tabular}{lcl} \hline
Symbol & Value & Description \\ \hline
$t_m$ & $\sim 250$ ms & average and read out current\\
$t_{dw}$ & 50 ms & wait after setting gun deflection\\
$t_{cw}$ & 10 s & wait after setting gun current\\
$t_{bw}$ & 60 s & wait after setting bias\\ \hline
\end{tabular}
\end{center}
\end{table}

The time line for the measurements and system modifications is
outlined in \autoref{T:hist}.  Significant improvements for Phase
IIa were made between January 2011 and August 2011 when the SEY
stations were out of the tunnel for hardware modifications, prior to
the beginning of Phase IIa measurements; the hardware modifications
are described in Sections~\ref{S:protrude}, \ref{S:LCmit}, and
\ref{S:trans}.  Further improvements in the measurement methods were
made between November 2011 and July 2012.  The measurement hardware
and techniques have been relatively stable since the start of Phase
IIb in August 2012.

\begin{figure}[htbp]
\centering
\includegraphics[width=\columnwidth]{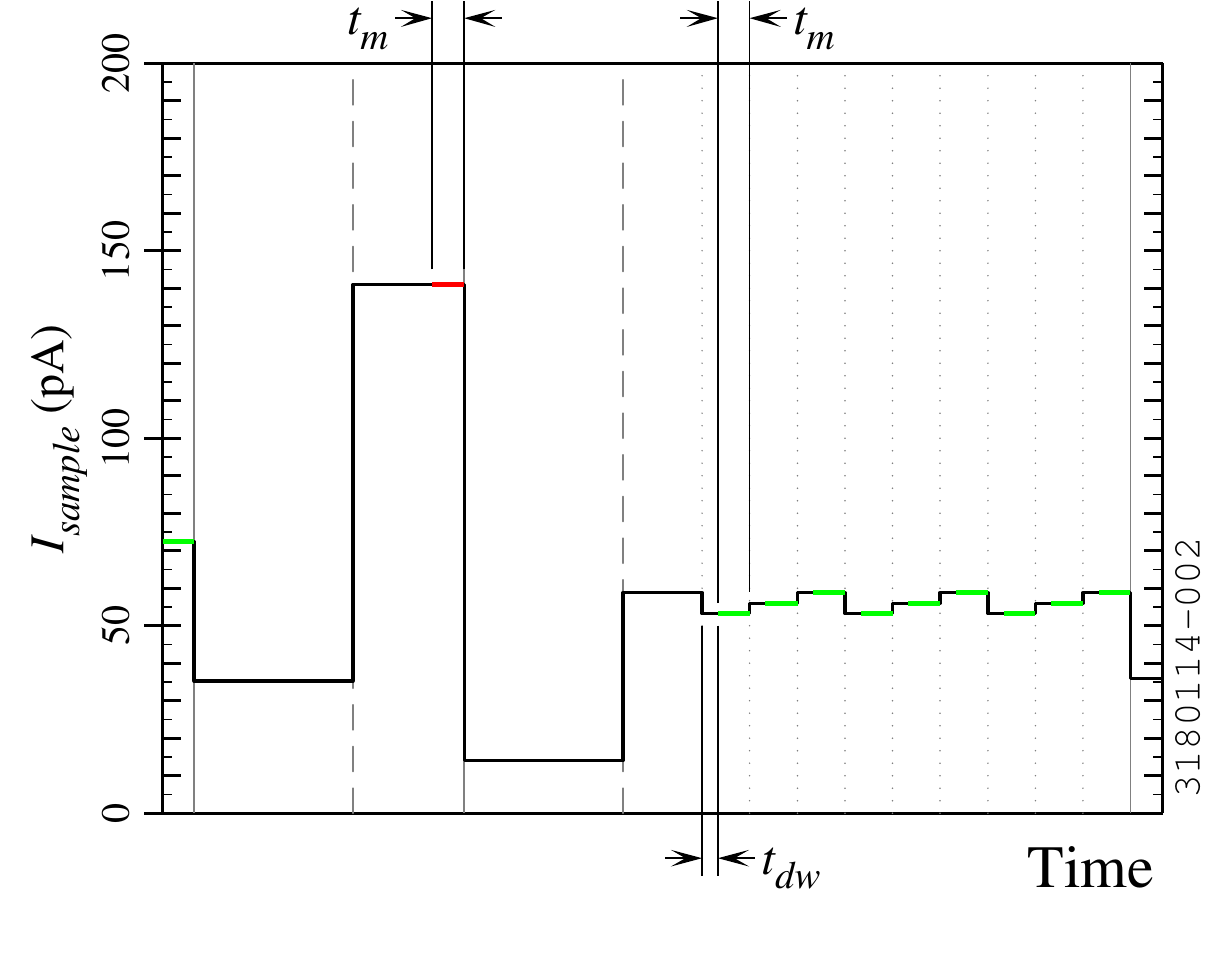}\\[-2ex]

\caption{Zoomed-in timing schematic of sample current for SEY scans in
  Phase IIb.  As in the previous figure, time intervals for averaging
  $I_p$ are shown in red and time intervals for averaging $I_t$ are
  shown in green.\label{F:SEYtimezoom}}

\end{figure}

The changes for Phase II led to a significantly different timing
algorithm for the SEY scans, shown diagrammatically in
\autoref{F:SEYtime}; a zoomed-in version of \autoref{F:SEYtime}g is
shown in \autoref{F:SEYtimezoom}.  For illustrative purposes, the
horizontal axes in Figures~\ref{F:SEYtime} and \ref{F:SEYtimezoom} are
not to scale and a simple $n_x = 3$ by $n_y = 3$ grid is shown ($n_x$
and $n_y$ are the number of horizontal and vertical grid points and $n
= n_x n_y$; see \ref{S:TimingIIb} for a version with a realistic
number of grid points and a realistic time axis).  Solid gray lines
indicate a bias change; dashed gray lines indicate an increase in the
gun current from the standby value to the full value; dotted gray
lines indicate a deflection change.  The final timing parameters used
for Phase IIb are given in \autoref{T:SEYtimepar}.  Selected features
of \autoref{F:SEYtime} will be described in this section as the issues
are discussed.

\begin{figure}
\centering
\GRAFwidth[0.925\narrowwidth]{490}{205}
\GRAFlabelcoord{490}{140}
\begin{tabular}{c}
\incGRAFlabel{fig_timing_a}{(a)}\\[-3.5ex]
\incGRAFlabel{fig_timing_b}{(b)}\\[-3.5ex]
\incGRAFlabel{fig_timing_c}{(c)}\\[-3ex]
\incGRAFlabel{fig_timing_d}{(d)}\\[-3ex]
\incGRAFlabel{fig_timing_e}{(e)}\\[-3ex]
\incGRAFlabel{fig_timing_f}{(f)}\\[-3ex]
\incGRAFlabel{fig_timing_g}{(g)}\\[-1.5ex]
\end{tabular}

\caption{Timing schematic for SEY scans in Phase IIb: (a) gun energy,
  (b) focus, (c) horizontal deflection, (d) vertical deflection, (e)
  gun emission current, (f) sample bias, and (g) sample current as a
  function of time for 2 iterations (75~eV, 100~eV) in the energy
  scan.  In (g), the averaging of $I_p$ is in red and the averaging of
  $I_t$ in green.\label{F:SEYtime}}

\end{figure}

\subsection{Ensuring Direct Photon Bombardment\label{S:protrude}}

With the closest bending magnet about 6~m from the samples, photons
radiated by the electron beam in the bend are nearly tangent to the
beam pipe wall (the angle is approximately 7~mrad from the tangent,
though it varies by a small amount depending on the electron beam
trajectory).  As the sample diameter is approximately 16 mm, a sample
which only slightly recessed from the beam pipe wall (by $\geq
0.1$~mm) is masked by the pipe and does not receive any direct SR
photons.

As shown in \autoref{F:grid}, the sample design includes a groove
in the front shoulder for a spring similar to that used in the back of
the sample; the edge of the front spring is visible in
\autoref{F:grid}d.  The front spring was used for the Phase I
measurements.  It is intended to ensure good electrical contact with
the beam pipe so that the image currents of the passing bunches are
minimally disrupted.  Very little difference was observed in the SEY
between the two samples in the Phase I measurements, which led to some
doubt as to whether the samples might be slightly recessed from the
beam pipe wall.

We replaced the beam pipe chamber after the end of Phase I due to a
vacuum leak in the original chamber.  We added view-ports to the beam
pipe opposite the samples to allow for inspection of the
samples and verification of the sample position.  Furthermore, we
omitted the front spring in Phase II (as it is not required for
typical CESR beam conditions, and, in fact, running beams with the
samples out of the beam pipe is not a problem).  We were able to
confirm that the samples were slightly protruding into the beam pipe
for all of the Phase II beam exposure periods.  In Phase II,
significant differences were observed in the early conditioning of the
samples.

\subsection{Mitigation of Electron Gun Current Drift\label{S:Istab}}

We observed in Phase I that the measured primary current ($I_p$)
changes slowly with time, in addition to being a function of energy.
For the Phase I measurements on aluminum and amorphous carbon-coated
samples, the ``before $I_t$'' and ``after $I_t$'' measurements of
$I_p$ differed by about 8\% on average and by about 16\% in the worst
case.  The first pair of measurements in \autoref{F:IpE} are close
to the typical reproducibility.

To reduce the systematic error due to current drift, a new measurement
procedure was developed for Phase II in which $I_p$ measurements are
interleaved with $I_t$ measurements.  As shown in
\autoref{F:SEYtime}, the Phase II measurement sequence is to set
the gun energy, apply a positive bias to the sample, move the beam to
the parking point and wait for the current to stabilize, measure $I_p$
at one grid point, apply a negative bias to the sample, park the beam
and wait for the current to stabilize, measure $I_t$ for all desired
grid points, and then proceed to the next energy.  As shown in
\autoref{T:SEYtimepar}, we wait for a time $t_{bw} = 60$~s after
changing the bias; this is a compromise between the need for a short
measurement time and the need to allow the transient current to
diminish (see \autoref{S:trans} below).  The longer waiting time
required us to reduce the number of energy steps
(\autoref{S:EngRes}).

With the Phase II method, we estimate that the error in the current
measurements due to gun current drift is $\lappeq 2$\%.

\subsection{Reduction of Charging and Parasitic Conditioning\label{S:charge}}

\begin{figure}
\centering
\includegraphics[width=\columnwidth]{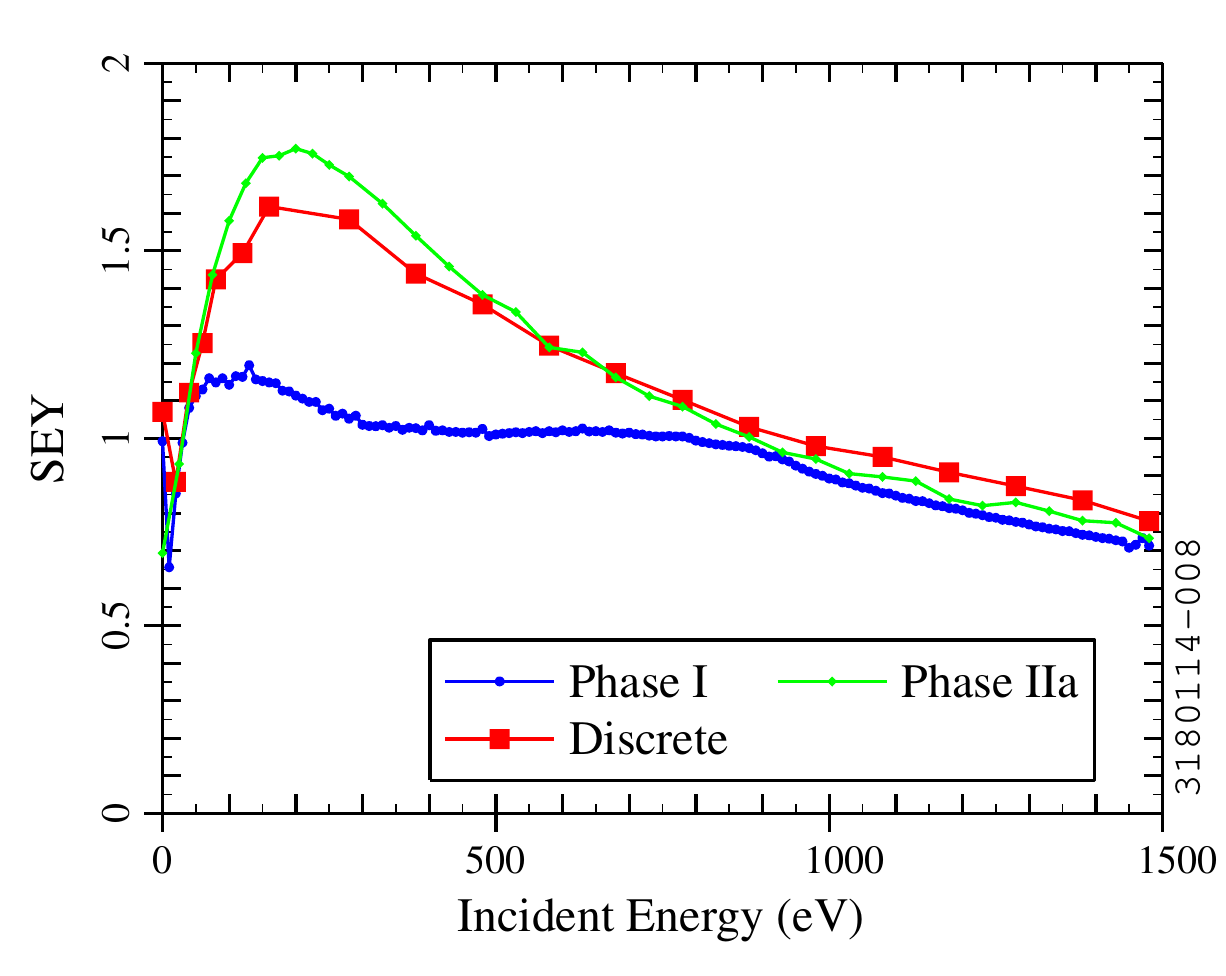}\\[-2ex]

\caption{SEY as a function of incident electron energy for a
  diamond-like carbon-coated aluminum sample, comparing different
  measurement methods.  The middle grid point is shown ($\theta =
  25\degree$).  Blue: Phase I method ($I_p \sim 2$~nA, 5 seconds for
  each energy, 9 grid points measured).  Red: ``discrete'' scan (large
  energy step, $I_p \sim 0.5$~nA, 3 to 4 minutes waiting period with
  the electron beam parked away from the middle point before each
  measurement; only 1 grid point measured).  Green: Phase IIa method
  ($I_p < 0.2$~nA, 9 grid points, beam parking away from middle grid
  point).\label{F:dlc}}

\end{figure}

SEY measurements on samples with diamond-like carbon (DLC) coatings on
aluminum were first done in 2011 in the off-line SEY station.  The DLC
coatings are being evaluated by KEK for SEY reduction and EC
mitigation.  Our DLC samples were provided by S. Kato (KEK\@).  A
measurement on DLC using the Phase I method is shown in
\autoref{F:dlc} (blue circles).  The SEY curve appears distorted.
We suspected that the distortion was due to charging of the DLC-coated
surface by the electron beam.  The charging is presumably due to the
DLC coating having insulator-like properties.  Qualitatively similar
effects have been reported in SEY measurements on other materials (for
example, measurements on MgO by Scholtz {\em et
  al.}\ \cite{APPSURFSCI111:259to264}).

To test the charging hypothesis, we remeasured the SEY with a long
wait time (3 to 4 minutes) between each energy step to allow the
surface to discharge, and with a smaller electron gun current ($\sim
0.5$~nA instead of $\sim 2$~nA) to reduce the supply of charge to the
sample.  To avoid charging during the waiting period, a single grid
point was measured, and the electron beam was parked at a different
grid point during the waiting period.  The results are shown in
\autoref{F:dlc} (red squares).  As can be seen, the additional
delay with the beam deflected and the reduction in current produced a
significant increase in the measured SEY\@.  The new curve is closer
to what one would expect based on measurements of other materials, as
well as being more consistent with other measurements on DLC
\cite{SHINKU48:118to120, ANTIECLOUDCOAT09:24, IPAC11:TUPS028}.

The results on DLC motivated us to reduce the electron gun current for
measurements in L3.  In Phase II, we used $I_p \sim 0.2$~nA for
standard measurements.  A side benefit of the current reduction was to
reduce unintended conditioning of the sample by the electron gun,
which, as discussed in \autoref{S:park}, should be minimised in
order to accurately measure the effect of the accelerator environment.
A complication is that, in Phase II, we switched the bias to measure
$I_p$ and $I_t$ at each energy (\autoref{S:Istab}), with a longer
waiting time to mitigate the transient current (see
\autoref{S:trans} below).  The longer waiting time increased the
integrated current per energy step; to shorten the measurement time
and reduce charging and conditioning, we adjusted the number of
energies measured (see \autoref{S:EngRes} below).  The net result
was an increase in the integrated flux for the parking point and a
decrease in integrated flux for other grid points.  A measurement on
the same DLC sample (in the $45\degree$ station) using the Phase IIa
method is also included in \autoref{F:dlc} (green diamonds).  The
differences between the discrete scan and the Phase IIa scan are
mainly due to the leakage correction (see Sections~\ref{S:LCmeas} and
\ref{S:LCC}) included in the Phase IIa case.

In Phase I, the parking point was between two grid points
(\autoref{S:phasei}).  In Phase II, due to the smaller deflection
step, the parking point was also a measurement point, as shown in
\autoref{F:GridSamXY}b below.  Hence, in Phase II, we measured the
SEY at the parking point, though there were issues with spatial
resolution in Phase IIa (see \autoref{S:hshake}).

In Phase IIb, an additional improvement was introduced, which was to
decrease the electron gun current by about a factor of 4 while waiting
for the sample current to stabilize after a change in the bias.  The
gun current is lowered for 50~s and then we return to the nominal gun
parameters for a time $t_{cw} = 10$~s to allow the gun emission
current to stabilise before the measurement (\autoref{F:SEYtime}).
The current modulation reduces the dose to the parking point by about
a factor of 3.5.  Details on how the current modulation was
implemented can be found in \autoref{S:gunmod}.  Though different in the
details, our method of current modulation is conceptually similar to
previously-used techniques for insulating materials (see
\cite{APPSURFSCI111:259to264}, for example).

With current modulation and Phase IIb parameters, one SEY scan
produces an integrated electron flux of $\sim 0.8$~$\mu$C/mm$^2$ for
the parking point and $\lappeq 12$~nC/mm$^2$ for the other grid
points.  In past studies on electron gun conditioning by other groups,
the peak SEY decreased by $\lappeq 10$\% for doses of order
1~$\mu$C/mm$^2$ for Cu \cite{EPAC00:THXF102, EPAC02:WEPDO014}, TiN
\cite{NIMA551:187to199}, and Al \cite{JVSTA23:1610to1618}.  Based on
this, we would expect to see a small amount of conditioning at the
parking point on unconditioned samples.  (However, there may be less
conditioning in our Phase II SEY scans because the electron energy is
low for a large fraction of the scan, and it has been found that
conditioning is less efficient at low energies \cite{PRL109:064801}).

We did not see a significant difference in the parking point's SEY
for Cu, stainless steel, TiN, or Al in Phase II\@.  An example is
shown in \autoref{S:exam}.

As described above, we found DLC to be more susceptible to charging.
Off-line measurements on an unconditioned DLC sample with Phase IIb
parameters showed a decrease in the measured SEY at the parking point
with current modulation ($\sim 7$\%) and a larger decrease without
current modulation ($\sim 24$\%).  On the other hand, a conditioned
and air-exposed DLC sample did not show a difference in measured SEY
at the parking point.  Additional measurements on DLC and amorphous C
are being done in the off-line station to better quantify their
susceptibility to charging and conditioning and check the
reproducibility of our observations.

\subsection{Leakage Current: Mitigation\label{S:LCmit}}

Ideally, the picoammeter measures only the current due to primary and
secondary electrons.  In reality, because the insulators are
imperfect, additional current flows through the picoammeter to ground
when the voltage bias is applied; this is generally referred to as
``leakage current.''  As has been pointed out previously, the leakage
current should be a small fraction of $I_p$ to avoid systematic errors
in the calculated SEY \cite{ECLOUD04:107to111}.  In Phase I, no
leakage corrections were applied.  As discussed in the preceding
section, initial measurements on DLC led us to reduce the electron gun
current in Phase II\@.  Because the relative contribution from the
leakage current increases as the gun current decreases, an effort was
made to quantify the leakage current and ascertain its effect while
preparing for Phase II.

Measurements indicated that the leakage current was strongly
correlated with the ambient humidity.  At high humidity, we found that
the leakage could be as high as several nA (hence exceeding
$I_p$) and could vary significantly in the time needed for an SEY
scan, which could produce large errors in the measurements.

As discussed in \autoref{S:elec}, we use a small shielded circuit
to apply the bias to the triaxial cable.  We found that there was
significant leakage in one of these circuits.  The circuits were
re-soldered more carefully and the exposed conductors were painted with
a silicone coating\footnote{Silicone Conformal Coating, 422-55ML, MG
  Chemicals, Surrey, BC, Canada.} to provide a moisture barrier.  After
this modification, the main leakage paths were found to be the
insulating stand-offs and the ceramic break (shown in green in
\autoref{F:ElecBias}).

\begin{figure}[b]
\centering
\includegraphics[width=\columnwidth]{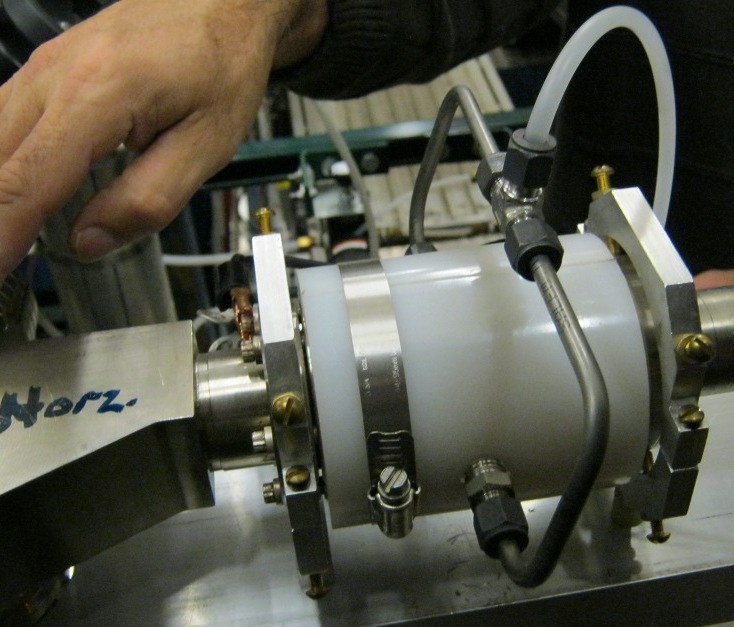}

\caption{Photograph of the horizontal SEY station after installation
  of a two-piece Teflon tube to establish a nitrogen gas blanket
  around the ceramic break.  The tube is connected to the grounded SEY
  chamber on the left.  There is a small gap on the right between the
  tube and the biased sample positioner arm.\label{F:ClamShells}}

\end{figure}

The decrease in resistivity of insulators due to moisture has been
extensively documented in the literature in the past century (see, for
example, Refs. \cite{BULLBSUS11:359to420, JAP17:318to325,
  IEEEEI11:76to80}).  In a humid environment, current is conducted
along the surface of an insulator, where there is a layer of moisture
from the ambient air.  These considerations led us to a redesign: (i)
the original G10 stand-offs were replaced by similar parts with a
smoother surface finish, more careful cleaning, and with blind holes
instead of through holes; (ii) a nitrogen gas ``blanket'' was made to
isolate the ceramic break from the ambient air.  The blanket was
established by covering the ceramic with a Teflon tube made from 2
halves.  As shown in \autoref{F:ClamShells}, the tube is attached to
the grounded flange of the ceramic break, with a small gap on the
biased flange of the ceramic break to avoid adding another path for
the leakage current.  (In \autoref{F:ElecBias}, the Teflon tube is
blue and the blanket region and gap are orange.)  The presence of the
gap required us to use a steady flow of nitrogen gas (about 2.5 SCFH
$\approx$ 20~mL/s per station) to establish the blanket.  The N$_2$
gas source is boil-off from the building's liquid nitrogen storage
Dewar, hence it has very low moisture content.

With high humidity, the nitrogen blanket alone did not produce a low
and stable leakage current; we had to first warm the ceramic with a
heat gun to remove the existing moisture.  We found that even a brief
interruption in the gas flow allowed moisture to return, necessitating
a reheat of the ceramic.  In a dry environment, the leakage current
with gas flow was stable without the need to heat the ceramic.  At a
low relative humidity, the leakage currents with and without gas flow
were comparable.  Additional information on the correlation between
humidity and leakage current can be found in \autoref{S:humid}.

After the modifications to the system, the typical leakage current was
of order 30~pA or less at $V_b = 150$~V\@.  This corresponds to an
error of $\lappeq 14$\% in the $I_p$ measurement for Phase II
parameters (not including the transient contribution, which is
discussed in \autoref{S:trans}).  Repeated measurements indicated that
the leakage current could still vary over time, even with the gas
blanket.  The variation can be as much as a factor of 2 over long
periods, as discussed in more detail in \autoref{S:LCtrend}.

\subsection{Leakage Current: Measurement\label{S:LCmeas}}

As discussed above, even with mitigation, the leakage current is not
negligible
relative to the Phase II gun current.  Consequently, in Phase II, we
added the step of measuring the leakage current prior to each SEY
measurement.  The leakage scan is done with the same data acquisition
method as the SEY scan, but with the electron gun turned off.  We
found that it is better to repeat several iterations of positive and
negative sample bias to allow the current to stabilise; however, we
perform fewer iterations for the leakage scan (16 typically) than for
the SEY scan (44 typically).

As will be discussed in \autoref{S:LCC}, the measured values of
$I_p$ and $I_t$ are corrected by subtracting the measured leakage
current with the corresponding sample bias before calculating SEY\@.
\autoref{S:LCtrend} includes more information on the measured
leakage as a function of time over the course of Phase II.

Time permitting, a second leakage scan is done after the SEY scan to
quantify the leakage current stability.  Typically, the leakage
currents before and after the SEY measurement agree within $\pm 2$~pA
with positive bias ($V_b = +150$~V) and within $\pm 0.5$~pA with
negative bias ($V_b = -20$~V\@).  Hence we estimate that the leakage
current drift contributes an error in the measured and corrected
currents of about 1\% of $I_p$ for the Phase II parameters.

\subsection{Transient Current: Mitigation\label{S:trans}}

A change in the sample bias produces a transient in the sample current
due to the stray capacitance of the system and the response of the
picoammeter.  The stray capacitance includes a contribution from the
triaxial cable and the SEY station, whose biased positioner arm is in
proximity to the grounded tube leading to the beam pipe and the
grounded support beam (\autoref{F:ElecBias}).  In the initial
measurements, the SEY stations made use of a special kapton-insulated
gasket instead of the ceramic break, and included bellows intended to
ease the alignment of the sample positioner with the beam pipe hole.
Drawings showing the original design can be found in an earlier paper
\cite{ECLOUD10:PST12}.  Prior to the start of Phase II, the insulated
gasket and bellows were replaced with a traditional gasket and ceramic
break (as shown in Figures~\ref{F:sys}, \ref{F:arm}, and
\ref{F:ElecBias}).  We estimate that the change from the insulated gasket
to the ceramic reduced the capacitance to ground from 1.4~nF to 10~pF.

After these hardware modifications, the transient signal was
nevertheless large, with a current spike peaking at about 0.5~nA
(hence exceeding the nominal $I_p$ for Phase II), and a decay time of
order 30~s (examples are included in \ref{S:derive}).
Ideally, one would wait for the current to reach its equilibrium value
before starting the measurement.  In Phase I SEY scans, the bias
voltage was switched only twice, so an extended waiting period after a
bias change was tolerable.  A wait time of several minutes was found
to be adequate.

On the other hand, with the Phase II procedure to mitigate the gun
current drift (\autoref{S:Istab}), the bias is switched twice for
each energy (\autoref{F:SEYtime}f), making a long wait time after
each bias change impractical, given the time available for the weekly
tunnel access.  Though the capacitance reduction associated with the
station redesign helped, the time to reach equilibrium was still too
long for practical measurements.  Hence a compromise solution was
necessary: waiting for time $t_{bw} = 60$~s after a bias change,
reducing the number of energy steps (\autoref{S:EngRes}), and
correcting for the residual effects from the transients.  Because the
leakage scans described above are done after switching the bias with
the same timing algorithm as is used for the SEY scans, the leakage
measurement also includes a contribution from the transient current
that has not vanished completely in the 1 minute wait time.  The
correction for the leakage current thus also corrects for the residual
transient current, which is about 4\% of $I_p$ for the Phase II
parameters.

Initial leakage measurements were done in conjunction with SEY scans
on 9 grid points.  Because of the relatively short time required for
the $I_t$ measurements, the leakage scans were done using only 1 grid
point.  When the number of grid points was increased for improved
spatial resolution (see \autoref{S:pts} below), we began using the
same number of grid points for the leakage scan as for the SEY scan.
The transient response of the system produces a change in the measured
leakage current over the time required to measure all of the grid
points (see \autoref{F:LCT} below).  When recording the current, we
also recorded the time stamp (with 0.1~ms resolution) in order to know
the time elapsed since the most recent change in sample bias, which
varies from one energy iteration to another for a given grid point due
to the variation in the number of grid points for higher energies
(\autoref{S:pts}) and the adjustments to the waiting time
(\autoref{S:cross}).  A time dependence was included in the
leakage correction to account for the change in current during the
$I_t$ measurements, as will be described in \autoref{S:LCC}.

\subsection{Energy Resolution and Segmentation\label{S:EngRes}}

As indicated above, Phase I measurements were done with a starting
electron gun energy of 20 eV, a final energy of 1500 eV, and an energy
step of 10 eV\@.  Because low-energy electrons are thought to be
important to the build-up of the electron cloud, a variable energy
step was introduced in Phase II to allow for smaller steps at lower
energies in routine measurements.  At the same time, the procedure was
adjusted so that both $I_p$ and $I_t$ were measured in a single energy
scan (\autoref{S:Istab}), with long waiting times at each energy
(\autoref{S:trans}).  To keep the overall measurement time short
enough to be compatible with the weekly access schedule, the energy
step was increased for higher energies (we were also motivated by the
need to minimise charging and conditioning by the electron beam, as
discussed in \autoref{S:charge}).  This resulted in a net decrease
in the number of energies measured.  The energy segments for the
variable-step scans are given in \autoref{S:EngSeg}.

\subsection{Improved Spatial Resolution and Range\label{S:pts}}

\begin{figure}
\centering
\includegraphics[width=\narrowwidth]{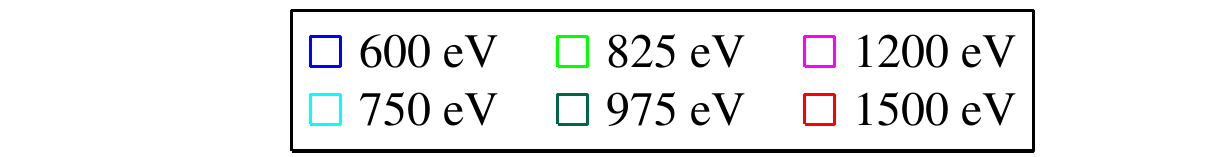}\\
\GRAFwidth[\narrowwidth]{490}{490}
\GRAFoffset{-65}{-65}
\GRAFlabelcoord{15}{355}
\incGRAFlabel{fig_grid_ang}{(a)}\\[-2ex]
\incGRAFlabel{fig_grid_xy}{(b)}\\[-2ex]

\caption{Grid points for double scans.  (a) Gun deflection angles for
  each grid point.  (b) Sample coordinates for each grid point ($x$
  and $y$ are the horizontal and vertical distance from the middle of
  the sample, respectively, in Cartesian coordinates).  Legend:
  maximum gun energy measured for each grid point.  Squares: first
  array; diamonds: second array.  Solid gray circles: estimated beam
  spot size at high gun energy.  Orange: sample face; yellow: sample
  shoulder.  Brown circles: edges of groove for the front spring and
  outer shoulder edge.  P: parking point.\label{F:GridSamXY}}

\end{figure}

As discussed above, the Phase I measurements were done over a 3 by 3
grid (\autoref{F:grid}a).  We found that more detailed information
would be useful to give us a more complete picture of the SEY's
dependence on position and angle.  Consequently, we implemented scans
with increased range and resolution in Phase II\@.  A uniformly-spaced
grid with high resolution and full range was not practical for weekly
measurements, so a compromise solution was developed: scanning with
high resolution and full range, but only over 3 horizontal segments
and 3 vertical segments.  (Occasional ``high definition'' scans are
done with high resolution and full range over the entire sample when
additional time is available.)  The grid points are shown in
\autoref{F:GridSamXY}.  The corresponding deflection parameters are
listed in \autoref{S:defl}.  One complication is that the largest
deflections cannot be reached at high energies, because the voltages
that can be applied to the electron gun deflecting electrodes are
limited to $\pm 150$~V\@.  The colors in \autoref{F:GridSamXY}
indicate the maximum gun energy measured for each grid point.  The
data acquisition software begins to skip some grid points when the gun
energy exceeds 600 eV, and measures only about half of the grid points
at the highest energy.  This complicates the timing of the
measurements, as will be discussed in \autoref{S:cross}.

For simplicity, the grid point layout shown in \autoref{F:GridSamXY}
is measured using two arrays of electron gun deflections.  As a
result, 9 of the grid points are measured twice (the repeated points
coincide with the points in the 3 by 3 grid of the Phase I
measurements).  This provides some additional information about
systematic and statistical errors.  In Phase IIa, the data acquisition
software allowed only one array, which required us to run the SEY
measurement twice.  In Phase IIb, we implemented the option of
specifying 2 arrays of gun deflections, so that all of the grid points
could be measured in one energy scan.  This allowed us to shorten the
measurement time significantly, since the majority of the time is
spent waiting for transients to settle after changing the sample bias.

To avoid additional complications in the data acquisition algorithms,
the electron gun deflection is varied linearly from one grid point to
the next, leading to the uniformly spaced deflection angles shown in
\autoref{F:GridSamXY}a (strictly speaking, it is the tangent of the
deflection angle that has a constant increment).  Because the electron
gun axis is at 25\degree{} relative to the sample axis, the grid point
spacing is not exactly symmetric between the left and right sides of
the sample.  Moreover, the sample face is curved, which produces some
distortion in the grid spacing between the middle of the sample and
the upper and lower portions.  These asymmetries in the grid layout
can be seen in \autoref{F:GridSamXY}b.  In the analysis of the SEY
measurements, the curvature of the sample and the angular offset of
the gun are taken into account when calculating grid point
coordinates.

In \autoref{F:GridSamXY}b, the sample's face is shown in orange and
the sample's shoulder is shown in yellow.  The brown circles represent
various features on the shoulder (which can be seen more clearly in
\autoref{F:grid}).  As can be seen, the grid point coordinates are
not all on the sample face.\footnote{For simplicity,
\autoref{F:GridSamXY}b does not show the actual point of impact
for grid points that are not on the sample's face---when the beam
does not hit the face, it travels a longer distance, resulting in
additional transverse motion.}  The solid gray circles indicate the
estimated beam size for a gun energy of 1500 eV (not taking into
account possible distortion in the size and shape of the beam spot for
large deflecting angles).  There is some overlap between adjacent grid
points over the majority of the sample.  As discussed in
\autoref{S:spot}, the estimated beam spot size is smaller at
intermediate energies; for the smaller spot size, none of the grid
points overlap.

The horizontal axes in \autoref{F:GridSamXY} are reversed in order
to show the grid points as viewed by an observer looking at the front
of the sample (the $xy$ coordinate system being based on the sign
convention for the electron gun deflection electrodes).

\subsection{Spatial Resolution: Time Control and Hand-Shaking\label{S:hshake}}

In the Phase I measurements, there was no ``hand-shaking'' operation
between the DAQP and picoammeter.  We unintentionally used
incompatible timing parameters between the picoammeter and the DAQP:
the picoammeter was set up to average the current over 1 second, but
the DAQP waited for only 0.2~s after a change in the deflection.  As a
result, $I_t$ measurements for grid points other than the first point
included significant averaging over more than one grid point.  The
DAQP used a waiting time of 1.5~s after an energy step, so there was
no unintentional mixing of different energies.

After measurements on the first few samples, the timing of
the current measurements was investigated more closely.  We realised
that the picoammeter and DAQP timing parameters were indeed
incompatible.  (Nevertheless, a statistically significant variation in
SEY as a function of grid point was observed in the early measurements
\cite{ECLOUD10:PST12, PAC11:TUP230}, in spite of the unintended
averaging over multiple grid points.)

After this problem was identified, alternative timing methods were
investigated.  We found that we could decrease the averaging time
without making the noise-to-signal ratio excessively large.
A hand-shaking algorithm was ultimately chosen: after adjusting the
energy, bias, and deflection, the DAQP waits for a settling time
$t_{dw}$, and then instructs the picoammeter to clear its buffer,
average the current, and return the averaged value.  The DAQP waits
for the picoammeter's value before proceeding to the next deflection
(or next combination of deflection, bias, and energy).  Because the
deflection is set prior to the start of the picoammeter's measurement,
the value does not include contributions from previous grid points.
With the new method, the current is averaged over $\frac{1}{6}$~sec.
Including the time for communication with the picoammeter and the wait
time after setting the deflection ($t_{dw}$), the net measurement time
per grid point is about 0.3~sec.  Additional information about the
picoammeter parameters is given in \autoref{S:pApar}.

\subsection{Time Control: Cross-Talk Avoidance\label{S:cross}}

As discussed in \autoref{S:elec}, low-noise triaxial cables connect
the sample positioner arms to the picoammeters, but the positioner
arms are not electrically shielded.  In L3, the stations are
relatively close to one another (when they are in the beam pipe, the
samples are about 0.4~m apart).  As shown in \autoref{F:SEYtime}f, the
Phase II SEY scan algorithm requires two steps in the bias voltage for
each energy.  We observed that a change in bias on one sample produces
a spike in the measured current of the other sample.  With Phase II
parameters, the current perturbation can be up to about 50\% of $I_p$.
However, the current spikes due to cross-talk have a short duration
($\lappeq 0.2$~sec), in contrast to the current transients due to a
bias change, which require a long waiting time.

\begin{figure}[htb]
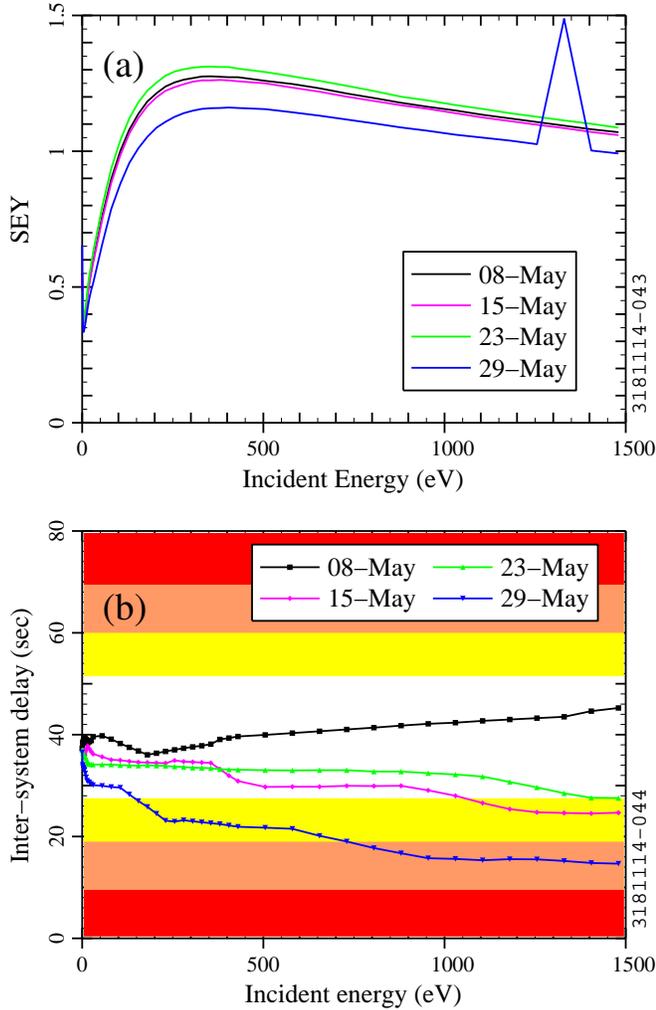

\centering
\GRAFwidth[\widewidth]{490}{390}
\GRAFoffset{-65}{-65}
\begin{tabular}{c}
\GRAFlabelcoord{15}{255}
\incGRAFlabel{fig_xtalk_exam}{(a)}\\[-2ex]
\GRAFlabelcoord{15}{235}
\incGRAFlabel{fig_delt_e_float}{(b)}\\[-2ex]
\end{tabular}

\caption{Examples of SEY scans without timing compensation using 63
  grid points (Cu, May 2012): (a) SEY as a function of incident energy
  for 45\degree{} sample ($V_x/V_{gsp} = -0.1$, $V_y/V_{gsp} = +0.05$).  (b)
  Timing delay of 45\degree{} system relative to horizontal system.
  Shaded areas: ``quiet zones'' for current measurements with 1 grid
  point (red), 63 grid points (orange), and 120 grid points
  (yellow).\label{F:XTalkFloat}}

\end{figure}

If the bias of one sample is changed while the current of the other
sample is being measured, this can produce a noise spike in the
measured SEY\@.  To illustrate, \autoref{F:XTalkFloat}a shows repeated
SEY measurements done on copper in the 45\degree{} system during Phase
IIa with 63 grid points.  A large spike in SEY is present for 1 out of
4 measurements.  The spike appears at different energies for different
grid points (a single grid point is shown in \autoref{F:XTalkFloat}a),
though always for $K \gappeq 1000$~eV\@.  In this case, the upward
spikes in SEY are the result of downward spikes in $I_t$.  No spikes
are seen for the horizontal sample.\footnote{For simplicity, the
  incident energy in \autoref{F:XTalkFloat} is not corrected to
  account for the gun deflection (see \autoref{S:CorEDefl}).}

\autoref{F:XTalkFloat}b shows the delay between current measurement
times for the two systems.  The delay values are based on the time
stamps for the $I_p$ measurements.  The shaded areas indicate ``quiet
zones'': if the time delay overlaps a shaded area, there is a risk of
a spike in the recorded current due to a bias change by the other
system.  For the present example with 63 grid points, the red and
orange areas are the relevant ones.  The spikes are observed when the
delay is in an orange zone (blue curve, $K \geq 750$~eV).  This is
consistent with our hypothesis that the spikes are due to cross-talk
between the systems.

To avoid the cross-talk, we implemented a delay between the start
times for the scans on the 2 samples to ensure that we did not switch
the bias for one system when measuring the current on the other
system.  With a waiting time after a bias change of $t_{bw} = 60$~s,
there is a timing margin of about $\pm 30$~s for measurements on one
grid point (indicated by the red zones in \autoref{F:XTalkFloat}b).
With the 120 grid points used for Phase IIb scans with improved range
and resolution, the $I_t$ measurements take about 35~s, which reduces
the timing margin to about $\pm 12$~s (yellow zones in
\autoref{F:XTalkFloat}b).  The basis for this timing margin is
discussed in \ref{S:TimingIIb}, which includes an example of current
measurements during a simultaneous SEY scan with both stations.

Even with control of the start time delay, spikes still occurred for a
significant fraction of the SEY scans.  Further investigation
indicated that occasionally the time to measure one grid point is
significantly longer than the nominal 0.3~s, sometimes being $\sim
1$~s.  When the longer delays are random, there is little cumulative
effect.  With the large number of grid points used in Phase II, we
found that a cluster of several longer delays sometimes occurs for one
system, accumulating enough time difference to produce cross-talk
again.  The problem can be seen in \autoref{F:XTalkFloat}b: the delay
between the two systems is initially about the same for all 4 scans,
but it varies during the scan, with 3 out of 4 cases getting near to
or crossing into a yellow area, and hence being marginal or unsuitable
for scans with 120 grid points.

We suspected that hand-shaking (\autoref{S:hshake}) was a source
of time variation.  However, we found that the hand-shake wait time
(time spent by the DAQP waiting for the picoammeter to return a value)
varies by only a few ms.  The evidence suggests that the time to write
values to the data file is the main cause for the occasional long
delays (at present the DAQP write the current to the data file after
each grid point, and generally the file is on a remote file system for
logistical reasons).

To eliminate the cross-talk problem in a reliable way, we modified the
timing algorithm.  When we initiate a measurement, the data
acquisition program uses the wait times and expected measurement time
per point to predict the overall time per energy step.  After each
energy step, the DAQP checks the time elapsed.  In the next energy
step, it adjusts the wait time before the $I_p$ measurement ($t_{bw}$
with $V_b = -150$~V) to compensate for the actual time of the previous
step being different from the desired time.  This prevents the timing
variations from producing a large cumulative time offset.  It has the
side effect that the wait time varies from one energy to another; this
is taken into account in the data analysis, as discussed in
\autoref{S:LCC}.

\begin{figure}[htb]
\centering
\includegraphics[width=\widestwidth]{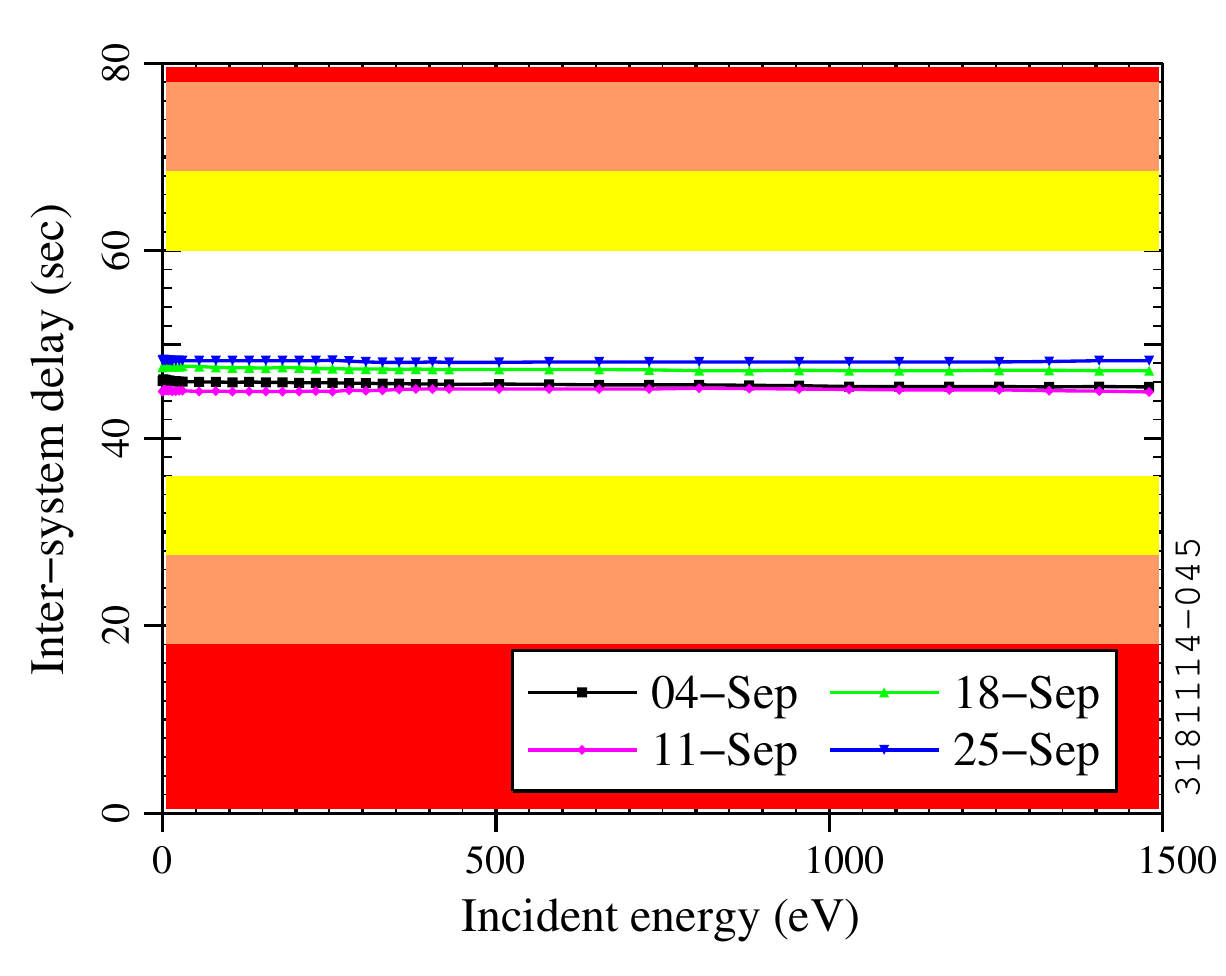}\\[0ex]

\caption{Examples of inter-system timing delay with timing
  compensation using 120 grid points (stainless steel, September
  2012).  Shaded areas: ``quiet zones'' for current measurements with
  1 grid point (red), 63 grid points (orange), and 120 grid points
  (yellow).\label{F:XTalkFix}}

\end{figure}

\autoref{F:XTalkFix} shows some examples of SEY scan timing with
compensation.  There is little cumulative change in the delay between
the two systems, and there is a comfortable timing margin for scans
with 120 grid points.  We did not observe any spikes in the SEY due to
cross-talk after implementing the compensation algorithm.

For scans with increased spatial range, there is the additional
complication that the number of grid points decreases with energy for
gun energies above 600 eV, as discussed in \autoref{S:pts}.  The
DAQP accounts for this by subtracting the number of skipped grid
points when calculating the expected time for a given iteration.  As a
result, the nominal wait times are the same for all energies.  As long
as their start times have the appropriate offset, the 2 systems remain
synchronised as the time per iteration decreases.

\subsection{Final Measurement Procedure}

Our procedure for SEY measurements in Phase IIb is as follows: (i)
move the samples from the beam pipe to the measurement position and
close the gate valves; (ii) do a leakage scan, switching between
positive and negative bias (\autoref{S:LCmeas}); (iii) warm up the
electron guns for 30 to 60 minutes and then adjust the gun parameters
to make $I_p$ approximately 200~pA (\autoref{S:warm}); (iv) do an
SEY scan; (v) repeat the leakage scan if time permits; (vi) return the
samples to the beam pipe.  With our standard parameters, the leakage
scan takes 40 minutes and the SEY scan takes 110 minutes.  Including
set up and removal of the equipment, the full measurement takes about
5 hours.  This requires us to measure the samples in parallel rather
than sequentially, since the access time is typically 6 hours or less.

\section{Data Analysis\label{S:analysis}}

The SEY is calculated from $I_p$ and $I_t$ using
\cref{E:seyt}.  The gun energy is corrected to account for
the effect of the electrostatic deflection.  The sample's voltage bias
is taken into account when associating the SEY with an incident
energy.  The measured values of $I_p$ and $I_t$ are corrected to
account for leakage current and current transients when calculating
the SEY.

\subsection{Energy Correction for Electrostatic Deflection\label{S:CorEDefl}}

The electron gun accelerates its electrons to the set point energy
with a longitudinal electrostatic field.  Paired deflection plates at
the exit of the gun deflect the electrons electrostatically to the
desired horizontal angle ($\alpha_x$) and vertical angle ($\alpha_y$)
relative to the axis of the gun.  Because the kicks are produced by an
electric field, they change the kinetic energy of the electrons, in
addition to changing their direction.  Hence the kinetic energy $K_g$
of the electrons is larger than the set point value $K_{gsp}$ when they
are deflected.  In the non-relativistic case,
\begin{equation}
K_g = K_{gsp} [1 + \tan^2(\alpha_x) + \tan^2(\alpha_y)]
\end{equation}
For the 3 by 3 grid used in Phase I, the gun energy is 2.6\% higher
than the set point energy in the worst case.  For the standard double
grid used in Phase IIb, the gun energy increases by 9.5\% in the worst
case.  The energy correction was not included in the preliminary
reports on the SEY measurements \cite{CLNS:12:2084, ECLOUD10:PST12,
PAC11:TUP230, IPAC13:THPFI087}, though most of the results were for
the middle of the sample, where no correction is needed.

\subsection{Correction for Sample Bias Voltage\label{S:BiasCor}}

The SEY is in general a function of the kinetic energy $K$ and angle
$\theta$ of the incident primary electron: SEY = SEY$(K, \theta)$.
The measurements of $I_p$ and $I_t$ are done while scanning the gun
energy (as indicated in \autoref{T:eng}).  Because the sample is
biased, the incident kinetic energy of electrons reaching the sample
is the sum of the electron gun energy ($K_g$) and the electron charge
magnitude ($q_e$) times the bias voltage ($V_b$):
\begin{equation}
K = K_g + q_e V_b\label{E:K}
\end{equation}
As a result, the incident energy is in principle smaller than the gun
energy by 20~eV when we measure $I_t$ with $V_b = -20$~V and larger by
150~eV when we measure $I_p$ with $V_b = 150$~V\@.  Ideally, the
negative bias for the $I_t$ measurement repels all of the secondary
electrons produced at the surface of the sample, while the positive
bias prevents the escape of any secondaries.  With the assumption that
the intrinsic primary current to the sample is independent of the bias
voltage and that no secondaries escape with $V_b = 150$~V for the
$I_p$ measurement, we may use $V_b = -20$~V in \cref{E:K} to
calculate the appropriate incident energy $K$ associated with the
measured SEY\@.  This correction is included for recent measurements
\cite{IPAC13:THPFI087}, though we plotted SEY simply as a function of
$K_g$ for early analyses \cite{CLNS:12:2084, ECLOUD10:PST12,
PAC11:TUP230}.

Because the primary electrons' incident angle is not normal to the
sample ($\theta \neq 0$), the sample bias can change not only the
kinetic energy of the incident electrons, but also their trajectory,
and hence can shift their incident angle and impact position before
the electrons reach the sample.  A simple analytic model suggests that
this effect will be significant for electron energies of order 100~eV
and lower (at 100~eV, the model predicts a shift in the incident angle
of $3\degree$ and a shift in the impact point by 1~mm for the middle
grid point).  The analytic model assumes a simplified electric field
and is likely to overestimate the deflection.  At present, our data
analysis does not account for the shift in angle and impact position
(we plan to develop a model that describes the SEY measurements in a
more complete way in the future).  As a result, the reader should be
cautious about making inferences about the SEY for energies lower than
about 100~eV based on our measurements.

\subsection{Time-Dependent Correction of Leakage Current and Current Transients\label{S:LCC}}

In Phase I, $I_p$ was about 2~nA; in Phase II, we decreased $I_p$ to
about 200~pA to reduce charging and conditioning
(\autoref{S:charge}).  As a result, the leakage current became a
larger fraction of $I_p$, and we took steps in Phase IIa to mitigate
(\autoref{S:LCmit}), measure (\autoref{S:LCmeas}), and correct
for the leakage current.  A static leakage correction was done
initially: the leakage current was measured with the same bias
voltages as for the SEY measurement and the average leakage current
for each bias was subtracted from $I_p$ and $I_t$.  In the leakage
measurement, the bias was switched with the same wait time as for the
$I_p$ and $I_t$ measurements, in order to account for the transient
response of the system (\autoref{S:trans}).

In the course of Phase IIa, we increased the number of grid points for
better spatial resolution (\autoref{S:pts}), and began to measure
the leakage current with the same timing of bias switching as done for
the SEY measurement; we observed that the leakage current was changing
slowly in the time required to measure all of the grid points.  This
led us to develop a model for the leakage current that includes a
contribution from the transient current.  The derivation of the model
is given in \ref{S:derive}.  We use the time-dependent
leakage model for the analysis of Phase II measurements with more than
9 grid points.

\begin{figure}
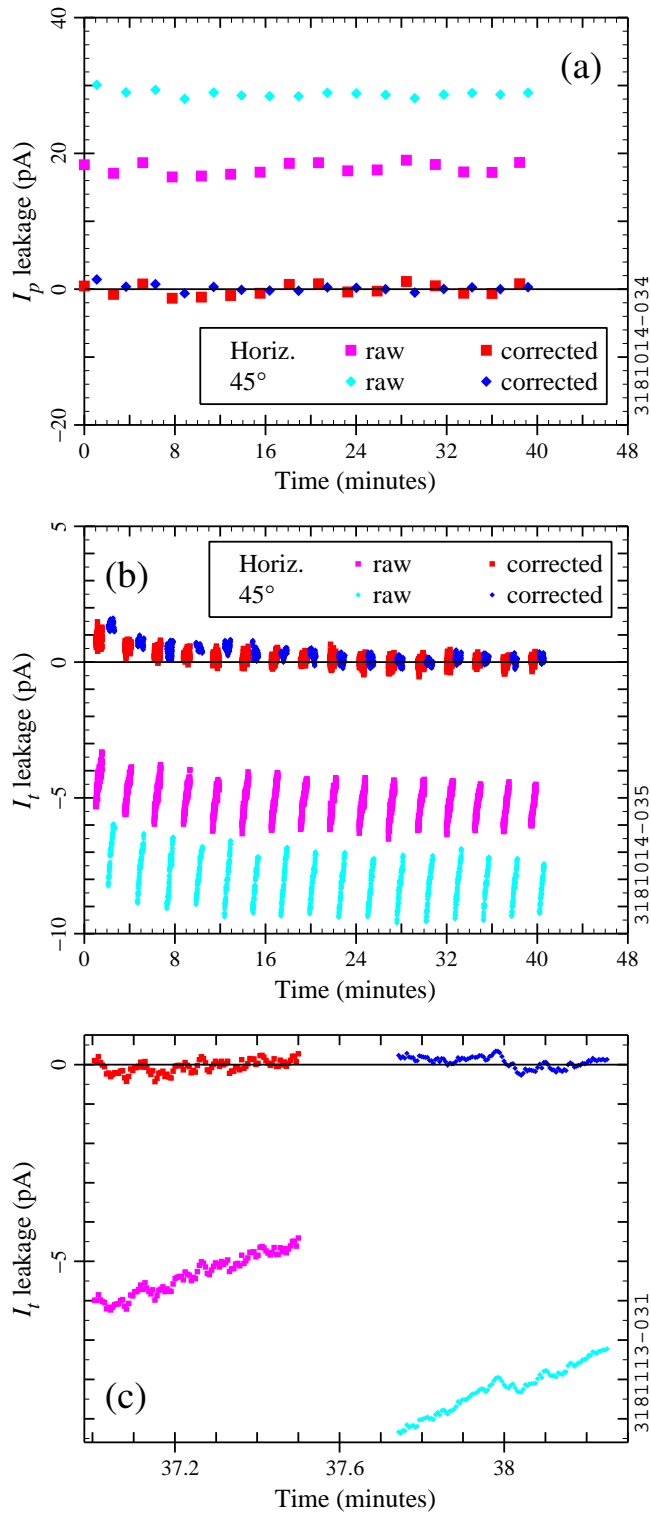

\centering
\GRAFwidth[\narrowwidth]{490}{390}
\GRAFoffset{-65}{-65}
\GRAFlabelcoord{350}{255}
\incGRAFlabel{fig_leak_t_a}{(a)}\\[-2ex]
\GRAFlabelcoord{15}{255}
\incGRAFlabel{fig_leak_t_b}{(b)}\\[-2ex]
\GRAFlabelcoord{15}{25}
\incGRAFlabel{fig_leak_t_c}{(c)}\\[-2ex]

\caption{Example of measurements of leakage current as a function of
  time while switching the bias voltage.  (a) Leakage measurements
  with $V_b = +150$~V for $I_p$ correction (1 point per iteration);
  (b) leakage measurements with $V_b= -20$~V for $I_t$ correction (120
  points per iteration); (c) same as (b), but zoomed in for a better
  view of the penultimate iteration.  Light colors (magenta and cyan)
  indicate uncorrected values and dark colors (red and blue) indicate
  corrected values.  The measurements were done in August 2012 with
  stainless steel samples.\label{F:LCT}}

\end{figure}

\autoref{F:LCT} shows an example of leakage scans on both stations.
The lightly-colored markers (magenta and cyan) indicate the measured
(``raw'') current as a function of time.  With $V_b = 150$~V
(\autoref{F:LCT}a), the leakage current is about 20~pA for the
horizontal station and about 30~pA for the 45\degree{} station.  With
$V_b = -20$~V (\autoref{F:LCT}b), the leakage current is about $-8$~pA
for the horizontal station and about $-5$~pA for the 45\degree{}
station.  The measurements with negative bias are repeated 120 times,
which takes about 35~s, following the same timing algorithm as for the
SEY scans.  The current changes by 2 to 3 pA during the measurement,
which is about 2\% of $I_p$ for Phase II parameters.

The darkly-colored markers (red and blue) in \autoref{F:LCT} show the
result of applying the time-dependent leakage correction described in
\autoref{S:lcmodel}.  Over most of the scan, the corrected current is
$\pm 1$~pA or less, which is about 1\% of $I_p$, comparable to the
leakage current drift (\autoref{S:LCmeas}).  There are larger
discrepancies during the first few minutes of the scan, which are
likely due to the need for iterations to reach a stable current
(\autoref{S:LCmeas}).

As can be seen in \autoref{F:LCT}b, the time-dependent correction
compensates for the transient behaviour reasonably well.  Zooming in
on one iteration (\autoref{F:LCT}c), we see that the corrected
current differs from zero, but the systematic differences are
comparable to the noise in the measurement.

The time-dependent correction is based on the recorded time stamp for
each current measurement, so that variations in the time per grid
point and the adjustments in the waiting time to avoid cross-talk
(\autoref{S:cross}) are taken into account.
This has the added advantage that we can still apply the leakage
correction procedure even if the timing and scanning parameters are
not exactly the same for the leakage scan and the SEY scan.

The measurements shown in \autoref{F:LCT} were taken over the same
time interval, with measurements at positive and negative bias
interleaved, and with a time offset between the horizontal and
45\degree{} stations.  The time is measured relative to the first
current measurement for the horizontal station.

\begin{figure}[tb]
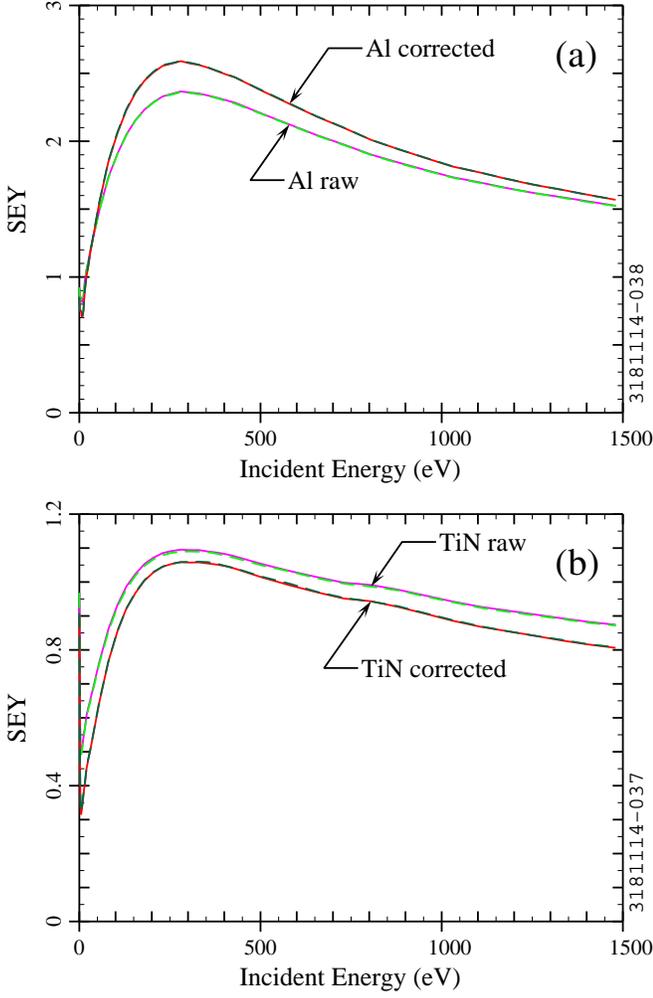

\centering
\GRAFwidth[\middlewidth]{490}{390}
\GRAFoffset{-65}{-65}
\GRAFlabelcoord{350}{255}
\incGRAFlabel{fig_sey00d_al2lcex}{(a)}\\[-2ex]
\incGRAFlabel{fig_sey00d_tin2lcex}{(b)}\\[-2ex]

\caption{Example of measurements of SEY as a function of energy
  without correction (light colors) and with correction for leakage
  current and transient response (dark colors): (a) unconditioned Al
  measured in January 2013; (b) reconditioned TiN measured in November
  2012.  Both the first measurement (solid curves) and second
  measurement (dashed curves) are shown.  All values are for the
  middle grid point of the 45\degree{} sample ($\theta =
  25\degree$).\label{F:SEYcor}}

\end{figure}

The SEY is calculated from $I_p$ and $I_t$ using \cref{E:seyt}.
Because the numerator and denominator are both corrected, the effect
on the SEY can be partially cancelled.  For example, if the
uncorrected and corrected $I_t$ values are small relative to $I_p$,
SEY is approximately 1 and the corrections to $I_p$ do not produce
much change in the SEY\@.  \autoref{F:SEYcor} shows examples of the
current correction's impact on the SEY values.  For unconditioned Al
with a peak SEY of approximately 2.5 (\autoref{F:SEYcor}a), the
correction increases the peak by about 10\%.  For reconditioned TiN
with a peak SEY close to 1 (\autoref{F:SEYcor}b), the correction
decreases the peak by about 5\%.

In \cref{F:SEYcor}, both the first (solid curves) and the repeated
(dashed curves) measurements are shown, since this grid point is
measured twice in the double scan; the first and second $I_t$
measurements are separated in time by $\sim 17$~s.  The $I_t$ values
are corrected by different amounts to account for the current
transient.  However, there is little difference in SEY, which
indicates that the variation in $I_t$ over the time required to scan
the grid points has little impact on SEY\@.

Thus, in the examples above, the magnitude of the leakage current is
approximately 15\% of $I_p$ or less, which is typical for the Phase II
measurements with leakage mitigation; the unmitigated leakage current
could be 100\% of $I_p$ or more under adverse conditions.  The
correction to the measured SEY due to the leakage current is
approximately 10\% or less, which is also typical for Phase II
measurements.  With 120 grid points, there is a clear time dependence
in the leakage current due to the bias switch transient.  The
time-dependent leakage correction accounts for this effectively, but
the impact on the measured SEY is small for the Phase II SEY scans, in
which we used a wait time after a bias switch of $t_{bw} = 60$~s.
Hence it may be possible to shorten the wait time in future
measurements.  (However, as discussed in \autoref{S:cross}, with the
Phase IIb parameters, there is little timing margin to avoid
cross-talk; hence a different cross-talk avoidance method would be
needed if we were to decrease $t_{bw}$ significantly.)

\subsection{Uncertainties\label{S:uncertain}}

As discussed above, a number of modifications for Phase II were
oriented toward reducing the systematic errors in the measurements.
\autoref{T:ErrSummary} summarises the estimated contributions to the
systematic error from various sources for various measurement
scenarios.  The values apply to both $I_p$ and $I_t$ measurements, but
they are expressed as a percentage of $I_p$ (it is not straightforward
to estimate the error as a fraction of $I_t$).

\begin{table}[htbp]

\caption{Summary of estimated current measurement errors as a
  percentage of $I_p$.  For errors due to leakage and transient
  currents, the Phase II value of $I_p \sim 200$~pA is assumed; the
  estimated errors would be smaller for Phase I, since $I_p$ was $\sim
  2$~nA\@.  The scenarios used for the final Phase II procedure are in
  bold type.  HH = high humidity, LH = low humidity (as quantified in
  \autoref{S:humid}), SC = static correction, TDC = time-dependent
  correction.\label{T:ErrSummary}}

\begin{center}
\begin{tabular}{l|ll|r}\hline
Source & Mitigate? & Correct for? & Error \\ \hline\hline
Gun current      & no (Ph. I) & no & $\lappeq 16$\% \\
drift (\S \ref{S:Istab})   & {\bf yes} (Ph. II) & {\bf no} & \boldmath $\lappeq 2\%$ \\ \hline
Leakage          & no (HH) & no & $\gappeq 100$\% \\
current          & no (LH) & no & $\lappeq 14$\% \\
(\S \ref{S:LCmit}--\ref{S:LCmeas})    & no (HH) & yes & $\gappeq 100$\% \\
                 & no (LH) & yes & $\lappeq 1$\% \\
                 & yes & no & $\lappeq 14$\% \\
                 & {\bf yes} & {\bf yes} & \boldmath $\lappeq 1\%$ \\ \hline
Transient        & no ($t_{bw} = 0$) & no & $\gappeq 100$\% \\
current          & yes ($t_{bw} = 60$ s) & no & $\lappeq 4$\% \\
(\S \ref{S:trans}, \S \ref{S:LCC}) & yes ($t_{bw} = 60$ s) & yes (SC) & $\lappeq 2$\% \\
                 & {\bf yes} ($t_{bw} = 60$ s) & {\bf yes} (TDC) & {\boldmath $\lappeq 1\%$} \\ \hline
Cross-talk       & no (Ph. IIa) & no & $\lappeq 50$\% \\
 (\S \ref{S:cross})       & {\bf yes} (Ph. IIb) & {\bf no} & {\bf none} \\ \hline
\end{tabular}
\end{center}
\end{table}

Using \cref{E:seyt}, one can infer the impact of errors in
the measurement of $I_p$ and $I_t$ on the calculated SEY\@.  For Phase
IIb, we expect the items listed in \autoref{T:ErrSummary} to produce
a systematic error in SEY of at most a few percent for $0 \leq$ SEY
$\leq 2$.

Estimates of errors due to charging and conditioning are not included
in \autoref{T:ErrSummary}.  From \autoref{F:dlc}, we infer that
the error in the calculated SEY due to charging was $\sim 45$\% for DLC
with the Phase I method.  With the Phase IIb method, as discussed in
\autoref{S:charge}, our observations indicate that there is some
charging or conditioning of unconditioned and susceptible materials at
the parking point (decreasing the measured SEY by $\lappeq 7$\% with
mitigation).  We will return to the discussion of errors due to
conditioning in \autoref{S:reproduce}.

Overall, we expect the items considered in \autoref{S:phaseii} to
contribute a few percent to the systematic error in the SEY for most
grid points (and $\lappeq 10$\% for the parking point) with the Phase
IIb method.  We estimate that the statistical errors are of the same
order.  A future paper will include more detailed results with a more
complete error analysis.

\section{Examples of SEY Results\label{S:exam}}

Some examples of SEY measurements with the in-situ stations are
presented in this section.  The beam dose to the samples is calculated
in terms of the integrated current of stored electron bunches; for a
beam energy of 5.3 GeV, 1 ampere$\cdot$hour corresponds to about $3
\cdot 10^{21}$~photons/m of direct synchrotron radiation at the
location of the SEY samples.  As discussed above, the horizontal
sample is oriented to receive direct SR photons, but the 45\degree{}
sample receives only scattered photons (though both samples may be
conditioned by electrons from the cloud).

\subsection{SEY as a Function of Energy}

\begin{figure}[htb]
\centering
\includegraphics[width=\widewidth]{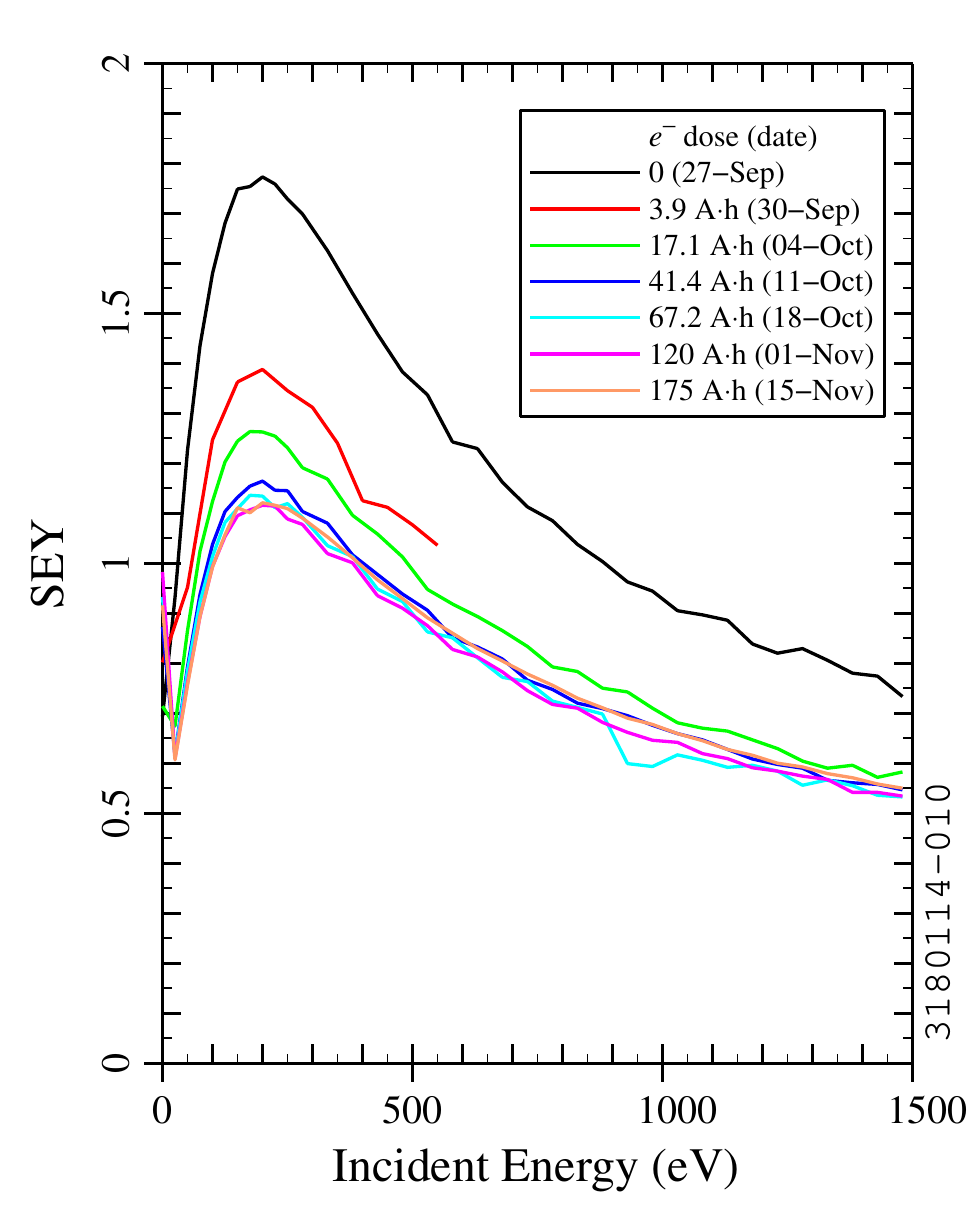}\\[-2ex]

\caption{SEY as a function of incident energy for the 45\degree{}
  diamond-like carbon sample.  The values shown are for the grid point
  in the middle of the sample, which has $\theta = 25\degree$.  The
  sample was exposed to the beam pipe environment during Phase IIa
  from September 2011 to November 2011.  The second measurement (red
  curve) was stopped early due to time constraints.\label{F:SeyEng}}

\end{figure}

\autoref{F:SeyEng} shows the measured SEY of the $45\degree$
diamond-like-carbon-coated sample as a function of energy for
different beam doses.  The measurements indicate a peak in the SEY at
about 200~eV\@.  There is a clear decrease in the SEY as a function of
beam dose.  Before conditioning, the peak SEY is about 1.8; for beam
doses greater than 20 ampere$\cdot$hours, the peak SEY is
significantly lower, in the range of 1.1 to 1.2.  The observed changes
due to conditioning are large compared to the estimated errors in the
measurement.

\subsection{Peak SEY as a Function of Vertical Position}

\begin{figure}
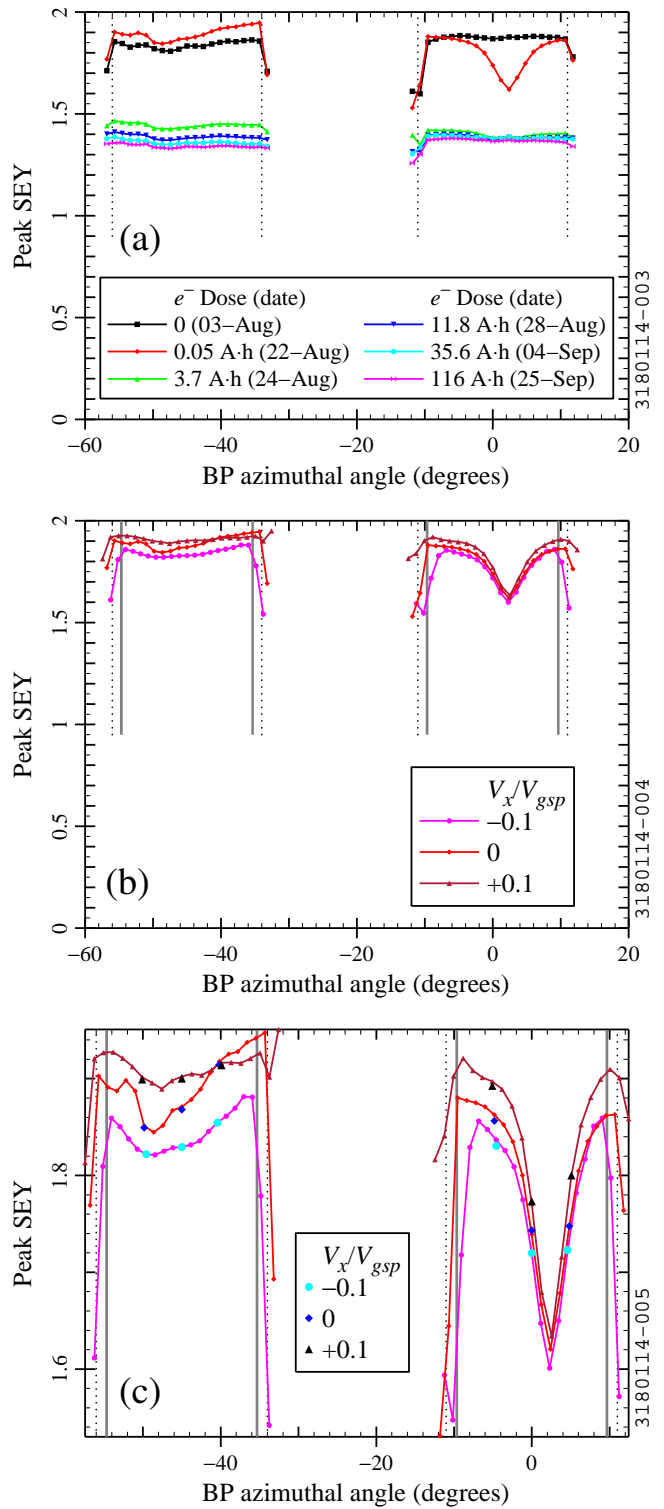

\centering
\GRAFwidth[\narrowwidth]{490}{390}
\GRAFoffset{-65}{-65}
\GRAFlabelcoord{25}{125}
\incGRAFlabel{fig_seyp_y_dose}{(a)}\\[-2ex]
\GRAFlabelcoord{15}{25}
\incGRAFlabel{fig_seyp_y_x}{(b)}\\[-2ex]
\GRAFlabelcoord{25}{25}
\incGRAFlabel{fig_seyp_y_zoom}{(c)}\\[-2ex]

\caption{Peak SEY of stainless steel samples as a function of position
  expressed in terms of the azimuthal angle along the beam pipe (BP).
  (a) Scans along the middle of the sample ($\theta \approx
  25\degree$) for different beam doses; (b) scans along the left,
  middle, and right of the sample for the 0.05 A$\cdot$h case; (c)
  zoomed-in version of (b) with repeated points from the double scan
  included.  The black dotted lines indicate the edges of the sample
  for the middle deflection scan; the solid gray lines correspond to
  the sample edges for the left and right deflection scans.  The
  measurements were done in Phase IIb from August 2012 to September
  2012.\label{F:SeyHeight}}

\end{figure}

\autoref{F:SeyHeight} shows measurements of the peak SEY as a function
of vertical position for stainless steel.  The gun deflection angle is
converted to azimuthal angle along the inside of the beam pipe.  The
coordinate system is such that the middle of the horizontal sample is
at zero and the middle of the 45\degree{} sample is at $-45\degree$.

\autoref{F:SeyHeight}a compares different beam doses.  Before beam
exposure (black), the peak SEY is about 1.8 and is approximately
constant.  After a small beam dose (red), a dip in the SEY appears
near the middle of the horizontal sample, presumably due to photon
bombardment from direct SR\@.  For high doses, the SEY decreases and
returns to being approximately independent of position.  The
measurements thus suggest that direct SR produces rapid conditioning,
while conditioning by scattered photons and/or electrons happens more
slowly.  In the stainless steel case, little difference is seen except
for the lowest dose (which is a small fraction of the typical weekly
beam dose with CHESS currents).  The observed differences due to beam
scrubbing are again large compared to the estimated errors in the
measurement.

As described in \autoref{S:pts}, double scans are done for 3 different
values of the horizontal and vertical deflection.
\autoref{F:SeyHeight}b compares the peak SEY as a function of position
for 3 different horizontal deflections (grid points with square
markers in \autoref{F:GridSamXY}) for the 0.05 A$\cdot$h case (the red
case in \autoref{F:SeyHeight}a).  The dip near the middle of the
horizontal sample is seen in all 3 scans.  \autoref{F:SeyHeight}c
shows a zoomed-in version of \autoref{F:SeyHeight}b with additional
values for repeated grid points included, as will be discussed in
\autoref{S:reproduce}.  The values in \autoref{F:SeyHeight}b and
\autoref{F:SeyHeight}c are labelled according to the relative
horizontal deflection, $V_x/V_{gsp}$, which is defined in
\autoref{S:defl}.

\subsection{Peak SEY as a Function of Incident Angle}

\begin{figure}[tb]
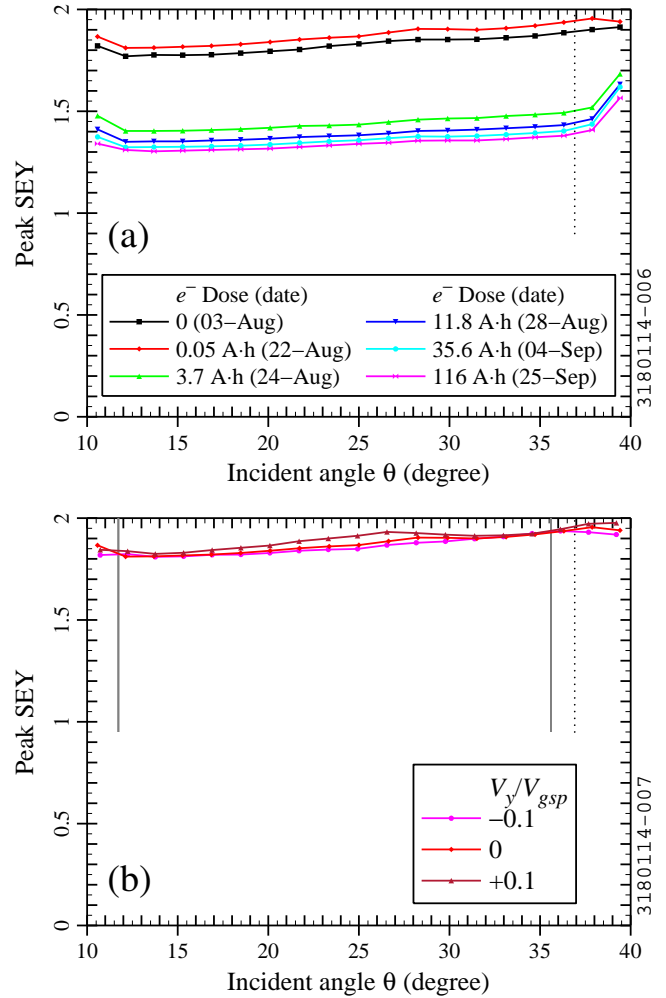

\centering
\GRAFwidth[\middlewidth]{490}{390}
\GRAFoffset{-65}{-65}
\GRAFlabelcoord{15}{125}
\incGRAFlabel{fig_seyp_x_dose}{(a)}\\[-2ex]
\GRAFlabelcoord{15}{25}
\incGRAFlabel{fig_seyp_x_y}{(b)}\\[-2ex]

\caption{Peak SEY of the 45\degree{} stainless steel sample as a
  function of incident angle relative to the surface normal.  (a)
  Scans along the middle of the sample for different beam doses; (b)
  scans along high, middle, and low lines along the sample for the
  0.05 A$\cdot$h case.  The black dotted lines indicate the edges of
  the sample for the middle deflection scan; the solid gray lines
  correspond to the sample edges for the high and low deflection scans.
  The measurements were simultaneous with the measurements of SEY as a
  function of position shown in the previous figure.\label{F:SeyAngle}}

\end{figure}

\autoref{F:SeyAngle} shows the peak SEY as a function of incident
angle for the measurements on stainless steel described in the
previous section.  The results are based on scanning the horizontal
deflection and converting to the angle of incidence $\theta$ relative
to the sample's surface normal.  (Hence, the position and the angle
are both varying.)  The incident angle is 25\degree{} in the middle of
the sample.  Only the 45\degree{} sample is shown.

\autoref{F:SeyAngle}a compares different beam doses.  Consistent
with the left side of \autoref{F:SeyHeight}a, the peak SEY
decreases with increasing beam dose.  There is a slight increase in
SEY with increasing angle, which is qualitatively consistent with what
one would expect---generally, the SEY has been observed to increase as
the primary beam's angle changes from normal incidence to grazing
incidence.

\autoref{F:SeyAngle}b compares the peak SEY as a function of angle
for 3 different vertical deflections (grid points with diamond markers
in \autoref{F:GridSamXY}) for the 0.05 A$\cdot$h case.  There is
some variation from one vertical deflection to another, but all cases
show a similar increase in SEY with increasing $\theta$.  The
consistency between different vertical positions and different beam
doses suggests that our measurements are able to resolve the angular
dependence of the SEY, though the differences are not very large
compared to the estimated errors in the measurement.

The curves in \autoref{F:SeyAngle}b are labelled according to the
relative vertical deflection, $V_y/V_{gsp}$ (defined in
\autoref{S:defl}).  As discussed in \autoref{S:charge}, the integrated
flux from the electron gun is highest for the parking point.  The
parking point has $V_x/V_{gsp} = +0.1$ (corresponding to $\theta
\approx 31.5\degree$) and $V_y/V_{gsp} = +0.1$.  As can been seen in
\autoref{F:SeyAngle}b, there is no evidence of additional conditioning
at the parking point for stainless steel.

\subsection{Reproducibility\label{S:reproduce}}

As discussed in \autoref{S:pts}, nine grid points are measured twice
in the double scan (shown by overlapping squares and diamonds in
\autoref{F:GridSamXY}). This provides a way to check the
reproducibility of the SEY measurement over short time intervals.  In
\autoref{F:SeyHeight}c, the scans along the 3 different horizontal
lines from \autoref{F:SeyHeight}b are shown in a zoomed-in view (in
shades of red), with additional values for the repeated grid points
included (in black and shades of blue).  In both cases, the colors
range from light to dark as the horizontal deflection ranges from
negative to positive.  The values from the repeated measurement are
reasonably consistent.  This suggests that the features in the
vertical position scans are reproducible and that the system is able
to properly resolve the dependence of SEY on position.

We are not able to repeat the in-situ SEY measurements routinely, but
we can occasionally during extended access periods.  Repeated
measurements on Al samples (1 or 2 days apart, without intervening
exposure to beam) in Phase II indicate that the measured SEY can vary
by up to 5\% or more for a few grid points.  For most of the 120 grid
points, the SEY varies by a few percent or less.  Thus the measured
changes with beam exposure are large compared to the day-to-day
reproducibility of the measurements.  The reproducibility is
consistent with what we expect based on the systematic uncertainties
discussed in \autoref{S:uncertain}.

\section{Conclusion}

We have developed an in-situ secondary electron yield measurement
system to observe conditioning of metal and coated samples by CESR
beams.  Our system allows for the measurement of SEY as a function of
incident electron energy, position on the sample, and incident
electron angle.  Our experience with the initial measurements led us
to implement improvements in the method to reduce charging and
conditioning by the electron gun; mitigate and correct for the leakage
current and transient current; eliminate cross-talk between the
adjacent SEY stations; and mitigate the slow drift in the electron gun
current.  We have reduced the contributions to the systematic error
from these effects to a few percent, allowing us to measure the
dependence of the SEY on beam dose, incident angle, and position with
better resolution.  In-situ measurements have been carried out on a
number of materials.  Preliminary results have been reported
previously, and will be presented in more detail in a future paper.

SEY models generally divide the secondaries into elastic, rediffused,
and true secondaries.  Our measurement technique is well-suited to the
low-energy true secondaries, but not as well-suited to the
higher-energy rediffused and elastic secondaries.  Hence, we expect to
be able to extract reasonable model parameters for the true secondary
contribution (as will be presented in a future paper), but we expect
more uncertainty in the model parameters of the rediffused and elastic
contributions.

There is room for additional improvements in the SEY measurement
techniques.  Sources of systematic error that we have not yet
accounted for include (i) the escape of elastic and rediffused
secondaries when the sample is positively biased to measure the
primary current, and (ii) the deflection of electrons by the sample
bias in the case of primary electrons with a small incident energy.
Our progress with (ii) is important to our goal of finding SEY model
parameters that are reliable for incident electrons of low energies.
A more direct measurement of the primary and/or secondary current may
help reduce some of the systematic error.  We would like to gather
additional information about the energy distribution of the secondary
electrons, in order to distinguish true secondaries, rediffused
secondaries, and elastic secondaries.  Additional improvements to the
measurement apparatus and techniques might allow us to reduce the
measurement time and decrease the incidence of noise spikes in the
current due to nearby activity.  Some of the improvements described
above may not be practical for our in-situ apparatus, and may have to
be implemented with a more advanced out-of-tunnel SEY measurement
system.

Our ultimate goal is to use the SEY measurements under realistic
conditions to constrain the SEY model parameters as much as possible;
this will help improve the predictive ability of models for electron
cloud build-up, allowing for more successful electron cloud mitigation
in future accelerators to help them achieve better performance and
higher reliability.

\section*{Acknowledgments}
\addcontentsline{toc}{section}{Acknowledgments}

We are grateful for the support of collaborators at SLAC, who provided
hardware, samples, and guidance for the SEY studies at \CesrTA.
Amorphous carbon coating and diamond-like carbon coating of samples
was done by CERN and KEK, respectively.  We thank our collaborators at
Fermilab for useful discussions.

Our work would not have been possible without the support of personnel
in the design, electronics, fabrication, information technology,
operations, survey, technical services, and vacuum groups.  We are
particularly thankful for the work by V. Medjidzade and the support
from W. J. Edwards, B. M. Johnson, J. A. Lanzoni, R. Morey, and
R. J. Sholtys.  We thank our \CesrTA{} collaborators for their
support, ideas, and encouragement, particularly J. R. Calvey,
J. A. Crittenden, G. F. Dugan, J. P. Sikora, and K. G. Sonnad.
S. T. Wang provided valuable help and guidance with our data
acquisition program development work.  We thank S. B. Foster for doing
off-line SEY measurements and helping with in-situ measurements.  We
appreciate the support from the CESR and laboratory management for our
studies, particularly from M. G. Billing, D. H. Rice, D. L. Rubin,
J. W. Sexton, and K. W. Smolenski.


This work was supported by the National Science Foundation through
Grants PHY-0734867 and PHY-1002467 and by the Department of Energy
through Grants DE-FC02-08ER-41538 and
DE-SC0006505.

\appendix

\section{Current Control and Measurement}

\subsection{Electron Gun Current Control\label{S:gunctl}}

The cathode power (adjusted via the cathode voltage, $V_{source}$)
provides the primary method of controlling the electron gun current.
The Wehnelt potential (grid bias, referred to as $G_1$ by the gun
manufacturer) and first anode potential (referred to as $G_2$ by the
manufacturer) provide additional control over the gun parameters,
including the current; the current decreases with $G_1$ and increases
with $G_2$.  For our SEY measurements, we establish the desired
electron gun current by applying a constant voltage across the cathode
filament ($V_{source} = 1.2$~V) and setting $G_1$ and $G_2$ values of
order 16~V and 100~V, respectively; the values of $G_1$ and $G_2$ vary
by a few percent from one measurement to another in order to get the
desired gun current.

The electron gun power supply has a beam current read-back and an
``emission control'' feature in which a feedback loop adjusts the
source voltage to make the gun current read-back equal to a set point
value.  However, for our parameters, we found that the feedback loop
went unstable after an energy step.  As a result, all of the SEY
measurements have been done with the feedback loop turned off.

For low-current SEY measurements (see \autoref{S:charge}), the gun
current is too low for an accurate read-back value.  Even at higher
currents, we sometimes observe that the gun current read-back is
inaccurate.  As a result, we generally rely on the sample current as
measured by the picoammeter for the SEY measurements rather than the
gun current read-back.

\subsection{Electron Gun Current Modulation\label{S:gunmod}}

As discussed in \autoref{S:charge}, in Phase IIb, we decreased the
electron gun current while waiting for the sample current to stabilize
after a change in the bias.  This is done by raising $G_1$, rather
than by decreasing $V_{source}$ (had we adjusted $V_{source}$, we
would have had to account for the thermal time constant of the cathode
and would have likely worsened the long-term stability of the cathode
emission characteristics).  During each 60-second waiting period after
a change in bias, we increase $G_1$ by 1~volt.  The increase in $G_1$
produces a decrease in the gun emission current by about a factor of
4.  The choice of $\Delta G_1 = 1$~V is a compromise between our
desire for a large change in $G_1$ to minimise the dose and our desire
to keep $\Delta G_1$ small for stable gun current.  With $\Delta G_1 =
1$~V, we found that the gun current stabilises within about 7~seconds
after the step in $G_1$.  This was the basis for returning to the
nominal value of $G_1$ for a time $t_{cw} = 10$~s before starting the
measurement (\autoref{F:SEYtime}).

\subsection{Current Measurement Parameters\label{S:pApar}}

As described in \autoref{S:hshake}, the picoammeter parameters
were adjusted in Phase II to avoid unintentional averaging over grid
points.  The picoammeters average the current internally and provide
the averaged value to the data acquisition program.  The picoammeters
can be set for either a ``moving average'' without hand-shaking or a
``repeated average'' with hand-shaking.  The time required per
measurement is the product of the integration time of the analog to
digital converter and the number of points to be averaged.  The 
parameters for the Phase I and Phase IIb measurements are shown in
\autoref{T:pApar}.

\begin{table}[htbp]

\caption{Picoammeter parameters for Phase I and Phase IIb.\label{T:pApar}}

\begin{center}
\begin{tabular}{lcc}\hline
Phase & I & IIb \\ \hline
Integration time & $\frac{1}{10}$ s (slow) & $\frac{1}{60}$ s (med.)\\
Average type & moving & repeated\\
Hand-shaking & no & yes\\
Measurements & 10 & 10\\
Time needed per grid point & 1 s & $\frac{1}{6}$ s\\ \hline
\end{tabular}
\end{center}
\end{table}

\section{Parameters for Scanning the Energy, Focus, and Deflection}

\subsection{Energy Segments\label{S:EngSeg}}

A variable energy step was used in Phase II for better energy
resolution at low energies (\autoref{S:EngRes}).  The energy
segments for variable-step SEY scans are listed in \autoref{T:eng}.

\begin{table}[htb]

\caption{Electron gun energy segments for SEY scans with a variable
  energy step.  A total of 44 energies are measured.\label{T:eng}}

\begin{center}
\begin{tabular}{lrrrr}\hline
Interval & \multicolumn{1}{c}{Start} & \multicolumn{1}{c}{End} & \multicolumn{1}{c}{Step} & Points \\ \hline
1st & 20 eV &  25 eV & 1 eV & 6\\
2nd & 25 eV &  35 eV & 2 eV & 5\\
3rd & 35 eV &  50 eV & 5 eV & 3\\
4th & 50 eV & 450 eV &25 eV &16\\
5th &450 eV &1500 eV &75 eV &14\\ \hline
\end{tabular}
\end{center}
\end{table}

\subsection{Focus as a Function of Energy\label{S:focus}}

\begin{figure}[htb]
\centering
\includegraphics[width=\columnwidth]{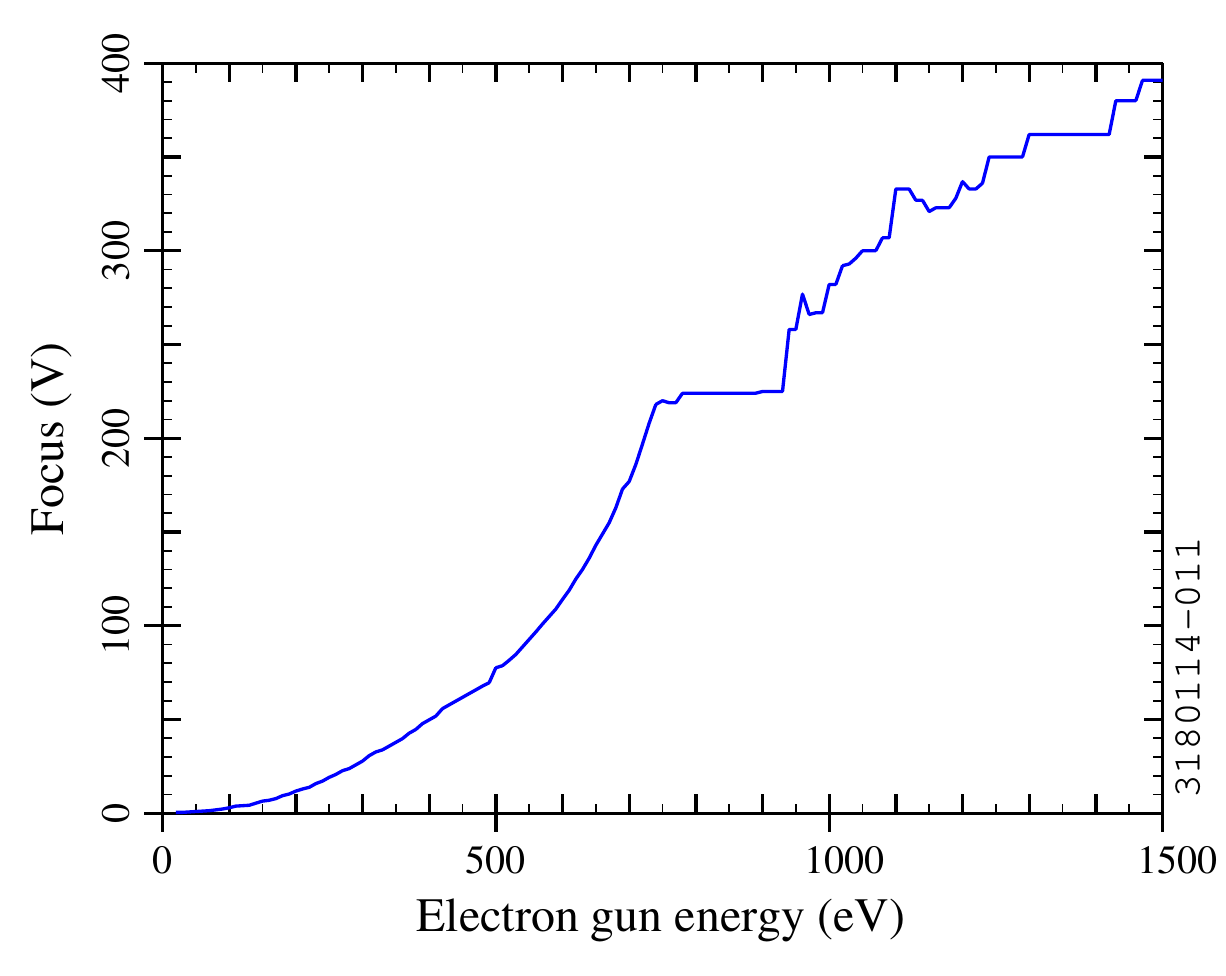}\\[-2ex]

\caption{Set point focussing voltage as a function of electron gun
  energy.\label{F:focus}}

\end{figure}

The electron gun focussing voltage is adjusted as a function of
electron gun energy to produce a minimum electron beam spot size on
the sample (\autoref{S:spot}).  The set point values are shown in
\autoref{F:focus}.

\subsection{Deflection Parameters\label{S:defl}}

Paired electrostatic plates deflect the electrons to the desired
horizontal angle ($\alpha_x$) and vertical angle ($\alpha_y$) relative
to the gun axis.  The deflection set point is specified by the
voltages $V_x$ and $V_y$ to be applied to the deflecting electrodes.
The $x$ deflection angle, $x$ deflection voltage, and gun energy set
point $K_{gsp} = q_e V_{gsp}$ are related by
\begin{equation}
\tan(\alpha_x) = g \frac{V_x}{V_{gsp}}\; ,\label{E:defl}
\end{equation}
where $g$ is a constant that depends on the spacing and length of the
deflecting electrodes ($q_e$ = the electron charge magnitude).  The
relationship between $\alpha_y$ and $V_y$ is analogous.  For our
electron gun model, $g = \frac{40}{3} \tan(4.9\degree)$.  We checked
the above relationship using a phosphor screen to view the electron
beam spot.

Two arrays of grid points were used for better spatial range and
resolution (\autoref{S:pts}).  The deflection parameters for the
double scans are given in \autoref{T:grids}.  Per \cref{E:defl}, the
deflecting voltages must be scaled with the set point energy $K_{gsp}
= q_e V_{gsp}$.  The relative horizontal deflection $V_x/V_{gsp}$ is
incremented from $-(V_x/V_{gsp})_{max}$ to $+(V_x/V_{gsp})_{max}$ with
a given step size; the relative vertical deflection $V_y/V_{gsp}$ is
incremented similarly.  The relative deflection step is 0.025 at high
resolution and 0.1 at low resolution.  The desired step size and range
determines the dimensions of the array ($n_x$ by $n_y$).  As shown in
\autoref{F:GridSamXY}, the arrays overlap, resulting in some grid
points being measured twice.

\begin{table}[tb]

\caption{Grid parameters for double scans.\label{T:grids}}

\begin{center}
\begin{tabular}{c|cc|cc}\hline
\multicolumn{5}{c}{\bf Standard Double Scan}\\ \hline
Array & $n_x$ & $n_y$ & $(V_x/V_{gsp})_{max}$ & $(V_y/V_{gsp})_{max}$ \\ \hline
1 &  3 & 21 & 0.1   & 0.25 \\
2 & 19 &  3 & 0.225 & 0.1  \\ \hline\hline
\multicolumn{5}{c}{\bf High Definition Double Scan}\\ \hline
Array & $n_x$ & $n_y$ & $(V_x/V_{gsp})_{max}$ & $(V_y/V_{gsp})_{max}$ \\ \hline
1 &  3 &  3 & 0.1   & 0.1  \\
2 & 19 & 21 & 0.225 & 0.25 \\ \hline
\end{tabular}
\end{center}
\end{table}

Some additional measurements are done with a ``high definition'' array
(along with a low definition array which provides the same duplication
of points as for the standard double scan).  The deflection parameters
for high definition scans are also included in \autoref{T:grids}.

\section{Collimation Slit Measurements\label{S:collim}}

As discussed in \autoref{S:spot}, collimation measurements were
done to determine the focus setting as a function of beam energy to
produce an approximate minimum in the beam spot size and estimate the
minimum spot size as a function of energy.  Furthermore, as described
in \autoref{S:shield}, the collimation measurements provided a way
to check the magnetic shielding of the system.

In the collimation measurements, the sample was biased with +20~V and
was used as a Faraday cup.  A collimator with a 1~mm slit was
electrically isolated from the sample and centered in front of the
sample, with the slit oriented in the vertical ($y$) direction.  With
the electron gun at the nominal distance from the sample (32.9~mm),
two picoammeters were used to measure the electron currents reaching
the collimator and the sample.

\subsection{Focus and Beam Spot Size}

At each electron beam energy, the gun's focusing voltage was
varied to find the focus to produce the maximum current to the sample
and the minimum current to the collimator.  These measurements
provided the basis for the set point focus values used in the SEY
measurements and shown in \autoref{F:focus}.

\begin{figure}[htb]
\centering
\includegraphics[width=\columnwidth]{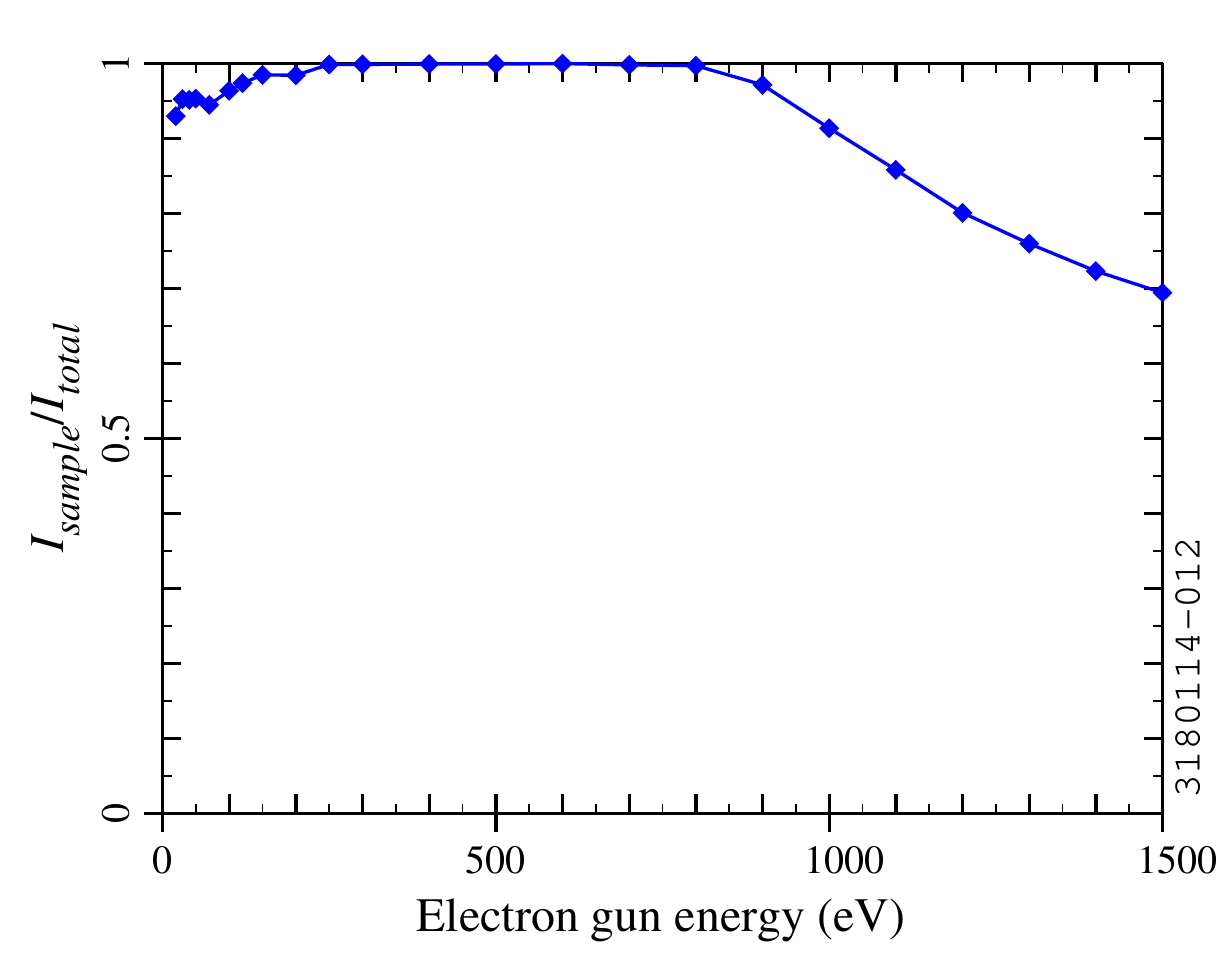}\\[-2ex]

\caption{Slit collimation measurements for the SEY system.  For the vertical
axis, $I_{sample}$ is the current reaching the sample and $I_{total}$
is the current reaching the sample plus the current reaching the
collimation electrode.\label{F:collim}}

\end{figure}

\autoref{F:collim} shows the current passing through the slit and
reaching the sample divided by the total current (current to sample
plus current to collimator) as a function of energy with the focus set
as described above and without $x$ deflection.  For beam energies
between 200~eV and about 800~eV, nearly all of the current reaches the
sample, indicating that the beam spot size is smaller than 1~mm.
Outside this energy range, some of the current is intercepted by the
collimator.  These results are the basis for the beam size estimates
given in \autoref{S:spot}; we assumed a Gaussian beam distribution
based on the specifications provided by the electron gun manufacturer.

We attempted to do a more direct measurement of the beam spot size
using a phosphor screen.  However, we found that the beam spot was not
visible except at high electron beam energies with high electron beam
current.  As a result, we relied on the method described above
instead.

\subsection{Deflection Check}

At each energy, the beam was scanned across the slit using the gun's
horizontal ($x$) deflection electrode to find the deflection value for
maximum current to the sample and minimum current to the collimation
electrode (corresponding to the beam passing through the middle of the
slit).  Over the full range of electron beam energy (0 to 1500~eV),
the value of the $x$ deflection voltage to center the beam spot on the
slit was zero, which confirms that the stray magnetic field is well
shielded.

\section{Model for Time-Dependent Correction of Leakage
  Current and Current Transients\label{S:derive}}

As indicated in \autoref{S:LCC}, with a large number of grid
points, we observed that the leakage current was changing slowly in
the time required to measure $I_t$ for all of the points.  This
provided the motivation to develop a model for the leakage current
that accounted for the transient current.

\subsection{Simple Measurements; Semi-Empirical Model\label{S:SEmodel}}

Our first approach was to develop a relatively simple circuit model
with a voltage source representing the picoammeter's biasing power
supply; a series resistance $R_s$ between the power supply and the
sample; a resistance $R_\parallel$ between the sample and ground to
produce a leakage current; and a capacitance $C_\parallel$ from the
sample to ground to produce a transient current.  Unsurprisingly, this
circuit's response to a step in the power supply voltage is a
transient current which decays exponentially in time to a steady-state
current.

\begin{figure*}[tb]
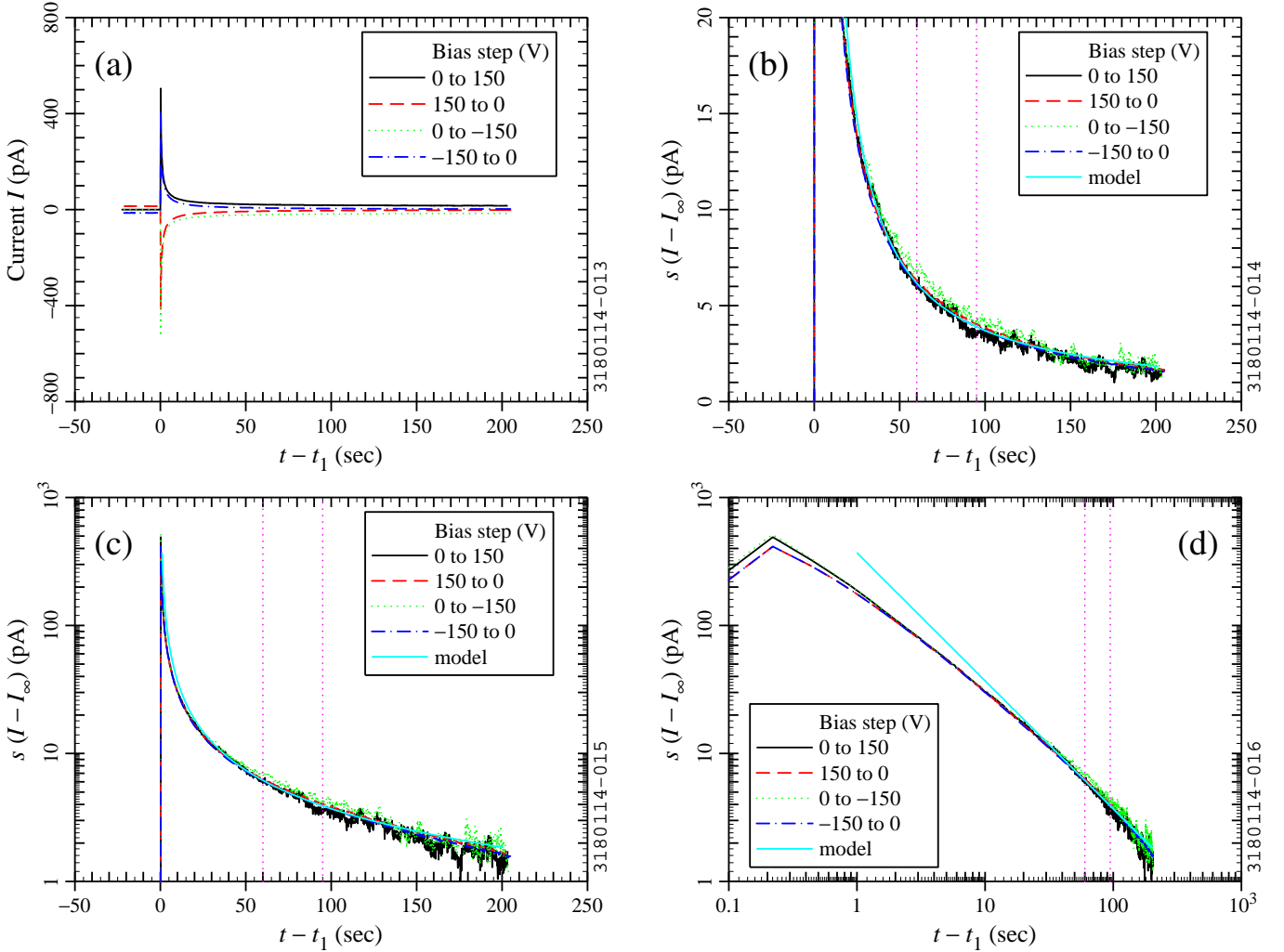

\centering
\GRAFwidth[0.49\textwidth]{490}{390}
\GRAFoffset{-65}{-65}
\GRAFlabelcoord{15}{255}
\incGRAFlabel{fig_trans_i_t_a}{(a)}%
\hspace*{\fill}%
\incGRAFlabel{fig_trans_i_t_b}{(b)} \\[-2ex]
\incGRAFlabel{fig_trans_i_t_c}{(c)}%
\hspace*{\fill}%
\GRAFlabelcoord{350}{255}
\incGRAFlabel{fig_trans_i_t_d}{(d)}
\\[-2ex]

\caption{(a) Measured current as a function of time for the
  45\degree{} SEY station with a step in the sample bias at time
  $t_1$.  Measured current as a function of time with subtraction of
  the steady state current $I_\infty$ and adjustment of the sign ($s =
  \pm1$), with (b) linear, (c) linear-log, and (d) log-log scale.
  The values of $I_\infty$ are 15~pA, 0, $-14$~pA, and 1~pA, in the
  same order as given in the legend.  The solid cyan curve represents
  the model described in the text.  The dotted vertical lines indicate
  the time range of interest for SEY measurements with a double
  scan.\label{F:dIvt}}

\end{figure*}

To evaluate the usefulness of the circuit model, we measured the
sample current $I$ as a function of time $t$ after stepping the bias
voltage.  \autoref{F:dIvt}a shows the current as a function of time
for the 45\degree{} SEY system for simple cases, stepping the bias
voltage between $V_b = 0$ and $V_b = \pm 150$~V\@.  The approximate
time at which the bias is stepped ($t_1$) is subtracted from $t$.  The
measured transient current is as high as about $\pm 500$~pA, which is
much larger than the steady-state current
of $\pm 15$~pA or less.

In \autoref{F:dIvt}b, we subtract the approximate steady-state current
($I_\infty$) from $I(t)$ and we zoom in for a more clear view of the
current for $t - t_1 > 20$~seconds.  Additionally, $I - I_\infty$ is
multiplied by a sign correction coefficient $s = \pm 1$ to compare the
transients associated with upward and downward steps in $V_b$ on the
same footing.  For our present method, the time interval of interest
for SEY measurements starts 60 seconds after the bias step and, in the
case of the double scan, lasts for 35 seconds.  The dotted vertical
lines in \autoref{F:dIvt}b delimit this time interval.  (For high
definition double scans, the measurement takes 118 seconds.)

The values of $I_\infty$ are given in the caption of
\autoref{F:dIvt}.  Ideally, we should have $I_\infty = 0$ when $V_2
= 0$ and the same value of $|I_\infty|$ when $V_2 = \pm 150$~V, but
the best-match results vary by 1~pA, which presumably is indicative of
small offsets in the system.

Since the circuit model predicts that the current $I$ should decay
exponentially towards its steady state value $I_\infty$, a linear-log
plot of $I(t) - I_\infty$ versus $t - t_1$ should produce a straight
line.  However, as shown in \autoref{F:dIvt}c, the measured current
has a distinctive curvature on linear-log scales.  This indicates that
the current does not decay exponentially to its steady-state value as
predicted by the circuit model.  Our inference is that the picoammeter
is an active element of the circuit, not a passive element as assumed
for the circuit model.\footnote{The picoammeter has settings for
``damping on'' and ``damping off'' which affect its time response.
We found that neither setting gives an exponential decay in the
transient current.  All of the SEY scans were done with the default
setting of damping on.}

\autoref{F:dIvt}d shows a log-log plot of $I(t) - I_\infty$ versus $t - t_1$.
For $t - t_1 > 30$~seconds, the relationship is approximately linear.
The solid cyan line of \autoref{F:dIvt}d follows the simple form
\begin{equation}
I(t) - I_\infty = \frac{A}{t - t_1} \; ,\label{E:IAIinf}
\end{equation}
where $A$ is a constant.  The cyan curves in \autoref{F:dIvt}b and
\autoref{F:dIvt}c represent the same function as the cyan line of
\autoref{F:dIvt}d.  Based on considerations from the circuit model,
we expect $A$ and $I_\infty$ to be constants for given values of the
initial and final bias, with $I_\infty$ being approximately
proportional to the final bias voltage and $A$ being approximately
proportional to the voltage step.

\autoref{F:dIvt}d shows that the measured current differs
significantly from the cyan line of \cref{E:IAIinf} for $t
-t_1 < 30$~seconds; the model in fact predicts an infinite current for
$t - t_1 \rightarrow 0$; however, we are searching for a model which
can be applied for $t - t_1 \geq 60$~seconds, so the discrepancies for
$t - t_1 < 30$~seconds are not a problem for us in practice.

Denoting the initial bias as $V_0$ and the final bias as $V_2$, and
taking into account the above comments about $A$ and $I_\infty$, we
can formulate \cref{E:IAIinf} as
\begin{equation}
I(t) = \Gamma_\parallel \left(\frac{V_2 - V_0}{t - t_1}\right) + \frac{V_2}{R_\parallel} \; ,\label{E:ICeRp}
\end{equation}
where $\Gamma_\parallel$ and $R_\parallel$ are constants.  For $t
\rightarrow \infty$, $I(t) \rightarrow I_\infty = V_2/R_\parallel$,
consistent with the simple circuit model with resistance $R_\parallel$
between the sample and ground (for $R_s << R_\parallel$).  The
constant $\Gamma_\parallel$ has dimensions of capacitance, and can be
thought of as being a ``capacitance-like'' quantity, even though
\cref{E:ICeRp} does not represent the transient behaviour of
an $RC$ circuit.

The values of the model parameters corresponding to the cyan curves in
\autoref{F:dIvt} are given in \autoref{T:TDpar}.  Measurements of
the current as a function of time after a voltage step were done on
the horizontal station and the off-line station, in addition to those
on the 45\degree{} station reported above.  The measured currents for
the other stations were found to be consistent with the semi-empirical
relation of \cref{E:ICeRp}, although the horizontal system's
current was a bit more noisy.  The model parameters for the other
systems are also included in \autoref{T:TDpar}.

\begin{table}[htbp]

\caption{Values of the semi-empirical model parameters for the SEY
  stations inferred from the measured current as a function of time
  after a voltage step.\label{T:TDpar}}

\begin{center}
\begin{tabular}{lcc}\hline
Station & $\Gamma_\parallel$ & $R_\parallel$ \\ \hline
Horizontal & 2.0 pF & 15.8 T$\Omega$ \\
45\degree{} & 2.5 pF & 10.4 T$\Omega$ \\
Off-line & 2.5 pF & 14.6 T$\Omega$ \\ \hline
\end{tabular}
\end{center}
\end{table}

\subsection{Realistic Measurements; Additional Considerations}

\begin{figure*}
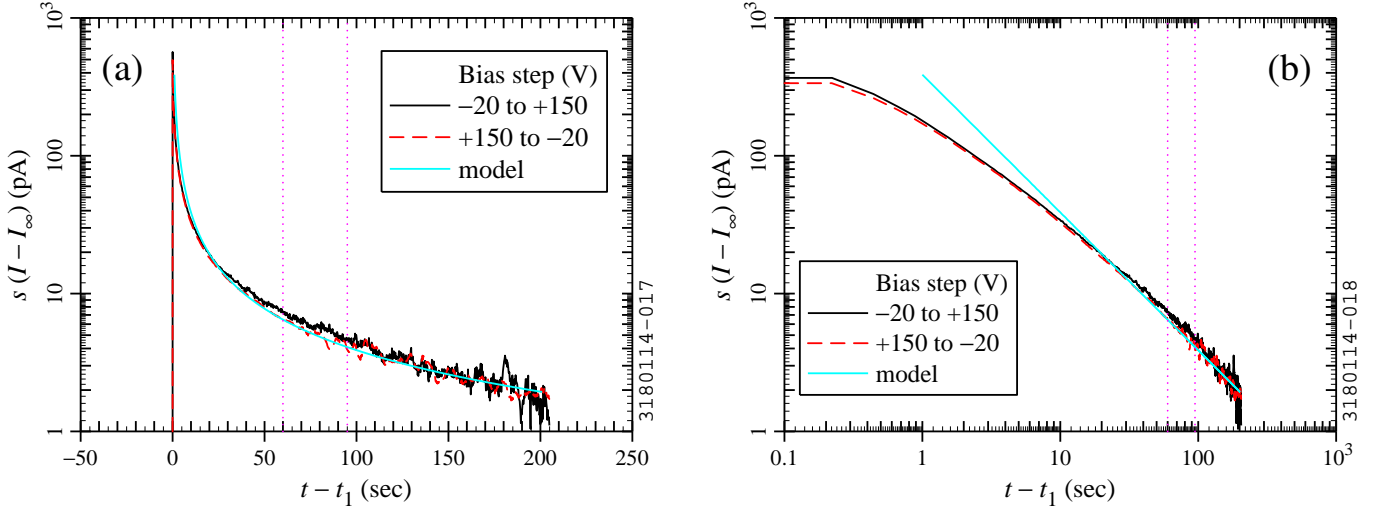

\centering
\GRAFwidth[0.49\textwidth]{490}{390}
\GRAFoffset{-65}{-65}
\GRAFlabelcoord{15}{255}
\incGRAFlabel{fig_trans170a}{(a)}%
\hspace*{\fill}
\GRAFlabelcoord{350}{255}
\incGRAFlabel{fig_trans170b}{(b)} \\[-2ex]

\caption{Measured current as a function of time for the 45\degree{}
  SEY station with a step in the sample bias at time $t_1$ between
  $-20$~V and $+150$~V, with (a) linear-log, and (b) log-log scale.
  The approximate step time $t_1$ is subtracted from $t$.  The steady
  state current $I_\infty$ is subtracted from $I(t)$ and the sign is
  adjusted ($s = \pm 1$) for comparison purposes.  The values of
  $I_\infty$ are 15~pA and $-2$~pA.  The solid cyan curve represents
  the model described in the text.  The dotted vertical lines indicate
  the time range of interest for SEY measurements with a double
  scan.\label{F:dIvtreal}}

\end{figure*}

To test the applicability of the semi-empirical model to a more
realistic case, we did additional measurements in which we switched
the bias between $-20$~V and $+150$~V, as is done for SEY scans.
\autoref{F:dIvtreal} shows results for the 45\degree{} system.  The
behaviour is similar to that of the simpler cases discussed above.
The cyan curves represent the semi-empirical model, which, as above,
fits the measurements reasonably well when $t - t_1 > 30$~seconds.
The best match corresponds to model parameters of $\Gamma_\parallel =
2.3$~pF and $R_\parallel = 10$~T$\Omega$, values that are slightly
different from those given in \autoref{T:TDpar}.

The results for the horizontal system and the off-line system are not
shown, but the behaviour was similar.  The best-match parameters were
similar to the values of \autoref{T:TDpar}.  In the case of the
off-line system, we oberved that a non-integer power relation of the
form $I(t) - I_\infty \sim (t - t_1)^{0.85}$ produced a better fit
than \cref{E:ICeRp}, but we chose to keep the semi-empirical
model simple rather than introduce additional parameters.

Although the model parameters obtained by switching between more
realistic bias values are slightly different from the values in
\autoref{T:TDpar} based on more simple measurements, they are all
consistent with the model parameters inferred from leakage scans,
which show some long-term variation in time (see \autoref{S:LCtrend} and
\autoref{F:LCtrend} below).

For the simple circuit model, the sample bias is not equal to the
power supply bias in general, but, if the series resistor $R_s$ is
small compared to the resistance $R_\parallel$ to ground, the sample
bias approaches the power supply voltage in steady state.  In the SEY
measurements with a double scan, however, the bias is switched from
low to high after 60 seconds and from high to low after 95 seconds.
\autoref{F:dIvtreal} indicates that these times are not long enough
for the system to reach steady state completely.  We do not attempt to
account for the possibility of ``long-term memory'' with the
semi-empirical model, as we found that the simple version is able to
correct for the transient effects reasonably well.  Typical leakage
scans show some transient behaviour during the first few iterations
(as was seen in \autoref{F:LCT}); it is possible that a more complete
model would be able to predict these effects at least in part.  The
approach we have taken is to keep the model relatively simple and scan
the leakage with enough iterations for the initial effects to settle
down.  Additionally, we make sure to set the bias to $-20$~V before
starting a leakage or SEY scan, so that all of the bias steps are
between that same voltages (as the initial step in the leakage and SEY
scans is to set the bias to $+150$~V to measure $I_p$).

The fact that we are accounting for the active response of the
picoammeter in an empirical way brings the question of how applicable
the model is for different situations.  In particular, when the
electron gun is turned on, the magnitude of the picoammeter current is
generally larger than it is for the leakage measurements.  To check
whether the electron gun affects the transient response of the system,
we did additional measurements of the sample current as a function of
time after a bias step with the electron gun on.  We used an Al sample
in the off-line station; we set the current for $I_p \approx 180$~pA
with $K_g = 300$~eV\@.  We found that $I(t) - I_\infty$ as a function
of time after a bias step (between $-20$~V and $+150$~V) was very
similar with the electron gun on and off, although $I_\infty$ was
quite different between the 2 cases.  We inferred from these
measurements that the picoammeter response is not qualitatively
different between the leakage scans and the SEY scans, and that it is
therefore reasonable to apply the semi-empirical model to correct for
transient effects in the SEY measurements.

\subsection{Inferring Model Parameters from a Leakage Scan\label{S:lcmodel}}

\cref{E:ICeRp} has two unknown parameters:
$\Gamma_\parallel$ and $R_\parallel$.  It is straightforward to infer
these parameters from a measurement of the current as a function of
time after stepping the bias (with known values of initial bias $V_0$,
final bias $V_2$, and step time $t_1$), as illustrated in
\autoref{F:dIvt} above.

However, our measurements have indicated that the leakage current can
vary by a small amount from day to day.  As described in
\autoref{S:LCmeas}, this led us to develop a procedure in which we
measure the leakage current prior to each SEY measurement, switching
between positive and negative sample bias in the same way as is done
for the SEY measurements.  We denote the positive and negative sample
biases as $V_{hi}$ and $V_{lo}$ (typically $V_{hi} = +150$~V and
$V_{lo} = -20$~V, as described in the text).  Our procedure is as
follows:
\begin{description}
\item[Step 1] We step the bias up from $V_{lo}$ to $V_{hi}$ at time $t_1 =
  t_{up}$, wait for time $\Delta t_{up}$, and measure the current:
\begin{equation}
I_{pl} = I(t = t_{up} + \Delta t_{up})\; .
\end{equation}
\item[Step 2] We step the bias down from $V_{hi}$ to $V_{lo}$ at time
  $t_{down}$, wait for time $\Delta t_{down}$, and measure the
  current:
\begin{equation}
I_{tl} = I(t = t_{down} + \Delta t_{down})\; .
\end{equation}
\item[Step 3] We infer the model parameters 
($\Gamma_\parallel$, $R_\parallel$)
from $I_{pl}$ and $I_{tl}$.
\item[Step 4] We use the model parameters to correct the values of
  $I_t$ and $I_p$ measured in the SEY scan.
\end{description}
Note that $I_{pl}$ is the leakage current with the sample bias used to
measure $I_p$ and $I_{tl}$ is the leakage current with the bias for
measuring $I_t$.  Likewise, $\Delta t_{up}$ and $\Delta t_{down}$
represent the waiting times between the change of bias and the
measurement of $I_p$ and $I_t$, respectively.  As indicated above, for
one grid point, $\Delta t_{up} = \Delta t_{down} = 60$~seconds
typically; for multiple grid points, we established the convention of
using the middle grid point of the first array, so $\Delta t_{down}
\approx 69$~seconds.

As described in \autoref{S:LCmeas}, the leakage scan is done in
the same way as the SEY scan, so that we typically alternate Step 1
and Step 2 over 16 iterations, and use the average values, discounting
initial transients.

For Step 3, we need to be able to infer the model parameters
$\Gamma_\parallel$ and $R_\parallel$ from the measured currents
$I_{pl}$ and $I_{tl}$.  Substituting the appropriate values of $t_1$,
$t$, $V_0$, and $V_2$ into \cref{E:ICeRp} for Step 1 and
Step 2 gives
\begin{eqnarray}
I_{pl} &=& \Gamma_\parallel \left(\frac{V_{hi} - V_{lo}}{\Delta t_{up}}\right) + \frac{V_{hi}}{R_\parallel}\\
I_{tl} &=& \Gamma_\parallel \left(\frac{V_{lo} - V_{hi}}{\Delta t_{down}}\right) + \frac{V_{lo}}{R_\parallel}
\end{eqnarray}
With 2 equations and 2 unknowns, we can solve for the model parameters
$\Gamma_\parallel$ and $R_\parallel$:
\begin{eqnarray}
\Gamma_\parallel &=& \frac{V_{lo} I_{pl} - V_{hi} I_{tl}}%
{\left(V_{hi} - V_{lo}\right)%
\left(\frac{V_{lo}}{\Delta t_{up}} + \frac{V_{hi}}{\Delta t_{down}}\right)}\label{E:Ce}\\
R_\parallel &=& \frac{V_{hi} \Delta t_{up} + V_{lo} \Delta t_{down}}%
{I_{pl} \Delta t_{up} + I_{tl} \Delta t_{down}}\label{E:Rp}
\end{eqnarray}
Thus, in Step 3, we use Equations (\ref{E:Ce}) and (\ref{E:Rp}) to
calculate $\Gamma_\parallel$ and $R_\parallel$ for a given leakage
scan.  Then, in Step 4, we use \cref{E:ICeRp} to calculate
the leakage current at the time of each $I_p$ and $I_t$ measurement in
the corresponding SEY scan, setting $t - t_1$ equal to the time
elapsed since the last change in sample bias, with appropriate values
for the initial bias ($V_0$) and final bias ($V_2$).  We subtract the
leakage current from each of the current measurements of the SEY scan
in order to calculate the corrected SEY.

\section{Leakage Current Measurements}

In the effort to mitigate leakage current while preparing for Phase
II, leakage current measurements with and without mitigation were done
in parallel with changes to the hardware and data acquisition
procedure.  Some of the measurements without mitigation were repeated
in Phase IIb to get a better understanding of the SEY stations'
behaviour in their final hardware state, as will be discussed in
\autoref{S:humid}.  The leakage measurements during Phase II
allowed us to get a more complete picture of the leakage current
stability with mitigation, as will be discussed in
\autoref{S:LCtrend}.

\subsection{Unmitigated Leakage Current: Correlation with Humidity;
  Time Dependence\label{S:humid}}

To better quantify the leakage current without mitigation from the dry
nitrogen gas blanket, we turned off the gas flow to the off-line SEY
station and did measurements at various humidities.  The humidity was
set by the outside air conditions as modified by the climate control
system.  We measured the humidity using a portable
hygrometer.\footnote{Chilled mirror hygrometer, Model 4189, Control
Company, Friendswood, TX.}  Over several weeks, the relative
humidity varied between 9\% and 46\%.

\begin{figure}[tb]
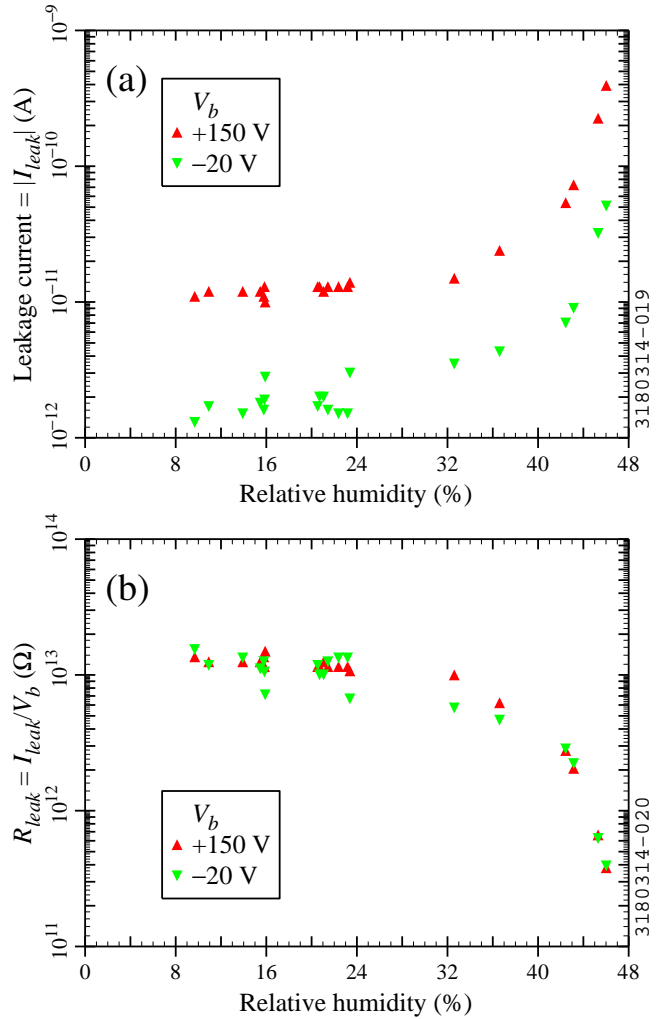

\centering
\GRAFwidth[\widewidth]{490}{390}
\GRAFoffset{-65}{-65}
\GRAFlabelcoord{15}{255}
\incGRAFlabel{fig_humid_a}{(a)}\\[-2ex]
\incGRAFlabel{fig_humid_b}{(b)}\\[-2ex]

\caption{Leakage measurements on the off-line SEY station without
  nitrogen gas flow: (a) leakage current as a function of humidity for
  positive and sample negative bias; (b) resistance to ground inferred
  from the leakage current.  The temperature was between 21.5 and
  23.5\degree C.\label{F:LChumid}}

\end{figure}

\autoref{F:LChumid}a shows the measured leakage current
($I_{leak}$) as a function of relative humidity for $V_b = 150$~V and
$-20$~V\@.  For relative humidities in excess of about 30\%, the
leakage current increases rapidly, changing by more than a factor of
10 between the lowest and highest humidities.  For relative humidities
below 30\% or so, the leakage current is low, though it still shows
some variation.  Some of this variation is due to not waiting long
enough after a change of bias (referring back to the leakage model of
\autoref{S:SEmodel}, ideally we should have $I_{leak} = I_\infty$,
but we waited for 1 to 6~minutes after setting the bias, rather than
an infinite time).  The low-humidity measurements may also be showing
some intrinsic variability in the electrical properties of the ceramic
breaks and stand-offs.  In some of the repeated low-current
measurements with $V_b = -20$~V, we observed that the initial
measurement had more leakage current than subsequent measurements made
after switching the bias to +150~V and back, which, as mentioned in
\autoref{S:LCmeas}, may be due to some insulator conditioning
effects (there is more noise at the lowest currents as well).

In \autoref{F:LChumid}b, we calculate the resistance to ground
($R_{leak}$) from the measured leakage current.  The calculated values
of $R_{leak}$ for positive and negative bias are roughly consistent;
again, referring back to the model of \autoref{S:SEmodel}, we should
ideally have $R_{leak} = R_\parallel$, but we can expect some
difference since our wait time was finite.  The highest values of
$R_{leak}$ are in the range of 10 to 20 T$\Omega$, consistent with the
values inferred for $R_\parallel$ in \ref{S:derive} (as well as
\autoref{S:LCtrend} below).  Hence, under low humidity conditions, the dry
nitrogen gas blanket is probably not required.

\autoref{F:LChumid} shows that the leakage correction will be large
for SEY measurements with low $I_p$ in a humid environment without a
gas blanket.  For example, with 46\% relative humidity, the leakage
current with $V_b +150$~V exceeds our nominal Phase II value of $I_p =
200$~pA\@.  The leakage current could be even higher: during the
preparations for Phase II, leakage current measurements were done at
relative humidities of up to 54\%.  The corresponding value of
$R_{leak}$ is of order $2 \cdot 10^{10}$~$\Omega$; at this humidity
level, the leakage current exceeds even the nominal value of $I_p =
2$~nA used for Phase I SEY measurements.

\begin{figure}[tb]
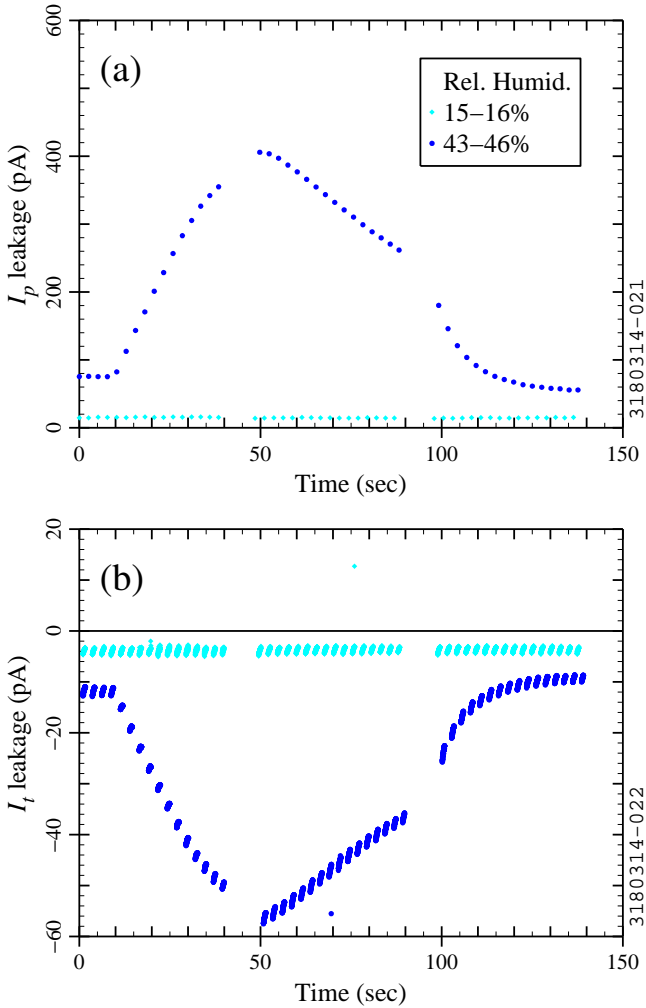

\centering
\GRAFwidth[\widewidth]{490}{390}
\GRAFoffset{-65}{-65}
\GRAFlabelcoord{15}{255}
\incGRAFlabel{fig_scan_rh_a}{(a)}\\[-2ex]
\incGRAFlabel{fig_scan_rh_b}{(b)}\\[-2ex]

\caption{Leakage scans on the off-line SEY station without nitrogen
  gas flow at different ambient humidities: (a) leakage measurements
  with $V_b = +150$~V for $I_p$ correction; (b) repeated leakage
  measurements with $V_b = -20$~V for $I_t$ correction.  The
  temperature was between 21.5 and 23\degree C.\label{F:LClowmedH}}

\end{figure}

Beyond the problem of high leakage current, there is the problem of
leakage current stability.  For relative humidities above 35\% or so,
\autoref{F:LChumid} shows that a small change in humidity
produces a large change in leakage current.  This implies that SEY
measurements with low $I_p$ in a humid environment without a gas
blanket are likely to have large systematic errors if there are small
variations in the humidity over the course of the measurement.  This
problem is illustrated by \autoref{F:LClowmedH}, which shows
leakage scans done on the off-line SEY station without gas flow, at
low and medium ambient humidity.  Three leakage scans were done for
each humidity level, for a total measurement time of about 140 minutes
(which is a bit longer than the 110 minutes needed for a Phase IIb SEY
scan).  In the low humidity measurements (shown in light blue), the
relative humidity decreased from 15.9\% to 15.5\% over the course of
the scans.  The leakage current remains low and stable; the behaviour
is similar to scans with the nitrogen gas blanket.  In the
measurements with high relative humidity (dark blue), the relative
humidity was initially 43.1\%, increasing to 46.0\% in the first scan,
decreasing to 45.3\% in the second scan, and further decreasing to
42.5\% in the last scan.  Correspondingly, the leakage current varies
by about a factor of 4 over the course of the measurement.  These
measurements support our inference that measurements in a humid
environment are prone to poor leakage current stability.  Such
variation in the leakage current during an SEY scan would introduce
large errors in the results.

We conclude that, in a dry environment in which the relative humidity
remains below 30\%, the leakage current for our SEY stations is
relatively stable, and a nitrogen gas blanket is not needed.  If the
relative humidity can exceed 30\%, a dry nitrogen blanket is useful to
ensure low and stable leakage current.  However, a dry environment may
not remove the need to measure the leakage current in conjunction with
SEY measurements; even with a nitrogen blanket, we found that it is
necessary to measure the leakage current, as will be discussed in the
next section.

\subsection{Mitigated Leakage Current: Long-Term Trends\label{S:LCtrend}}

\begin{figure}
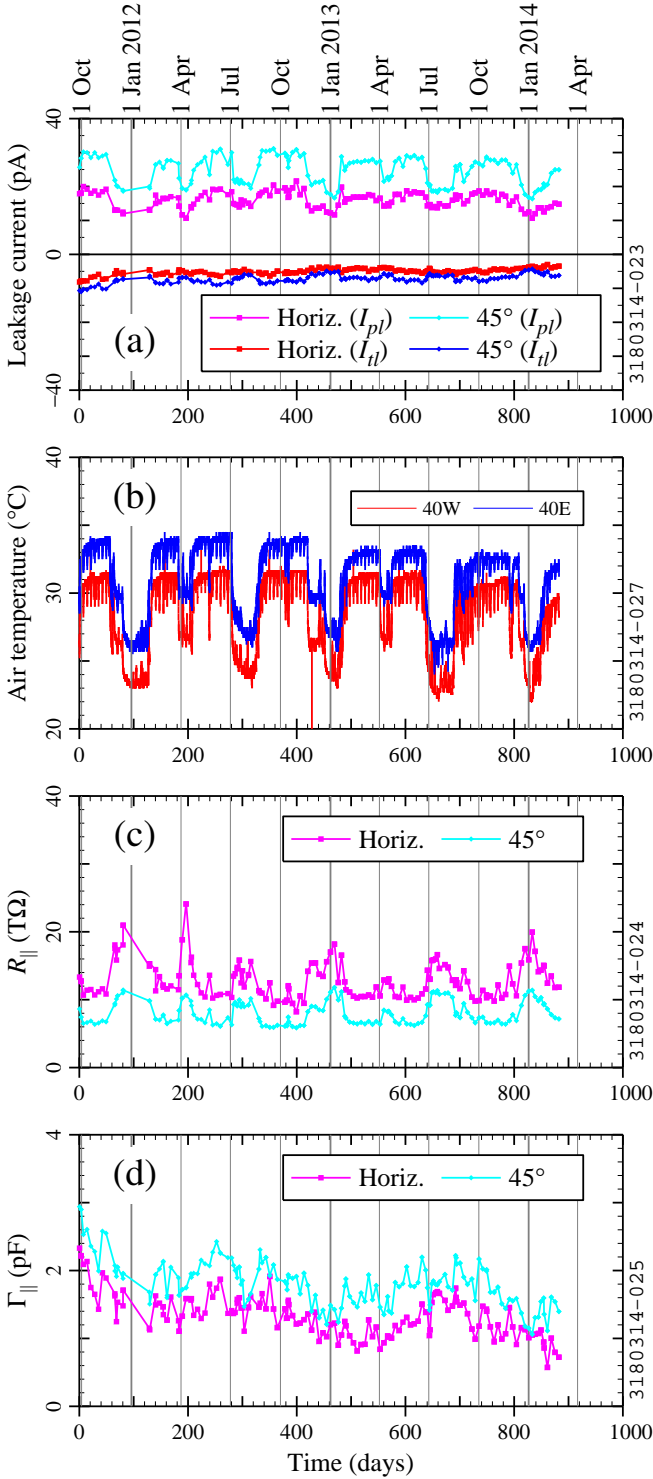

\centering
\GRAFwidth[\narrowwidth]{490}{313}
\GRAFoffset{-65}{-25}
\GRAFlabelcoord{15}{20}
\GRAFlabelbox{40}{30}
\incGRAFboxlabel{fig_trend_a}{(a)}\\[-0.3ex]
\GRAFwidth[\narrowwidth]{490}{250}
\GRAFlabelcoord{15}{155}
\incGRAFboxlabel{fig_trend_b}{(b)}\\[-0.3ex]
\incGRAFboxlabel{fig_trend_c}{(c)}\\[-0.3ex]
\GRAFwidth[\narrowwidth]{490}{290}
\GRAFoffset{-65}{-65}
\incGRAFboxlabel{fig_trend_d}{(d)}\\[-2ex]

\caption{Comparison of long-term trends: (a) measured leakage
  currents, (b) tunnel air temperature, (c, d) leakage model
  parameters as a function of time.  The gray lines correspond to
  quarterly calendar dates.\label{F:LCtrend}}

\end{figure}

As described in \autoref{S:LCmit}, we use a nitrogen gas blanket
to shield the SEY stations' ceramic breaks from ambient moisture in
Phase II\@.  In addition to using this leakage mitigation blanket, we
do a leakage scan on both in-situ SEY stations prior to each SEY scan,
as discussed in \autoref{S:LCmeas}.  \autoref{F:LCtrend}a shows
the measured leakage currents during Phase II\@.  The measurements
were done over approximately 2 years and 5 months, starting on 27
September 2011 (time = 0) and ending on 25 February 2014 (time = 882
days).  Using the procedure and notation of \autoref{S:lcmodel}, the
values labelled $I_{pl}$ were measured with a bias of $V_{hi} =
+150$~V; the values labelled $I_{tl}$ were measured with a bias of
$V_{lo} = -20$~V\@.  As can be seen in \autoref{F:LCtrend}a, the
leakage current has varied by roughly a factor of 2 over the time of
Phase II measurements.  However, the leakage current does not increase
or decrease steadily, as we would expect if the ceramic was
deteriorating or cleaning itself up.  The leakage current changes
enough over time to make repeated leakage scans necessary---a
constant-leakage assumption would introduce significant systematic
errors into the SEY calculation.

Although the leakage currents do not show a clear seasonal dependence,
they do not vary randomly.  \autoref{F:LCtrend}a suggests that the
leakage current tends to be higher when the accelerator is running and
lower during summer and winter down periods.  One thing that changes
significantly between high-current operation and down periods is the
air temperature in the tunnel (including the L3 area where the SEY
stations are located).  This is illustrated in
\autoref{F:LCtrend}b, which shows the air temperature measured with
thermocouple gauges at two locations in the tunnel (40E is
approximately 10~m East of the L3 area and 40W is approximately 10~m
West of L3).  The air temperature increases by about 8\degree C when
CHESS currents are stored due to ohmic losses in the magnets and
synchrotron radiation power.  Comparison of \autoref{F:LCtrend}a and
\autoref{F:LCtrend}b shows that the leakage current and the tunnel
temperature are indeed correlated.  This correlation could come about
if the leakage properties of the ceramic or stand-offs are
temperature-dependent, if the moisture content of the nitrogen gas
blanket is temperature-sensitive, or through some other mechanism.

As discussed in \autoref{S:trans}, the Phase II measurement
procedure is such that the waiting time after a bias change is not
long enough for the current to reach its steady state value.  As a
result, the leakage current values in \autoref{F:LCtrend}a are
affected by the waiting time.  The wait time for the $I_{pl}$
measurement ($\Delta t_{up}$) was about 60~s throughout Phase II\@.
As discussed in \autoref{S:lcmodel}, the $I_{tl}$ wait time
($\Delta t_{down}$) increased from about 60~s to about 69~s as we
increased the number of grid points in Phase II\@.  To remove the
effect of the transient current, we can express the leakage current in
terms of the leakage model parameters of \cref{E:ICeRp}:
$R_\parallel$ (resistance to ground) and $\Gamma_\parallel$ (the
capacitance-like parameter).

\autoref{F:LCtrend}c and \autoref{F:LCtrend}d show the model
parameters for Phase II\@.  These are calculated from the measured
leakage currents of \autoref{F:LCtrend}a using the procedure of
\autoref{S:lcmodel}.  The resistance to ground varies between 5
T$\Omega$ and 25 T$\Omega$, and shows a clear inverse correlation with
the measured leakage currents, as one would expect; $\Gamma_\parallel$
also shows a time dependence, though it is more difficult to
interpret.  \autoref{F:LCtrend}d suggests that $\Gamma_\parallel$
might have some seasonal correlation and possibly a slight downward
trend.  We would not expect the capacitance to ground to vary
significantly over time---the variation in $\Gamma_\parallel$ might be
an artifact of the semi-empirical nature of the model for the
transient response of the system.

\section{Inter-System Timing for Simultaneous SEY Scans\label{S:TimingIIb}}

As discussed in \autoref{S:cross}, a bias change for one station
produces a current spike for the other station, and hence it is
important to ensure that bias changes do not happen when we are
measuring the current.  Timing details for the case of Phase IIb
parameters are provided in this section.

\begin{figure}[tbh]
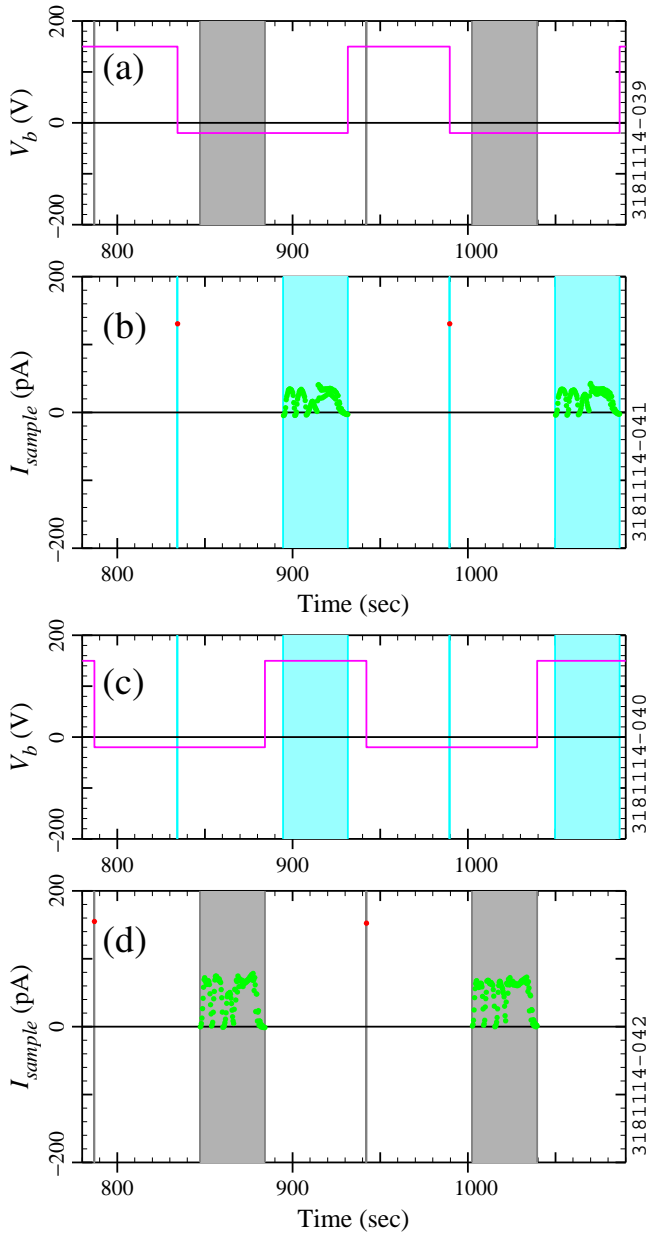

\centering
\begin{tabular}{c}
\GRAFwidth[\narrowwidth]{490}{205}
\GRAFoffset{-65}{-65}
\GRAFlabelcoord{15}{70}
\incGRAFlabel{fig_exam_vb_t_al2d}{(a)}\\[-2.5ex]
\GRAFwidth[\narrowwidth]{490}{290}
\GRAFoffset{-65}{-65}
\GRAFlabelcoord{15}{155}
\incGRAFlabel{fig_exam_i_t_al2d}{(b)}\\[-4ex]
\GRAFwidth[\narrowwidth]{490}{205}
\GRAFoffset{-65}{-65}
\GRAFlabelcoord{15}{70}
\incGRAFlabel{fig_exam_vb_t_al2h}{(c)}\\[-2.5ex]
\GRAFwidth[\narrowwidth]{490}{290}
\GRAFoffset{-65}{-65}
\GRAFlabelcoord{15}{155}
\incGRAFlabel{fig_exam_i_t_al2h}{(d)}\\[-2ex]
\end{tabular}

\caption{Inter-system timing schematic for SEY scans in Phase IIb: (a)
  bias and (b) current measurements for 45$\degree$ system; (c) bias
  and (d) current measurements for horizontal system.  Two iterations
  in the energy scan are shown.  Red markers: $I_p$ measurements;
  green markers: $I_t$ measurements.  Shaded areas: ``quiet zones''
  for current measurements (light blue: times of current measurements
  on $45\degree$ sample; gray: times of current measurements on
  horizontal sample).\label{F:IVvsTDH}}

\end{figure}

\autoref{F:IVvsTDH} shows an example of simultaneous SEY scans with
both stations using Phase IIb parameters.  \autoref{F:IVvsTDH}a shows
the bias on the $45\degree$ sample as a function of time.  A time
$t_{bw} = 60$~s after a change in bias, the $45\degree$ sample current
is measured, as shown in \autoref{F:IVvsTDH}b.  The $I_p$ measurements
are done once per energy iteration (red markers), while $I_t$ is
measured for 120 grid points (green markers).  The $I_t$ values vary
between 0 and 40 pA as a function of grid point for the $45\degree$
sample in this example.

\autoref{F:IVvsTDH}c shows the bias on the horizontal sample as a
function of time.  To ensure that there is no cross-talk, the bias on
the horizontal sample must be constant during a ``quiet zone'' when
current measurements are done on the $45\degree$ sample, indicated by
light blue shading in \autoref{F:IVvsTDH}b and \autoref{F:IVvsTDH}c.
A time $t_{bw}$ after a change in bias on the horizontal sample,
current measurements are done, as shown in \autoref{F:IVvsTDH}d.  The
bias on the $45\degree$ sample must be constant when current
measurements are done on the horizontal sample, indicated by gray
shading in \autoref{F:IVvsTDH}d and \autoref{F:IVvsTDH}a.  The
best-case scenario is a start delay of about 47~s between the two
systems, which is the case shown in \autoref{F:IVvsTDH}.  As can be
seen, with this start delay, the timing margin is about $\pm 12$~s.
(For high definition scans, the $I_t$ measurements take 118~s, so
measurements on the horizontal and 45\degree{} samples are done
sequentially instead of in parallel.)



\bibliographystyle{medium}
\bibliography{cta_sey_im_long}


\end{document}